\documentclass[a4paper,12pt]{article}
 
\PassOptionsToPackage{original}{pict2e}
\usepackage{amsmath}
\usepackage{dsfont}
\usepackage{amsfonts}
\usepackage{amssymb}
\usepackage{color}
\usepackage{mathrsfs}
\usepackage{calc}
\usepackage{colonequals}
\usepackage{lmodern}
\usepackage[english]{babel}
\usepackage{xcolor}
\usepackage{upgreek}
\usepackage{babel,blindtext,scrpage2,eso-pic}
\usepackage{graphicx,pspicture}
\usepackage{colortbl}
\usepackage{mathtools}
\usepackage{array}
\usepackage{wrapfig}
\usepackage[makeroom]{cancel}
\usepackage{parcolumns}
\usepackage{framed}
\usepackage{booktabs}
\usepackage{psfrag}
\usepackage[margin=10pt,font=small,labelfont=bf,
labelsep=period]{caption}
\usepackage[bottom]{footmisc} 
\usepackage{indentfirst} 
\usepackage{upref} 
\usepackage{cite} 
\usepackage[rightcaption]{sidecap}
\usepackage{psfrag}
\usepackage{subfig}
\usepackage{float}
\usepackage{hyperref}

\definecolor{rhsFormColor}{rgb}{0.05,0.05,0.35}
\definecolor{Formel}{rgb}{0.05,0.05,0.35}
\definecolor{Formel2}{rgb}{0.40,0.40,0.55}
\definecolor{SideNote}{rgb}{0.35,0.35,0.4}
\definecolor{fSubDeriv}{rgb}{0.1,0.1,0.1}
\definecolor{fSubDerivA}{rgb}{0.35,0.35,0.50}
\definecolor{Text}{rgb}{0,0,0}
\definecolor{can1}{rgb}{0.7,0.1,0.1}
\definecolor{can2}{rgb}{0.45,0.45,0.15}
\definecolor{can3}{rgb}{0.05,0.3,0.05}
\definecolor{can4}{rgb}{0.5,0.1,0.1}
\definecolor{can5}{rgb}{0.1,0.5,0.1}
\definecolor{can6}{rgb}{0.15,0.45,0.45}
\definecolor{can7}{rgb}{0.3,0.05,0.05}
\definecolor{can8}{rgb}{0.35,0.1,0.35}
\definecolor{can9}{rgb}{0.5,0.3,0.4}
\definecolor{SumWeylCol}{rgb}{0.05,0.05,0.35}
\definecolor{SumGeoCol}{rgb}{0.05,0.05,0.35}
\colorlet{TableR1}{Formel}
\colorlet{TableR2}{Formel!75}
\colorlet{TableC1}{SideNote}
\definecolor{TableC2}{rgb}{0.8,0.8,0.8}
\colorlet{TableC3}{SideNote!10}
\definecolor{TableNot}{rgb}{0.8,0.8,0.8}
\definecolor{TableYes}{rgb}{1,1,1}
\definecolor{pageNumberC}{rgb}{0,0,0}

\numberwithin{equation}{section}

\renewcommand{\baselinestretch}{1.248} 
\DeclareMathOperator*{\SumInt}{%
\mathchoice%
  {\ooalign{$\displaystyle\sum$\cr\hidewidth$\displaystyle\int$\hidewidth\cr}}
  {\ooalign{\raisebox{.14\height}{\scalebox{.7}{$\textstyle\sum$}}\cr\hidewidth$\textstyle\int$\hidewidth\cr}}
  {\ooalign{\raisebox{.2\height}{\scalebox{.6}{$\scriptstyle\sum$}}\cr$\scriptstyle\int$\cr}}
  {\ooalign{\raisebox{.2\height}{\scalebox{.6}{$\scriptstyle\sum$}}\cr$\scriptstyle\int$\cr}}
}

\usepackage{tikz}
\usetikzlibrary{fit}

\newcommand\raalign[2]{
   \tikz[overlay,every node/.style={inner sep=0pt,outer sep=0pt}]%
      \node[anchor=base west](g){\phantom{$\displaystyle #1\null=\null#2$}}%
         node[draw=blue!50!black,fit=(g),inner sep=5pt]{};%
   #1} 

\newcommand{\crg}{{\rm CR}}
\newcommand{\nBound}{\mathcal{N}}
\newcommand{\scon}{\text{\rm sc}}

\newcommand{\vol}{\text{\rm{Vol}}}

\makeatletter
\renewcommand*\env@matrix[1][*\c@MaxMatrixCols c]{%
  \hskip -\arraycolsep
  \let\@ifnextchar\new@ifnextchar
  \array{#1}}
\makeatother

\makeatletter
\def\hlinewd#1{%
\noalign{\ifnum0=`}\fi\hrule \@height #1 %
\futurelet\reserved@a\@xhline}
\makeatother

\renewcommand{\epsilon}{\varepsilon}

\newcommand{\cF}{\text{\rm Y}}
\newcommand{\spsym}{{\rm \text{split-sym}}}

\renewcommand{\vec}[1]{\boldsymbol{#1}}

\newcommand{\cix}{I}

\makeatletter

\newcommand{\Rmnum}[1]{\expandafter\@slowromancap\romannumeral #1@}
\makeatother

\newcommand{\noR}{\mathds{R}}

\newcommand{\noC}{\mathds{C}}

\newcommand{\MaFs}{\mathcal{M}}  

\newcommand{\Ric}{R}     
\newcommand{\SR}{R}     
\newcommand{\SRb}{\bar{\SR}}     
\newcommand{\Z}{D}      

\newcommand{\bg}{\bar{g}}   
\newcommand{\bZ}{\bar{\Z}}   

\newcommand{\Tr}{\text{Tr}}   
\newcommand{\Order}[1]{\mathcal{O}(#1)}  

\newcommand{\md}{\text{d}}
 
\newcommand{\NMW}{1cm} 
\newcommand{\FEW}{0.3cm} 
 
\newcommand{\const}{\text{const}}    
 
\definecolor{shadecolor}{rgb}{0.98,.98,1}
\definecolor{framecolor}{rgb}{0.5,0.5,1}
\definecolor{shadecolorII}{rgb}{0.98,0.98,0.98}
\definecolor{framecolorII}{rgb}{0.6,0.6,0.6}

\newenvironment{bshaded}{%
 \MakeFramed {\FrameRestore}}%
 {\endMakeFramed}

 {\endMakeFramed}

\newcommand{\EAA}{\Gamma}  
 
\newcommand{\Ghx}{\xi}   
\newcommand{\GhAx}{\bar{\xi}}  

\newcommand{\CosmConst}{\lambda}
\newcommand{\Kk}{\CosmConst}  
\newcommand{\Kkbar}{\Lambda}  

\newcommand{\KkbarD}{\Kkbar_k^{\dyn}}  
\newcommand{\KkbarB}{\Kkbar_k^{\background}}  

\newcommand{\background}{\text{B}}
\newcommand{\dyn}{\text{Dyn}}

\newcommand{\sm}{\text{sm}}

\newcommand{\KkB}{\CosmConst^{\background}}  
\newcommand{\KkD}{\CosmConst^{\dyn}} 

\newcommand{\attr}{\bullet}

\newcommand{\IR}{{\rm IR} }
\newcommand{\UV}{{\rm UV} }
 
\newcommand{\tg}{g}

\newcommand{\nkD}{G_k^{\dyn}}
\newcommand{\nkbB}{G_k^{\background}}

\newcommand{\flcb}{h}

\newcommand{\cfunc}{\mathscr{C}}

\begin{document}

\addtolength{\topmargin}{-10mm} 
\begin{titlepage}
\enlargethispage{1cm}
\renewcommand{\baselinestretch}{1.1}  
\title{\begin{flushright}
\normalsize{MITP\slash 14-083}
\vspace{1cm}
\end{flushright}
Towards a $C$-function in 4D quantum gravity}
\date{}
\author{Daniel Becker and Martin Reuter\\
{\small Institute of Physics, University of Mainz}\\[-0.2cm]
{\small Staudingerweg 7, D-55099 Mainz, Germany}}
\maketitle\thispagestyle{empty}

\begin{abstract} 
We develop a generally applicable method for constructing functions, $\cfunc$, which have properties similar to Zamolodchikov's $C$-function, and are geometrically natural objects related to the theory space explored by  non-perturbative functional renormalization group (RG) equations.
Employing the Euclidean framework of the Effective Average Action (EAA), we propose a $\cfunc$-function which can be defined for arbitrary systems of gravitational, Yang-Mills, ghost, and bosonic matter fields, and in any number of spacetime dimensions.
It becomes stationary both at critical points and in classical regimes, and decreases monotonically along RG trajectories provided the breaking of the split-symmetry which relates  background and quantum fields is sufficiently weak.
Within the Asymptotic Safety approach we test the proposal for Quantum Einstein Gravity in $d>2$ dimensions, performing detailed numerical investigations in $d=4$.
We find that the bi-metric Einstein-Hilbert truncation of theory space introduced recently is general enough to yield perfect monotonicity along the RG trajectories, while its more familiar single-metric analog fails to achieve this behavior which we expect on general grounds.
Investigating generalized crossover trajectories connecting a fixed point in the ultraviolet to a classical regime with positive cosmological constant in the infrared, the $\cfunc$-function is shown to depend on the choice of the gravitational instanton which constitutes the  background spacetime.
For de Sitter space in 4 dimensions, the Bekenstein-Hawking entropy is found to play a role analogous to the central charge in conformal field theory.
We also comment on the idea of a `$\Lambda$-$N$ connection' and the `$N$-bound' discussed earlier. 
\end{abstract}

\end{titlepage}
\addtolength{\topmargin}{10mm} 
\tableofcontents
\section{Introduction}
One of the most remarkable results in 2-dimensional conformal field theory is  Zamolodchikov's $c$-theorem \cite{Zamol-c}.
It states that every 2D Euclidean quantum field theory with reflection positivity, rotational invariance, and a conserved energy momentum tensor possesses a function $C$ of its coupling constants, which is non-increasing along the renormalization group trajectories and is stationary at fixed  points where it equals the central charge of the corresponding conformal field theory.

After the advent of this theorem many authors tried to find a generalization that would be valid also in dimensions greater than two \cite{Cardy88,Osborn,Neto-Fradkin-93,Klebanov-Pufu-Safdi-11,Cappelli-Friedan-Latorre,Shore,Bastianelli,Anselmi98}.
This includes, for instance, suggestions by Cardy \cite{Cardy88} to integrate the trace anomaly of the energy-momentum tensor $\langle T_{\mu\nu} \rangle$ over a 4-sphere of unit radius, $C\propto \int_{S^4}\md^4 x\sqrt{g}\, \langle T^{\mu}_{\mu} \rangle$, the work of Osborn \cite{Osborn}, and ideas based on the similarity of $C$ to the thermodynamical free energy \cite{Neto-Fradkin-93}, leading to a conjectural `$F$-theorem' which states that, under certain conditions, the finite part of the free energy of 3-dimensional field theories on $S^3$ decreases along RG trajectories and is stationary at criticality \cite{Klebanov-Pufu-Safdi-11}.
Cappelli, Friedan and Latorre \cite{Cappelli-Friedan-Latorre} proposed to define a $C$-function on the basis of the spectral representation of the 2-point function of the energy-momentum tensor.

While these investigations led to many important insights into the expected structure of the hypothetical higher-dimensional $C$-function, the search was successful only recently \cite{Komargo-Schwimmer,Lut-Pol-Rattaz} with the proof of the `$a$-theorem' \cite{Cardy88,Anselmi98}.
According to the $a$-theorem, the coefficient of the Euler form term in the induced gravity action of a 4D theory in a curved, but classical, background spacetime is non-increasing along RG-trajectories.

Clearly theorems of this type are extremely valuable as they provide non-perturbative information about quantum field theories or statistical systems in the strong coupling domain.
They constrain the structure of possible RG flows on theory space, and they rule out exotic behavior such as limit cycles, for instance (at least for a suitable class of beta-functions \cite{Curt-Jin-Zach}).

In this paper we describe and test a broadly applicable search strategy by means of which generalized `$C$-like' functions could be identified  under rather general conditions, in particular in cases where the known $c$- and the $a$-theorems do not apply.
Our main motivation is in fact theories which include quantized gravity, in particular those based upon the Asymptotic Safety construction \cite{wein,mr,NJP,livrev,alfio-As-infl}.

In a first step, we try to generalize only one specific feature of Zamolodchikov's $C$-function for a generic field theory in any number of dimensions, namely its `counting property': the sought-after function should roughly be equal to (or at least in a known way be related to) the number of degrees of freedom that are integrated out along the RG trajectory when the scale is lowered from the ultraviolet (UV) towards the infrared (IR).
Technically, we shall do this by introducing a higher-derivative mode-suppression factor in the underlying functional integral which acts as an IR cutoff.
We can then take advantage of the well established framework of the effective average action (EAA) to control the scale dependence \cite{wett-mr}.

In a generic theory comprising a certain set of dynamical fields, $\Phi$, and corresponding background fields, $\bar{\Phi}$, the EAA is a `running' functional $\Gamma_k[\Phi,\bar{\Phi}]$  similar to the standard effective action, but with a built-in IR cutoff at a variable mass scale $k$.
Its $k$-dependence is governed by an exact functional RG equation (FRGE).

The specific property of the EAA which will play a crucial role in our approach is the following:
For a broad class of theories, those that are `positive' in a sense we shall explain, the very structure of the FRGE implies that the EAA is a monotonically increasing function of the IR cutoff:
\begin{align}
\partial_k \Gamma_k[\Phi,\bar{\Phi}]\geq 0 \quad \forall\, (\Phi,\bar{\Phi}) 
\label{eqn:cm_in_001}
\end{align}
We shall refer to this property as the {\it pointwise monotonicity} of the EAA since it applies at all points  $(\Phi,\bar{\Phi})$ of field space independently.
Thus the EAA provides us with even infinitely many monotone functions of the RG scale $k$, one for each field configuration $(\Phi,\bar{\Phi})$.
So one might wonder if those functions, or a combination thereof, deserve being considered  a generalization of Zamolodchikov's $C$-function.

Unfortunately the answer is negative, for the following reason.
The pointwise monotonicity property refers to $\Gamma_k$ when it is evaluated at {\it dimensionful} field arguments, and is parametrized by  dimensionful running coupling constants.
However, the $c$-theorem and its  generalizations apply to the RG flow on {\it theory space,} $\mathcal{T}$, a manifold which is coordinatized by the {\it dimensionless} couplings.
The latter differ from the dimensionful ones by explicit powers of $k$ fixed by the canonical scaling dimensions.
As a consequence, when rewritten in terms of dimensionless fields and couplings, the property \eqref{eqn:cm_in_001} does not precisely translate into a monotonicity statement about the de-dimensionalized theory space analog of $\Gamma_k$, henceforth denoted $\mathcal{A}_k$.
Rather, when the derivative $\partial_k$ hits the explicit powers of $k$, additional canonical scaling terms arise which prevent us from concluding simply `by inspection' that $\mathcal{A}_k$ is monotone along RG trajectories.

Nevertheless, our main strategy will be to take maximum advantage of the pointwise monotonicity of $\Gamma_k$ as the primary input, and then try to get a handle on the  monotonicity violations that  occur in going from $\Gamma_k$ to $\mathcal{A}_k$.
We shall see that by evaluating the action functionals on special field configurations this difficulty can be reduced to a far more tractable level.

A related issue is that $\Gamma_k$, while having attractive monotonicity  properties, does not in addition also become stationary at fixed points, as a sensible generalization of the $C$-function in 2D should.
But, again for the above reason, this is not necessarily a drawback since at an RG fixed point it is anyhow not the dimensionful couplings but rather the dimensionless coordinates of theory space that are supposed to assume stationary values, i.e. actually it is  $\mathcal{A}_k$ that approaches a `fixed functional', $\mathcal{A}_*$.

When we look at $\Gamma_k$ and $\mathcal{A}_k$ pointwise, or equivalently, at the infinitely many $k$-dependent couplings  they parametrize them independently, then there is a clash between stationarity at fixed points and monotonicity along the RG flow:
$\mathcal{A}_k$ is stationary at fixed points, but not monotone, while for $\Gamma_k$ the situation is the other way around.
However, by adopting the pointwise perspective we are expecting by far too much, namely that all dimensionless couplings individually  behave like a $C$-function.
Presumably we can hope to find at best a single, or perhaps a few, real valued quantities with all desired properties.
We shall denote such a hypothetical function by $\cfunc_k$ in the following.
Assuming it exists, the quantity $\cfunc_k$ is {\it one} function depending on {\it infinitely many} running couplings along the RG trajectory, so the transition from $\EAA_k$ to $\cfunc_k$ amounts to a tremendous `data reduction'.

Thus, within the EAA framework, the central question is whether there exists a kind of `essentially universal' map from $k$-dependent functionals $\Gamma_k$ to a function $\cfunc_k$ that is monotone along the flow and stationary at fixed points.
Here the term `universal' is to indicate that we would require only a few general properties to be satisfied, comparable to reflection positivity, rotational invariance, etc. in the case of Zamolodchikov's theorem.
The reason why we believe that there should exist such a map is that the respective monotonicity properties of $\Gamma_k$ and the $C$-function in 2D have essentially the same  simple origin.
They both `count' in a certain way the degrees of freedom (more precisely: fluctuation modes) that are integrated out  at a given RG scale intermediate between the UV and the IR.

We begin by considering a class of $\cfunc_k$-candidates which are obtained by evaluating $\Gamma_k[\Phi,\bar{\Phi}]$ at a particularly chosen pair of arguments $(\Phi,\bar{\Phi})$ that will have an explicit dependence on $k$ in general. 
In the present paper we propose to use {\it self-consistent background field configurations} for this purpose.
We evaluate the EAA at a scale dependent point in field space, namely $\Phi=\bar{\Phi}\equiv \bar{\Phi}_k^{\scon}$.
By definition, a background field $\bar{\Phi}\equiv \bar{\Phi}_k^{\scon}$ is said to be self-consistent (`$\scon$') if the equation of motion for the dynamical field $\Phi$ that is implied by $\Gamma_k$ admits the solution $\Phi=\bar{\Phi}$.
With other words, if the system is put in a  background which  is  self-consistent, the fluctuations of the dynamical field, $\varphi\equiv \Phi-\bar{\Phi}$, have zero expectation value and, in this sense, do not modify this special background.
As we shall demonstrate in detail in section \ref{sec:02}, the proposal $\cfunc_k=\Gamma_k[\bar{\Phi}_k^{\scon},\bar{\Phi}_k^{\scon}]$ is indeed a quite promising candidate for a generalized $C$-function.
It is stationary at fixed points and it is `close to' being monotonically decreasing along the flow.

The phrase `close to' requires an explanation.
Especially in quantum gravity, Background Independence is a central requirement \cite{Kiefer,ARR}.
While in the causal dynamical triangulation approach \cite{CDT} or in loop quantum gravity \cite{A,R,T}, for instance, this requirement is met by strictly not using a background spacetime structure at all, the EAA framework uses the background field technique \cite{dewittbook}.
At the intermediate steps of the quantization one does introduce a background spacetime, equipped with a non-degenerate background metric in particular, but at the same time one  makes sure that no observable prediction will depend on it.\footnote{The construction in section \ref{sec:02} will force us to deal with a background metric and include it into the set $\bar{\Phi}$ even when analyzing pure matter theories on a classical (for instance, flat) spacetime.}
This can be done by means of the Ward identities pertaining to the  split-symmetry \cite{gluco, elisa2, creh2} which governs the interrelation between $\varphi$ and $\bar{\Phi}$.
This symmetry, if intact, ensures that the physical contents of a theory is independent of the chosen background structures.
Usually, at the `off-shell' level of $\Gamma_k$, in particular when $k>0$, the symmetry is broken by the gauge fixing and cutoff terms in the bare action.
Insisting on unbroken split-symmetry in the physical sector restricts the admissible RG trajectories the EAA may follow \cite{daniel2}.

The wording `close to' used above has the precise meaning that {\it in the idealized situation of  negligible split-symmetry violation the monotonicity of $\cfunc_k$ is manifest for the entire class of theories with pointwise monotonicity of the EAA.}
This can indeed be seen without embarking on any complicated analysis, whose outcome could then possibly depend on the type of theory under consideration.
Instead, in reality where the breaking of split-symmetry often is an issue, such an analysis is indeed necessary in order to check whether or not the split-symmetry violation is  strong enough to destroy the monotone behavior of $\cfunc_k$.
We believe that under weak conditions whose precise form needs to be found, the monotonicity is not destroyed so that the proposed $\cfunc_k$ indeed complies both with the stationarity and the monotonicity requirement.

In the present paper, rather than attempting a general proof we investigate a concrete system, asymptotically safe gravity in $d$ dimensions, $d=4$ in particular, determine its RG flow, and explore the properties of the resulting function $\cfunc_k$.
Clearly, for practical reasons we can study the flow only on a truncation of the a priori infinite dimensional theory space.
Nevertheless, we shall be able to demonstrate that, provided the truncation is sufficiently refined, the corresponding approximation to the exact $\cfunc_k$ is indeed a non-decreasing function of $k$.
It will be quite impressive to see how non-trivially the various components of the truncation ansatz for $\EAA_k$ must, and actually do conspire in order to produce this result.

Concretely we shall explore Quantum Einstein Gravity (QEG) both within the familiar (`single-metric') Einstein-Hilbert-truncation \cite{mr} in which the background metric appears only in the gauge fixing part of  the action, as well as within a more refined `bi-metric truncation' \cite{elisa2,MRS1} where the EAA can depend on it via a second Einstein-Hilbert term, over and above the one for the dynamical metric. The beta-functions for this `bi-metric Einstein-Hilbert truncation'  have been derived recently in \cite{MRS2, daniel2} and form the basis of the present analysis.

We emphasize that the goal of the present investigation is not, or at least not primarily, to reanalyze or reformulate the known $c$- and $a$-theorems within the EAA approach. 
(Work along these lines has been reported in \cite{liouv, codello-pagani-c}.)
Rather, we want to investigate the properties of a candidate for a generalized $C$-function which is distinguished and natural in its own right, namely from the perspective of the EAA.
As such it can be tentatively defined under conditions that are far more general than those leading to the $a$- and $c$-theorems (with respect to dimensionality, field contents, symmetries, etc.).
We expect that  after restricting this generality appropriately $\cfunc_k$ can be given properties similar to a $C$-function.
The long-term objective of this research program is to find out which restrictions precisely lead to interesting properties of $\cfunc_k$.
In the present paper we provide a first example where this strategy can be seen to actually work, namely (truncated) QEG in 4 dimensions.

The structure of this paper is as follows.
In section 2 we develop the general theory for using the EAA and $\cfunc_k$ as a counting device\footnote{A brief discussion and an application of the method to the example of black hole physics appeared already in \cite{daniel1}. (See also ref. \cite{Koch-Saueressig}.)}, after recalling some necessary background material.
Then, in section 3 we apply the resulting framework to the particularly important case of asymptotically safe Quantum Einstein Gravity.
Our setting will apply to an arbitrary dimensionality of spacetime; the numerical calculations needed to verify the claimed properties of $\cfunc_k$ are performed for the most interesting case of 4 dimensions though.

\section{The effective average action as a `$\boldmath{C}$-function'} \label{sec:02}
In this section we develop a generally applicable framework for constructing functions $\cfunc_k$ which have properties similar to a $C$-function, and at the same time are `geometrically natural' objects from the perspective of the theory space explored by the EAA.
To prepare the ground, and to fix various notations, it is unavoidable to embark on some special aspects of the EAA technique first.

\subsection{Counting field modes} \label{subsec:02_01}
{\bf\noindent (A)}
We consider a  general quantum field theory on a $d$ dimensional Euclidean spacetime, either rigid or fluctuating, that is governed by a functional integral
$Z=\int \mathcal{D}\hat{\Phi}\, e^{- S[\hat{\Phi},\bar{\Phi}]}$.
The bare action $S$ depends on a set of commuting and anticommuting dynamical fields, $\hat{\Phi}$, and on a corresponding set of background fields, $\bar{\Phi}$.
(Here and in the following we use a compact notation, leaving implicit all field indices and possible sign factors depending on the Grassmann parity of the field components.)
We assume that the functional integral is regularized in the UV in some way; we shall be more precise about this point in a moment.

In the case of a Yang-Mills theory, $\hat{\Phi}$ contains both the gauge field and the Faddeev-Popov ghosts, and $S$ is understood to include gauge fixing and ghost terms.
Furthermore, the corresponding background fields are part of $\bar{\Phi}$.
As a rule, the  {\it fluctuation field} $\hat{\varphi}\equiv \hat{\Phi}-\bar{\Phi}$ is always required to gauge-transform homogeneously, i.e. like a matter field.
Henceforth we regard $\hat{\varphi}$ rather than $\hat{\Phi}$ as the true dynamical variable and interpret $Z$ as an integral over the fluctuation variables: $Z=\int \mathcal{D}\hat{\varphi}\, \exp\left(-S[\hat{\varphi};\bar{\Phi}]\right)$.

For conceptual reasons that will become apparent below, the set of background fields, $\bar{\Phi}$, always contains a classical spacetime metric $\bg_{\mu\nu}$.
In typical particle physics applications on a rigid (flat, say) spacetime for instance one would not be interested in how $Z$ depends on the background metric and one might set $\bg_{\mu\nu}=\delta_{\mu\nu}$ throughout.
But in quantum gravity, when Background Independence is an issue, one wants to know $Z\equiv Z[\bg_{\mu\nu}]$ for {\it any} background.
In fact, employing the background field technique to implement Background Independence \cite{dewittbook} one represents  the dynamical metric as $\hat{g}_{\mu\nu}=\bg_{\mu\nu}+\hat{h}_{\mu\nu}$ and requires split-symmetry at the level of observable quantities \cite{daniel2}.
When the spacetime is dynamical, $\hat{g}_{\mu\nu}$ and $\hat{h}_{\mu\nu}$ are special  components of $\hat{\Phi}$ and $\hat{\varphi}$, respectively.

Next we pick a basis in field space, $\{\varphi_{\omega}\}$, and expand the fields that are integrated over.
Then, symbolically, $\hat{\varphi}(x)=\sum_{\omega} a_{\omega}\, \varphi_{\omega}(x)$ where $\sum_{\omega}$ stands for a summation and\slash or integration over all labels carried by the basis elements, and $\int \mathcal{D}\hat{\varphi}$ is now interpreted as the integration over all possible values that can be assumed by the expansion coefficients $a\equiv \{a_{\omega}\}$.
Thus, $Z=\prod_{\omega} \int_{-\infty}^{\infty}\md a_{\omega}\, \exp\left(-S[a;\bar{\Phi}]\right)$.

To be more specific, let us assume that the basis $\{\varphi_{\omega}\}$ is constituted by the eigenfunctions of a certain differential operator, $\mathcal{L}$, which may depend on the background fields $\bar{\Phi}$, and which has properties similar to the negative Laplace-Beltrami operator, $-\bZ^2$, appropriately  generalized for the types of (tensor, spinor, $\cdots$) fields present in $\hat{\varphi}$.
We suppose that $\mathcal{L}$ is built from covariant derivatives involving $\bg_{\mu\nu}$ and the background Yang-Mills fields, if any, so that it is covariant under spacetime diffeomorphism and gauge-transformations.
We assume an eigenvalue equation $\mathcal{L}\varphi_{\omega} =\Omega^2_{\omega} \varphi_{\omega}$ with positive spectral values $\Omega_{\omega}^2>0$.
The precise choice of $\mathcal{L}$ is arbitrary to a large extent.

The only property of $\mathcal{L}$ we shall need is that it should associate small (large) distances on the rigid spacetime equipped with the metric $\bg_{\mu\nu}$ to large (small) values of $\Omega_{\omega}^2$.
A first (but for us not the essential) consequence is that we can now easily install a UV cutoff by restricting the ill-defined infinite product $\prod_{\omega}$ to only those $\omega$'s  which satisfy $\Omega_{\omega}<\Omega_{\text{max}}$.
This implements a UV cutoff at the mass scale $\Omega_{\text{max}}$.

More importantly for our purposes, we also introduce a smooth IR cutoff at a variable scale $k\leq \Omega_{\text{max}}$ into the integral, replacing it with
\begin{align}
Z_k= {\prod_{\omega}}^{\prime} \int_{-\infty}^{\infty}\md a_{\omega}\, e^{-S[a;\bar{\Phi}]}e^{-\Delta S_k}
\label{eqn:cm_mot_001}
\end{align}
where the prime indicates the presence of the UV cutoff, and 
\begin{align}
\Delta S_k \equiv \frac{1}{2} \sum_{\omega} R_k(\Omega_k^{2})\, a_{\omega}^2
\label{eqn:cm_mot_002}
\end{align}
implements the IR cutoff.
The extra piece in the bare action, $\Delta S_k$, is designed in such a way that those $\varphi_{\omega}$-modes which have eigenvalues $\Omega_{\omega}^2 \ll k^2$ get suppressed by a small factor $e^{-\Delta S_k}\ll 1$ in eq. \eqref{eqn:cm_mot_001}, while $e^{-\Delta S_k}=1$ for the others.
The function $R_k$ is essentially arbitrary, except for its interpolating behavior between $R_k(\Omega_{\omega}^2)\sim k^2$ if $\Omega_{\omega} \ll k$ and $R_k(\Omega_{\omega}^2)=0$ if $\Omega_{\omega}\gg k$.

The operator $\mathcal{L}$ defines the precise notion of `coarse graining' field configurations.
We regard the $\varphi_{\omega}$'s with $\Omega_{\omega}>k$ as the `short wavelength' modes, to be integrated out first, and those with small eigenvalues $\Omega_{\omega}<k$ as the `long wavelength' ones whose impact on the fluctuation's dynamics is not yet taken into account.
This amounts to a diffeomorphism and gauge covariant generalization of the standard Wilsonian renormalization group, based on standard Fourier analysis on $\noR^d$, to situations with arbitrary background fields $\bar{\Phi}=(\bg_{\mu\nu},\bar{A}_{\mu},\cdots)$.

While helpful for the interpretation, for most practical purposes it is often unnecessary to perform the expansion of $\hat{\varphi}(x)$ in terms of the $\mathcal{L}$-eigenfunctions explicitly.
Rather, one thinks of \eqref{eqn:cm_mot_001} as a `basis independent' functional integral 
\begin{align}
Z_k = \int \mathcal{D}^{\prime}\hat{\varphi}\, e^{-S[\hat{\varphi};\bar{\Phi}]}  e^{-\Delta S_k[\hat{\varphi};\bar{\Phi}]}
\label{eqn:cm_mot_003}
\end{align}
for which the $\mathcal{L}$-eigen-basis plays no special role, while the operator $\mathcal{L}$ as such does so, of course.
In particular the cutoff action $\Delta S_k$ is now rewritten with $\Omega_{\omega}^2$ replaced by $\mathcal{L}$ in the argument of $R_k$:
\begin{align}
\Delta S_k[\hat{\varphi};\bar{\Phi}] = \frac{1}{2} \int \md^d x \sqrt{\bg}\, \hat{\varphi}(x)\, R_k(\mathcal{L})\, \hat{\varphi}(x)
\label{eqn:cm_mot_004}
\end{align}
Note that at least when $k>0$ the modified partition function $Z_k$ depends on the respective choices for $\mathcal{L}$ and $\bar{\Phi}$ separately.

{\bf\noindent (B)}
In this paper we propose to use the one-parameter family of partition function $k\mapsto Z_k $, for $k\in [0,\infty)$, as a diagnostic tool to investigate the RG flow between different quantum field theories.
This is  a quite general and flexible framework.
There is considerable freedom in choosing the cutoff operator, and even when $\mathcal{L}\equiv \mathcal{L}(\bar{\Phi})$ is fixed we may still choose $\bar{\Phi}$ in a large variety of different ways so as to `project out' different information from the partition function.
However, as we shall discuss in the next subsection, there exists a natural, almost `canonical' choice of background configurations $\bar{\Phi}$.

{\bf\noindent (C)}
Our discussion in the following sections is based upon the key observation that $Z_k$ enjoys a simple property  which is strikingly reminiscent of the $C$-theorem in 2 dimensions.
 
Let us assume for simplicity that all component fields constituting $\hat{\varphi}$ are commuting, and that $\bar{\Phi}$ has been chosen $k$-independent.
Then \eqref{eqn:cm_mot_003} is a (regularized, and convergent for appropriate $S$) purely bosonic integral with a positive integrand which, thanks to the suppression factor $e^{-\Delta S_k}$, decreases with increasing $k$.
Therefore, $Z_k$ and the `entropy' $\ln Z_k$, are monotonically decreasing functions of the scale:
\begin{align}
\partial_k \ln Z_k <0
\label{eqn:cm_mot_005}
\end{align}
The interpretation of \eqref{eqn:cm_mot_005} is clear:
Proceeding from the UV to the IR by lowering the infrared cutoff scale, an increasing number of field modes get un-suppressed, thus contribute to the functional integral, and as a consequence the value of the partition function increases.
Thus, in a not too literal sense of the word, $\ln Z_k$ `counts' the number of field modes that have been integrated out already.

Before we can make this intuitive argument more  precise and explicit we must introduce a number of technical tools in the following subsections.

\subsection{Running actions and self-consistent backgrounds}
{\bf\noindent (A)}
Introducing a source term for the fluctuation fields turns the partition functions into the generating functional
\begin{align}
Z_k[J;\bar{\Phi}] \equiv e^{W_k[J;\bar{\Phi}]}= \int \mathcal{D}^{\prime}\hat{\varphi}\, \exp \left(-S[\hat{\varphi};\bar{\Phi}]  -\Delta S_k[\hat{\varphi};\bar{\Phi}] + \int \md^d x \sqrt{\bg}\, J(x)\hat{\varphi}(x)\right)
\label{eqn:cm_mot_006}
\end{align}
Hence the $\bar{\Phi}$- and $k$-dependent expectation value $\langle \hat{\varphi} \rangle\equiv \varphi$ reads
\begin{align}
\varphi(x)\equiv \langle \hat{\varphi}(x)\rangle=\frac{1}{\sqrt{\bg(x)}} \frac{\delta W_k[J;\bar{\Phi}]}{\delta J(x)}
\label{eqn:cm_mot_007}
\end{align}
If we can solve this relation for $J$ as a functional of $\bar{\Phi}$, the definition of the Effective Average Action (EAA), essentially the Legendre transform of $W_k$, may be written as
\begin{align}
\Gamma_k[\varphi;\bar{\Phi}] = \int\md^d x\sqrt{\bg}\, \varphi(x)J(x)-W_k[J;\bar{\Phi}] -\Delta S_k[\varphi;\bar{\Phi}]
\label{eqn:cm_mot_008}
\end{align}
with the solution to \eqref{eqn:cm_mot_007} inserted, $J\equiv J_k[\varphi;\bar{\Phi}]$.
(In the general case, $\Gamma_k$ is the Legendre-Fenchel transform of $W_k$, with $\Delta S_k$ subtracted.)

The EAA gives rise to a source-field relationship which includes an explicit cutoff term linear in the fluctuation field:
\begin{align}
\frac{1}{ \sqrt{\bg}}\frac{\delta \Gamma_k[\varphi;\bar{\Phi}]}{\delta \varphi(x)} +\mathcal{R}_k[\bar{\Phi}] \varphi(x)=J(x)
\label{eqn:cm_mot_009}
\end{align}
Here and in the following we write $\mathcal{R}_k\equiv R_k(\mathcal{L})$, and the notation $\mathcal{R}_k[\bar{\Phi}]$ is used occasionally to emphasize that the cutoff operator may depend on the background fields.
The solution to \eqref{eqn:cm_mot_009}, and more generally all fluctuation correlators $\langle \hat{\varphi}(x_1)\cdots \hat{\varphi}(x_n)\rangle$ obtained by multiple differentiation of $\Gamma_k$, are functionally dependent on the background, e.g. $\varphi(x)\equiv \varphi_k[J;\bar{\Phi}](x)$.

{\bf\noindent (B)}
For the expectation value of the full, i.e. un-decomposed field $\hat{\Phi}=\bar{\Phi}+\hat{\varphi}$ we employ the notation 
\begin{align}
\Phi=\bar{\Phi}+\varphi \quad \text{ with }\,\, \Phi\equiv \langle \hat{\Phi}\rangle\, \text{ and }\, \varphi\equiv \langle \hat{\varphi}\rangle\,.
\label{eqn:cm_mot_010}
\end{align}
Using the complete field $\Phi$ instead of $\varphi$ as the second independent variable, accompanying  $\bar{\Phi}$, entails the `bi-field' variant of the EAA,
\begin{align}
\Gamma_k[\Phi,\bar{\Phi}]\equiv \Gamma_k[\varphi; \bar{\Phi}]\big|_{\varphi=\Phi-\bar{\Phi}}
\label{eqn:cm_mot_011}
\end{align}
which, in particular, is always `bi-metric': $\Gamma_k[g_{\mu\nu},\cdots,\bg_{\mu\nu},\cdots]$.

{\bf\noindent (C)}
Later on it will often be helpful to organize the terms contributing to $\Gamma_k[\varphi;\bar{\Phi}]$ according to their {\it level} which, by definition, is their degree of homogeneity in the $\varphi$'s.
The underlying assumption is that the EAA admits a {\it level expansion} of the form
\begin{align}
\Gamma_k[\varphi;\bar{\Phi}]=\sum_{p=0}^{\infty}\, \check{\Gamma}_k^{p}[\varphi;\bar{\Phi}]
\label{eqn:cm_mot_012}
\end{align}
where $\check{\Gamma}_k^p[c\,\varphi;\bar{\Phi}]=c^p \, \check{\Gamma}_k^p [\varphi;\bar{\Phi}]$ for $c>0$.
If $\Gamma_k[\varphi;\bar{\Phi}]$ admits a Taylor expansion in $\varphi$ about $\varphi=0$, this expansion exists, of course, with the level-($p$) contribution $\check{\Gamma}_k^p$ being its $p$-derivative term, but this is not guaranteed in general.

{\bf\noindent (D)}
We are interested in how the dynamics of the fluctuations $\hat{\varphi}$ depends on the environment they are placed in, the background metric $\bg_{\mu\nu}$, for instance, and the other classical fields collected in $\bar{\Phi}$.
It would be instructive to know if there exist special backgrounds in which the fluctuations are particularly `tame' such that, for vanishing external source, they amount to only small oscillations about a stable equilibrium, with a vanishing mean: $\varphi\equiv\langle \hat{\varphi}\rangle=0$.
Such distinguished backgrounds $\bar{\Phi}\equiv\bar{\Phi}^{\scon}$ are referred to as {\it self-consistent} ($\scon$) since, if we pick one of those, the expectation value of the field $\langle \hat{\Phi} \rangle=\Phi=\bar{\Phi}$ does not get changed by any violent $\hat{\varphi}$-excitations that, generically, can shift the point of equilibrium.

From eq. \eqref{eqn:cm_mot_009} we obtain the following condition $\bar{\Phi}^{\scon}$ must satisfy (since $J=0$ by assumption):
\begin{align}
\frac{\delta}{\delta \varphi(x)} \Gamma_k[\varphi;\bar{\Phi}]\big|_{\varphi=0, \bar{\Phi}=\bar{\Phi}_k^{\scon}}=0
\label{eqn:cm_mot_013}
\end{align}
This is the {\it tadpole equation} from which we can compute the self-consistent background configurations, if any.
In general $\bar{\Phi}^{\scon}\equiv \bar{\Phi}^{\scon}_k$ will have an explicit dependence on $k$.
A technically convenient feature of \eqref{eqn:cm_mot_013} is that it  no longer contains the somewhat disturbing $\mathcal{R}_k \varphi$-term that was present in the general field equation \eqref{eqn:cm_mot_009}.
Self-consistent backgrounds are equivalently characterized by eq. \eqref{eqn:cm_mot_007},
\begin{align}
\frac{\delta}{\delta J(x)} W_k[J;\bar{\Phi}]\big|_{J=0, \bar{\Phi}=\bar{\Phi}_k^{\scon}}=0
\label{eqn:cm_mot_014}
\end{align}
which again expresses the vanishing of the fluctuation's one-point function.

Note that provided the level expansion \eqref{eqn:cm_mot_012} exists we may replace \eqref{eqn:cm_mot_013} with 
\begin{align}
\raalign{
\frac{\delta}{\delta \varphi(x)} \check{\Gamma}^1_k[\varphi;\bar{\Phi}]\big|_{\varphi=0, \bar{\Phi}=\bar{\Phi}_k^{\scon}}=0}{}
\label{eqn:cm_mot_015}
\end{align}
which involves only the level-(1) functional $\check{\Gamma}^1_k$.
Later on in the applications this trivial observation has the important consequence that {\it self-consistent background field configurations $\bar{\Phi}^{\scon}_k(x)$ can contain only running coupling constants of level $p=1$}, that is, the couplings parameterizing the functional  $\check{\Gamma}_k^1$ which is linear in $\varphi$.\footnote{Notice that the $k$-derivative of $\bar{\Phi}^{\scon}_k(x)$ is in general governed also by higher level couplings due to their  appearance in the beta-functions of the level $p=1$ couplings.}

{\bf\noindent (E)}
In our later discussions the value of the EAA at $\varphi=0$ will be of special interest.
While it is still a rather complicated functional for a generic background where $\Gamma_k[0;\bar{\Phi}]=-W_k[J_k[0;\bar{\Phi}];\bar{\Phi}]$, the source which is necessary to achieve $\varphi=0$ for self-consistent backgrounds is precisely $J=0$, implying
\begin{align}
\raalign{\Gamma_k[0;\bar{\Phi}_k^{\scon}]\equiv \check{\Gamma}_k^0[0;\bar{\Phi}_k^{\scon}]=-W_k[0;\bar{\Phi}_k^{\scon}]}{}
\label{eqn:cm_mot_016}
\end{align}
Here we also indicated that in a level expansion  only the $p=0$ term of $\Gamma_k$ survives putting $\varphi=0$.

So we can summarize saying that {\it the value of $\Gamma_k[0;\bar{\Phi}_k^{\scon} ]$ can contain only the running couplings of the levels $p=0$ and $p=1$, respectively, the former entering via $\check{\Gamma}_k^0$, the latter via $\bar{\Phi}_k^{\scon}$.}

\subsection{FIDE, FRGE, and WISS}
The EAA satisfies a number of important exact functional equations which include a functional integro-differential equation (FIDE), the functional RG equation (FRGE),  the Ward identity for the Split-Symmetry (WISS), and the BRS-Ward identity.

\noindent {\bf(A) FIDE:} 
The FIDE is obtained by substituting \eqref{eqn:cm_mot_008} in \eqref{eqn:cm_mot_006}, using \eqref{eqn:cm_mot_009}, and reads
\begin{align}
e^{-\Gamma_k[\varphi;\bar{\Phi}]} &= \int \mathcal{D}^{\prime}\hat{\varphi}\, \exp\left(-S[\hat{\varphi};\bar{\Phi}]-\Delta S_k[\hat{\varphi};\bar{\Phi}] + \int \md^d x\, \hat{\varphi}(x) \frac{\delta \Gamma_k}{\delta \varphi(x)}[{\varphi};\bar{\Phi}] \right)
\label{eqn:cm_mot_017}
\end{align}
Here, as always, summation over field components and their tensor, spinor, internal symmetry, etc. indices is understood.
For the purposes of the present paper, the most important property of \eqref{eqn:cm_mot_017} is that the last term on its RHS, the one linear in $\hat{\varphi}$, vanishes if the background happens to be self-consistent, and at the same time the argument $\varphi=0$ is inserted on both sides of the FIDE:
\begin{align}
\raalign{\exp\left(-\Gamma_k[0;\bar{\Phi}_k^{\scon}]\right) = \int \mathcal{D}^{\prime}\hat{\varphi}\, \exp\left(-S[\hat{\varphi};\bar{\Phi}_k^{\scon}]-\Delta S_k[\hat{\varphi};\bar{\Phi}_k^{\scon}] \right)}{}
\label{eqn:cm_mot_018}
\end{align}
We shall come back to this identity soon.

\noindent {\bf(B) FRGE:}
Another important exact relation satisfied by the EAA is the functional RG equation (FRGE),
\begin{align}
k\partial_k \Gamma_k[\varphi;\bar{\Phi}]=\frac{1}{2}\text{STr}\left[\left(\Gamma_k^{(2)}[\varphi;\bar{\Phi}]+\mathcal{R}_k[\bar{\Phi}] \right)^{-1} k \partial_k \mathcal{R}_k[\bar{\Phi}]\right]
\label{eqn:cm_mot_019}
\end{align}
with the Hessian matrix of the fluctuation derivatives $\Gamma_k^{(2)}\equiv \delta^2 \Gamma_k \slash \delta \varphi^2$.
The supertrace `$\text{STr}$' in \eqref{eqn:cm_mot_019} provides the additional minus sign which is necessary for the $\varphi$-components with odd Grassmann parity, Faddeev-Popov ghosts and fermions.

\noindent {\bf(C) WISS:}
The EAA, written as $\Gamma_k[\Phi,\bar{\Phi}]$, satisfies the following exact functional equation which governs the `extra' background dependence it has over and above the one which combines with the fluctuations to form the full field $\Phi$:
\begin{align}
\frac{\delta}{\delta \bar{\Phi}(x)} \Gamma_k[\Phi,\bar{\Phi}]=\frac{1}{2}\text{STr}\left[\left(\Gamma_k^{(2)}[\Phi,\bar{\Phi}]+\mathcal{R}_k[\bar{\Phi}] \right)^{-1} \frac{\delta}{\delta \bar{\Phi}(x)} S_{\text{tot}}^{(2)}[\Phi,\bar{\Phi}]\right]
\label{eqn:cm_mot_020}
\end{align}
Here $S_{\text{tot}}^{(2)}$ is the Hessian of $S_{\text{tot}}=S+\Delta S_k$ with respect to $\Phi$, where $S$ includes gauge fixing and ghost terms.
The equation \eqref{eqn:cm_mot_020} is the Ward identity induced by the split-symmetry transformations $\delta \varphi=\epsilon$, $\delta\bar{\Phi}=-\epsilon$, hence the abbreviation WISS.
It was first obtained in \cite{gluco} in the context of Yang-Mills theory.
In quantum gravity, extensive use has been made of \eqref{eqn:cm_mot_020} in ref. \cite{elisa2} were it served as a tool to assess the degree of split-symmetry breaking and, related to that, the reliability of certain truncations of Quantum Einstein Gravity (QEG).

For a discussion of the modified BRS-Ward identity enjoyed by the EAA we refer to \cite{gluco} and \cite{mr}.

\subsection{Pointwise monotonicity} \label{subsec:2-03}
Our search for a generalized $C$-type counting function which depends monotonically on $k$ along the RG trajectories will be based upon the following structural property of the FRGE \eqref{eqn:cm_mot_019}.
From the very definition of the EAA by a Legendre transform it follows that for all $\bar{\Phi}$ the sum $\Gamma_k+\Delta S_k$ is a convex functional of $\varphi$, and that $\Gamma_k^{(2)}+\mathcal{R}_k$ is a strictly positive definite operator therefore which can be inverted at all scales $k\in(0,\infty)$.
Now let us suppose that the theory under consideration contains Grassmann-even fields only.
Then the supertrace in \eqref{eqn:cm_mot_019} amounts to the ordinary, and convergent trace of a positive operator so that the FRGE implies
\begin{align}
k \partial_k \Gamma_k[\varphi;\bar{\Phi}]\geq 0\quad \text{ at all fixed }\, \varphi,\, \bar{\Phi}.
\label{eqn:cm_mot_021}
\end{align}
Thus, at least in a class of distinguished theories the EAA, evaluated at any fixed pair of arguments $\varphi$ and $\bar{\Phi}$, is a monotonically increasing function of $k$.
With other words, {\it lowering $k$ from the UV towards the IR the value of $\Gamma_k[\varphi;\bar{\Phi}]$ decreases monotonically}.

We refer to this property as {\it pointwise monotonicity} in order to emphasize that it applies at all points of field space, $(\varphi,\bar{\Phi})$, separately.
In particular this means that the argument of $\EAA_k[\varphi,\bar{\Phi}]$ is assumed to have no $k$-dependence of its own here.

In presence of fields with odd Grassmann parity, fermions and Faddeev-Popov ghosts, the RHS of the FRGE is no longer obviously non-negative.
However,  if the only Grassmann-odd fields are ghosts the pointwise monotonicity \eqref{eqn:cm_mot_021} can still be made a general property of the EAA, the reason being as follows.
At least when one implements the gauge fixing condition strictly, it cuts-out a certain subspace of  the space of fields $\hat{\Phi}$ to be integrated over, namely the gauge orbit space.
Hereby the integral over the ghosts represents the measure on this subspace, the Faddeev-Popov determinant.
The subspace and its geometrical structures are invariant under the RG flow, however.
Hence the EAA pertaining to the manifestly Grassmann-even integral {\it over the subspace} is of the kind considered above, and the argument implying \eqref{eqn:cm_mot_021} should therefore be valid again.
For a more explicit version of this reasoning we refer to appendix \ref{app:cmA}.

\subsection{Monotonicity vs. stationarity}
The EAA evaluated at fixed arguments shares the monotonicity property with a $C$-function.
One of the problems is however that $\Gamma_k[\varphi;\bar{\Phi}]$ is not stationary at fixed points of the RG flow.
In order to see why, and how to improve the situation, some care is needed concerning the interplay of dimensionful and dimensionless variables, to which we turn next.

{\bf\noindent (A)}
Let us assume that the space constituted by the functionals of $\varphi$ and $\bar{\Phi}$ admits a basis $\{I_{\alpha}\}$ so that we can expand the EAA as 
\begin{align}
\Gamma_k[\varphi;\bar{\Phi}]=\sum_{\alpha} \bar{u}_{\alpha}(k)\,I_{\alpha}[\varphi;\bar{\Phi}]
\label{eqn:cm_mot_022}
\end{align}
with dimensionful running coupling constants $\bar{u}\equiv (\bar{u}_{\alpha})$.
They obey a FRGE in component form, $k\partial_k \bar{u}_{\alpha}(k)=\bar{b}_{\alpha}(\bar{u}(k);k)$, whereby the functions $\bar{b}_{\alpha}$ are defined by the expansion of $\text{STr}[\cdots]$ with respect to the basis:
\begin{align}
\frac{1}{2}\text{STr}\left[\left(\sum_{\alpha} \bar{u}_{\alpha}(k)\, I_{\alpha}^{(2)}[\varphi;\bar{\Phi}]+\mathcal{R}_k\right)^{-1} k\partial_k \mathcal{R}_k\right]=: \sum_{\alpha} \bar{b}_{\alpha}(\bar{u}(k);k)\, I_{\alpha}[\varphi;\bar{\Phi}]
\label{eqn:cm_mot_023}
\end{align}
Note that the statement of monotonicity in \eqref{eqn:cm_mot_021}, when it holds true, translates into the pointwise positivity of the sum $\sum_{\alpha} \bar{b}_{\alpha}I_{\alpha}$.

{\bf\noindent (B)}
Denoting the canonical mass dimension\footnote{Our conventions are as follows. We use dimensionless coordinates, $[x^{\mu}]=0$. Then $[\md s^2]=-2$ implies that all components of the various metrics have $[\hat{g}_{\mu\nu}]=[\bg_{\mu\nu}]=[g_{\mu\nu}]=-2$, and likewise for the fluctuations: $[\hat{h}_{\mu\nu}]=[\flcb_{\mu\nu}]=-2$.} of the running couplings by $[\bar{u}_{\alpha}]\equiv d_{\alpha}$, their dimensionless counterparts are defined by 
$u_{\alpha}\equiv k^{-d_{\alpha}} \bar{u}_{\alpha}$.
In terms of the dimensionless couplings the expansion of $\Gamma_k$ reads
\begin{align}
\Gamma_k[\varphi;\bar{\Phi}]=\sum_{\alpha} u_{\alpha}(k) k^{d_{\alpha}} I_{\alpha}[\varphi;\bar{\Phi}]
\label{eqn:cm_mot_024}
\end{align}
Now observe that since $\Gamma_k$ is dimensionless the basis elements have dimensions $\left[I_{\alpha}[\varphi;\bar{\Phi}]\right]=-d_{\alpha}$.
Purely by dimensional analysis, this implies that\footnote{We use the notation $c^{[\varphi]}\varphi \equiv \{c^{[\varphi_i]}\varphi_i\}$ for the set in which each field is rescaled according to its individual canonical dimension.}
\begin{align}
I_{\alpha}[c^{[\varphi]} \varphi; c^{[\bar{\Phi}]}\bar{\Phi}]=c^{-d_{\alpha}} I_{\alpha}[\varphi;\bar{\Phi}] \quad \text{ for any constant $c>0$.}
\label{eqn:cm_mot_025}
\end{align}
This relation expresses the fact that the nontrivial dimension of $I_{\alpha}$ is entirely due to that of its field arguments; there are simply no other dimensionful quantities available after the $k$-dependence has been separated off.
Using \eqref{eqn:cm_mot_025} for $c=k^{-1}$ yields
\begin{align}
k^{d_{\alpha}} I_{\alpha}[\varphi;\bar{\Phi}]&= I_{\alpha}[k^{-[\varphi]}\varphi;k^{-[\bar{\Phi}]}\bar{\Phi}]
\equiv I_{\alpha}[\tilde{\varphi};\tilde{\bar{\Phi}}]
\label{eqn:cm_mot_026}
\end{align}
Here we introduced the sets of dimensionless fields,
\begin{align}
\tilde{\varphi}(x)\equiv k^{-[\varphi]} \varphi(x),\,\quad \tilde{\bar{\Phi}}(x)\equiv k^{-[\bar{\Phi}]}\bar{\Phi}(x)
\label{eqn:cm_mot_027}
\end{align}
which include, for instance, the dimensionless metric and its fluctuations:
\begin{align}
\tilde{\flcb}_{\mu\nu}(x)\equiv k^2 \flcb_{\mu\nu}(x),\,\quad \tilde{\bg}_{\mu\nu}(x)\equiv k^2 \bg_{\mu\nu}(x)
\label{eqn:cm_mot_028}
\end{align}
Note that the quantity \eqref{eqn:cm_mot_026}, in whatever way we write it, is dimensionless.
When we insert the dimensionless fields rather than $\varphi$ and $\bar{\Phi}$ into the basis functionals, the latter loose their nonzero dimension: $\left[I_{\alpha}[\tilde{\varphi};\tilde{\bar{\Phi}}]\right]=0$.

Exploiting \eqref{eqn:cm_mot_026} in \eqref{eqn:cm_mot_024} we obtain the following representation of the EAA which is entirely in terms of dimensionless quantities\footnote{Here one should also switch from $k$ to the manifestly dimensionless `RG time' $t\equiv \ln (k)+\const$, but we shall not indicate this notationally.}
\begin{align}
\raalign{
\Gamma_k[\varphi;\bar{\Phi}]=\sum_{\alpha} u_{\alpha}(k)\, I_{\alpha}[\tilde{\varphi};\tilde{\bar{\Phi}}]
\,\equiv\, \mathcal{A}_k[\tilde{\varphi};\tilde{\bar{\Phi}}]}{}
\label{eqn:cm_mot_029}
\end{align}
Alternatively, one might wish to make its $k$-dependence explicit, writing,
\begin{align}
\Gamma_k[\varphi;\bar{\Phi}]=\sum_{\alpha} u_{\alpha}(k)\, I_{\alpha}[k^{-[\varphi]}\varphi;k^{-[\bar{\Phi}]}\bar{\Phi}]
\label{eqn:cm_mot_030}
\end{align}
In the second equality of \eqref{eqn:cm_mot_029} we introduced the new functional $\mathcal{A}_k$ which, by definition, is numerically equal to $\Gamma_k$, but its independent variables (arguments) are the dimensionless fields $\tilde{\varphi}$ and $\tilde{\bar{\Phi}}$.
Hence the $k$-derivative of $\mathcal{A}_k[\tilde{\varphi},\tilde{\bar{\Phi}}]$ is to be performed at fixed $(\tilde{\varphi},\tilde{\bar{\Phi}})$, while the analogous derivative of $\Gamma_k[\varphi;\bar{\Phi}]$ refers to fixed dimensionful arguments.
This leads to the following two trivial but momentous equations:
\begin{subequations}
\begin{align}
k\partial_k \mathcal{A}_k[\tilde{\varphi};\tilde{\bar{\Phi}}]&=\sum_{\alpha} k\partial_k u_{\alpha}(k)\, I_{\alpha}[\tilde{\varphi};\tilde{\bar{\Phi}}] \label{eqn:cm_mot_031A}\\
k\partial_k \Gamma_k[\varphi;\bar{\Phi}]&=\sum_{\alpha} \big\{ k\partial_k u_{\alpha}(k) + d_{\alpha} u_{\alpha}(k) \big\}\,k^{d_{\alpha}}\, I_{\alpha}[\varphi;\bar{\Phi}] \label{eqn:cm_mot_031B}
\end{align}
\label{eqn:cm_mot_031}
\end{subequations}
The extra term $\propto d_{\alpha} u_{\alpha}(k)$ in eq. \eqref{eqn:cm_mot_031B} arises by differentiating the factor $k^{d_{\alpha}}$ in \eqref{eqn:cm_mot_024}.
It leads to the well-known canonical scaling term in the $\beta$-functions of the dimensionless couplings.

{\bf\noindent (C)}
For the following it is crucial to recall that it is the {\it dimensionless} couplings $u\equiv (u_{\alpha})$ that serve as local coordinates on theory space, henceforth denoted $\mathcal{T}$.
Its points are functionals $\mathcal{A}$ which depend on dimensionless arguments: $\mathcal{A}[\tilde{\varphi};\tilde{\bar{\Phi}}]=\sum_{\alpha} u_{\alpha} \, I_{\alpha}[\tilde{\varphi};\tilde{\bar{\Phi}}]$.
The RG trajectories are curves $k\mapsto \mathcal{A}_k=\sum_{\alpha} u_{\alpha}(k)\, I_{\alpha}\in \mathcal{T}$ that are everywhere tangent to 
\begin{align}
k\partial_k \mathcal{A}_k=\sum_{\alpha} \beta_{\alpha}(u(k)) \, I_{\alpha}
\label{eqn:cm_mot_032}
\end{align}
The functions $\beta_{\alpha}$, components of a vector field $\vec{\beta}$ on $\mathcal{T}$, are obtained by translating $k\partial_k \bar{u}_{\alpha}(k)=\bar{b}_{\alpha}(\bar{u}(k);k)$ into the dimensionless language.
This leads to the autonomous system of differential equations
\begin{align}
k\partial_k u_{\alpha}(k)\equiv \beta_{\alpha}(u(k))=-d_{\alpha} u_{\alpha}(k) + b_{\alpha}(u(k))
\label{eqn:cm_mot_033}
\end{align}
Here $b_{\alpha}$, contrary to its dimensionful precursor $\bar{b}_{\alpha}$, has no explicit $k$-dependence, thus defining an RG-time independent vector field, the `RG flow' $(\mathcal{T},\vec{\beta})$.

If it has a fixed point at some $u^*$ then $\beta_{\alpha}(u^*)=0$, and the `velocity' of any trajectory passing this point vanishes there\footnote{To keep the notation simple, we assume here that among the $u_{\alpha}$'s there are no `inessential', aka `redundant', couplings that would not have to approach fixed point values.}, $k\partial_k u_{\alpha}=0$.
Hence by \eqref{eqn:cm_mot_032} the redefined functional $\mathcal{A}_k$ becomes stationary there, that is, its scale derivative vanishes pointwise,
\begin{align}
k\partial_k \mathcal{A}_k[\tilde{\varphi};\tilde{\bar{\Phi}}]=0 \,\quad \text{ for all fixed}\, \tilde{\varphi},\, \tilde{\bar{\Phi}}\,.
\label{eqn:cm_mot_034}
\end{align}
So the entire functional $\mathcal{A}_k$ approaches a limit, $\mathcal{A}_*=\sum_{\alpha} u_{\alpha}^*\, I_{\alpha}$.
The standard EAA instead keeps running in the fixed point regime:
\begin{align}
\Gamma_k[\varphi;\bar{\Phi}]=\sum_{\alpha} u_{\alpha}^*\, k^{d_{\alpha}}\, I_{\alpha}[\varphi;\bar{\Phi}]\quad \text{ when }\quad u_{\alpha}(k)=u_{\alpha}^*\,.
\label{eqn:cm_mot_035}
\end{align}

{\bf\noindent (D)}
This brings us back to the `defect' of $\Gamma_k$ we wanted to repair:
While $\Gamma_k[\varphi;\bar{\Phi}]$ was explicitly seen to decrease monotonically along RG trajectories,  it does not come to a halt at fixed points in general.
The redefined functional $\mathcal{A}_k$, instead, approaches a finite limit $\mathcal{A}_*$ at fixed points, but can we  argue that it is monotone along trajectories?

Unfortunately this is not the case, and the culprit is quite obvious, namely the $d_{\alpha}u_{\alpha}$-terms present in the scale derivative of $\Gamma_k$, but absent for $\mathcal{A}_k$:
The  positivity of the RHS of eq. \eqref{eqn:cm_mot_031B} does not imply the positivity of the RHS of eq. \eqref{eqn:cm_mot_031A}, and {\it there is no obvious structural reason for $k\partial_k \mathcal{A}_k[\tilde{\varphi};\tilde{\bar{\Phi}}]\geq 0$ at fixed $\tilde{\varphi}$, $\tilde{\bar{\Phi}}$.}
The best we can get is the bound 
\begin{align}
k\partial_k \mathcal{A}_k[\tilde{\varphi};\tilde{\bar{\Phi}}]\geq - \sum_{\alpha} d_{\alpha} u_{\alpha}(k) \, I_{\alpha}[\tilde{\varphi};\tilde{\bar{\Phi}}]
\label{eqn:cm_mot_036}
\end{align}
which follows by subtracting the two eqs. \eqref{eqn:cm_mot_031} and making use of \eqref{eqn:cm_mot_026}.

\subsection{The proposal} \label{subsec:2-05}
The complementary virtues of $\mathcal{A}_k$ and $\Gamma_k$ with respect to monotonicity along trajectories and stationarity at critical points suggest the following strategy for finding a $C$-type function with better properties:
Rather than considering the functionals pointwise, i.e. with fixed configurations of either the dimensionless or dimensionful fields inserted, one should evaluate them at {\it explicitly scale dependent arguments}:
$\cfunc_k \stackrel{?}{=} \Gamma_k[\varphi_k;\bar{\Phi}_k]\equiv\mathcal{A}_k[\tilde{\varphi}_k;\tilde{\bar{\Phi}}_k]$.

The hope is that the respective arguments
$\varphi_k\equiv k^{[\varphi]}\tilde{\varphi}_k$, and  $\bar{\Phi}_k\equiv k^{[\bar{\Phi}]}\tilde{\bar{\Phi}}_k$
can be given a $k$-dependence which is intermediate between the two extreme cases $(\varphi,\bar{\Phi})=\const$ and  $(\tilde{\varphi},\tilde{\bar{\Phi}})=\const$, respectively, so as to preserve as much as possible of the monotonicity properties of $\Gamma_k$, while rendering $\cfunc_k$ stationary at fixed points of the RG flow.

The most promising candidate of this kind which we could find is
\begin{align}
\raalign{\cfunc_k=\Gamma_k[0;\bar{\Phi}^{\scon}_k]=\mathcal{A}_k[0;\tilde{\bar{\Phi}}^{\scon}_k]}{}
\label{eqn:cm_mot_039}
\end{align}
Here the fluctuation argument is set to zero, $\varphi_k\equiv 0$, and for the background we choose a self-consistent one, $\bar{\Phi}^{\scon}_k$, a solution to the tadpole equation \eqref{eqn:cm_mot_013}, or equivalently its dimensionless variant
\begin{align}
\frac{\delta}{\delta \tilde{\varphi}(x)} \mathcal{A}_k[\tilde{\varphi};\tilde{\bar{\Phi}}]\big|_{\tilde{\varphi}=0,\, \tilde{\bar{\Phi}}=\tilde{\bar{\Phi}}^{\scon}_k}=0
\label{eqn:cm_mot_040}
\end{align}
The function $k\mapsto \cfunc_k$ defined by eq. \eqref{eqn:cm_mot_039} has a number of interesting properties to which we turn next.

\noindent {\bf(A) Stationarity at critical points.}
When the RG trajectory approaches a fixed point, $\mathcal{A}_k[\tilde{\varphi};\tilde{\bar{\Phi}}]$ approaches $\mathcal{A}_*[\tilde{\varphi};\tilde{\bar{\Phi}}]$ pointwise.
Furthermore, the tadpole equation \eqref{eqn:cm_mot_040} becomes $(\delta \mathcal{A}_*\slash \delta \tilde{\varphi})[0;\tilde{\bar{\Phi}}_*]=0$.
It is completely $k$-independent, and so is its solution, $\tilde{\bar{\Phi}}_*$.
Thus $\cfunc_k$ approaches a well defined, finite constant:
\begin{align}
\cfunc_k \xrightarrow{\text{FP}}{}\cfunc_*=\mathcal{A}_*[0;\tilde{\bar{\Phi}}_*]
\label{eqn:cm_mot_041}
\end{align}
Of course we can write this number also as $\cfunc_*=\Gamma_k[0;k^{[\bar{\Phi}]}\tilde{\bar{\Phi}}_*]$ wherein the explicit and the implicit scale dependence of the EAA cancel exactly when a fixed point is approached.

\noindent {\bf(B) Stationarity at classicality.}
In a classical regime (`$\crg$'), by definition, $\bar{b}_{\alpha}\rightarrow 0$, so that it is now the {\it dimensionful} couplings whose running stops: $\bar{u}_{\alpha}(k)\rightarrow \bar{u}_{\alpha}^{\crg}=\const$.
Thus, by \eqref{eqn:cm_mot_022}, $\Gamma_k$ approaches $\Gamma_{\crg}=\sum_{\alpha} \bar{u}_{\alpha}^{\crg}\, I_{\alpha}$ pointwise.
Hence the dimensionful version of the tadpole equation, \eqref{eqn:cm_mot_013}, becomes $k$-independent, and the same is true for its solution, $\bar{\Phi}^{\scon}_{\crg}$.
So, when the RG trajectory approaches a classical regime, $\cfunc_k$ looses its $k$-dependences and approaches a constant:
\begin{align}
\cfunc_{k} \xrightarrow{\crg}{} \cfunc_{\crg}=\Gamma_{\crg}[0;\bar{\Phi}_{\crg}^{\scon}]
\label{eqn:cm_mot_042}
\end{align}
Alternatively we can write $\cfunc_{\crg}=\mathcal{A}_k[0;k^{-[\bar{\Phi}]} \bar{\Phi}^{\scon}_{\crg}]$ where it is now the explicit and implicit $k$-dependence of $\mathcal{A}_k$ which cancel mutually.

We observe that there is a certain analogy between `criticality' and `classicality', in the sense that dimensionful and dimensionless couplings exchange their roles.
The difference is that the former situation is related to special {\it points} of theory space, while the latter concerns  extended {\it regions} in $\mathcal{T}$.
In those regions, $\mathcal{A}_k$ keeps moving as $\mathcal{A}_k[\,\cdot\,]=\sum_{\alpha} \bar{u}_{\alpha}^{\crg}\,k^{-d_{\alpha}}\, I_{\alpha}[\,\cdot\,]$.
Nevertheless it is thus plausible, and of particular interest in quantum gravity, to apply a (putative) $C$-function not only to crossover trajectories in the usual sense which connect two fixed points, but also to {\it generalized crossover transitions} where one of the fixed points, or even both, get replaced by a classical regime.

\noindent {\bf(C) Monotonicity at exact split-symmetry.}
If split-symmetry is exact in the sense that $\Gamma_k[\varphi;\bar{\Phi}]$ depends on the single independent field variable $\bar{\Phi}+\varphi\equiv \Phi$ only, and the theory is such that pointwise monotonicity \eqref{eqn:cm_mot_021} holds true, then $k\mapsto \cfunc_k$ is a monotonically increasing function of $k$.
In fact, differentiating \eqref{eqn:cm_mot_039} and using the chain rule yields
\begin{align}
\raalign{
\partial_k \cfunc_k = \left(\partial_k \Gamma_k\right)[0;\bar{\Phi}_k^{\scon}] + \int\md^d x\left(\partial_k \bar{\Phi}_k^{\scon}(x)\right) \left(\frac{\delta}{\delta \bar{\Phi}(x)} - \frac{\delta}{\delta \varphi(x)}\right) \Gamma_k[\varphi;\bar{\Phi}]\big|_{\varphi=0,\, \bar{\Phi}=\bar{\Phi}_k^{\scon}\!\!\!\!\!\!\!\!\!\!\!\!\!\!\! }}{}
\label{eqn:cm_mot_043}
\end{align}

In the first term on the RHS of \eqref{eqn:cm_mot_043} the derivative $\partial_k$ hits only the explicit $k$-dependence of the EAA.
By eq. \eqref{eqn:cm_mot_021} we know that this contribution is non-negative.
The last term, the $\delta\slash \delta \varphi$-derivative, is actually zero by the tadpole equation \eqref{eqn:cm_mot_013}.
Including it here it becomes manifest that the integral term in \eqref{eqn:cm_mot_043} vanishes when $\Gamma_k$ depends on $\varphi$ and $\bar{\Phi}$ only via the combination $\varphi+\bar{\Phi}$.
Thus we have shown that
\begin{align}
\partial_k \cfunc_k \geq 0 \,\quad \text{ at exact split-symmetry}
\label{eqn:cm_mot_044}
\end{align}

Note that the result for $\partial_k \cfunc_k$ in \eqref{eqn:cm_mot_043} is much closer to what one needs to prove in order to rightfully call $\cfunc_k$ a `$C$-function' than the inequality \eqref{eqn:cm_mot_036}. 
In theories that require no breaking of split-symmetry, for instance, the integral term in \eqref{eqn:cm_mot_043} is identically zero and we know that $\partial_k \cfunc_k\geq 0$ holds true.

The degree of split-symmetry violation varies over theory space in general.
Split-symmetry is unbroken at points $u=(u_{\alpha})$ where at most those coordinates $u_{\alpha}$ are non-zero that belong to basis functionals $I_{\alpha}[\varphi;\bar{\Phi}]$ which happen to depend on $\bar{\Phi}+\varphi$ only.

The breaking of split-symmetry is best discussed in terms of the functional $\Gamma_k[\Phi,\bar{\Phi}]\equiv \Gamma_k[\Phi-\bar{\Phi};\bar{\Phi}]$ for which perfect symmetry amounts to independence of the second argument: $\frac{\delta}{\delta \bar{\Phi}}\Gamma_k[\Phi,\bar{\Phi}]=0$.
In this language, $\cfunc_k$ is written as  $\cfunc_k=\Gamma_k[\bar{\Phi}_k^{\scon},\bar{\Phi}_k^{\scon}]$, and its scale derivative assumes the form
\begin{align}
\raalign{
\partial_k \cfunc_k = \left(\partial_k \Gamma_k\right)[\bar{\Phi}_k^{\scon},\bar{\Phi}_k^{\scon}] + \int\md^d x\left(\partial_k \bar{\Phi}_k^{\scon}(x)\right) \left.\frac{\delta \Gamma_k[\Phi,\bar{\Phi}]}{\delta \bar{\Phi}(x)} \right|_{\Phi=\bar{\Phi}=\bar{\Phi}_k^{\scon}}}{}
\label{eqn:cm_mot_045} 
\end{align}

Whether or not $\partial_k \cfunc_k$ is always non-negative depends on the size of the split-symmetry breaking the EAA suffers from.
To prove monotonicity of $\cfunc_k$ one would have to show on a case-by-case basis that the second term on the RHS of \eqref{eqn:cm_mot_045} never can override the first one, known to be non-negative, so as to render their sum negative.

In fact, for the derivative $\delta \Gamma_k \slash \delta \bar{\Phi}$ in \eqref{eqn:cm_mot_045} we have an exact formal identity  at our disposal, the WISS of eq. \eqref{eqn:cm_mot_020}.
The equations \eqref{eqn:cm_mot_045} and \eqref{eqn:cm_mot_020} together with the FRGE for the $(\partial_k \EAA_k)$-term  could be the starting point of future work on exact estimates. 

In the next section of the present paper we shall investigate the monotonicity properties of $\cfunc_k$ in certain truncations of pure Quantum Einstein Gravity using a different strategy.
In this case it is easier to work directly with the definition of $\cfunc_k$, eq. \eqref{eqn:cm_mot_039}, rather then using the WISS.

\subsection{The mode counting property revisited}\label{subsec:2-06}

Equipped with the EAA machinery, we now return to the heuristic argument about the mode `counting' property of $Z_k$ that was presented at the end of subsection \ref{subsec:02_01}.
Trying to make it more precise, we shall now demonstrate that our candidate $\cfunc_k\equiv \Gamma_k[0;\bar{\Phi}_k^{\scon}]$ is really a measure for the `number' of field modes, since it is closely related to a spectral density.

In fact, after our preparations in the previous subsections this should not be too much of a surprise.
By virtue of the general identity \eqref{eqn:cm_mot_018}, a special case of the FIDE satisfied by the EAA, the exponential $e^{-\EAA_k[0,\bar{\Phi}_k^{\scon}]}$ equals precisely the  partition function considered $Z_k$ in subsection \ref{subsec:02_01}, provided the latter is specialized for a self-consistent background.

In order to present a picture which is as clear as possible let us make a number of specializations and approximations.
In particular we consider purely bosonic theories again, and invoke the idealization of perfect split-symmetry.
Then, by eq. \eqref{eqn:cm_mot_043}, we have $\partial_k \cfunc_k=\left(\partial_k\Gamma_k\right)[0;\bar{\Phi}_k^{\scon}]$, and upon expressing $\left(\partial_k \Gamma_k\right)$ via the FRGE we arrive at\footnote{Note that the equation \eqref{eqn:cm_mot_046} is of course insufficient to {\it determine} the function $k\mapsto \cfunc_k$. We rather use it to {\it interpret} a given $\cfunc_k$ which was derived from a known solution to the full-fledged FRGE.}
\begin{align}
k \partial_k \cfunc_k &= \frac{1}{2}\Tr \left[\left(\Gamma_k^{(2)}[0;\bar{\Phi}_k^{\scon}]+R_k(\mathcal{L})\right)^{-1} k\partial_k R_k(\mathcal{L})\right]
\label{eqn:cm_mot_046}
\end{align}
Now we consider a situation where the Hessian appearing in this equation itself qualifies as a cutoff operator.
When we choose $\mathcal{L}=\Gamma_k^{(2)}[0;\bar{\Phi}_k^{\scon}]$ we obtain
\begin{align}
k \partial_k \cfunc_k &= \frac{1}{2}\Tr \left[\left(\mathcal{L}+R_k(\mathcal{L})\right)^{-1} k\partial_k R_k(\mathcal{L})\right]
= \frac{1}{2} \SumInt_{\Omega^2} \frac{k\partial_k R_k(\Omega^2)}{\Omega^2 + R_k(\Omega^2)}
\label{eqn:cm_mot_047}
\end{align}
Here $\Omega^2$ denotes the eigenvalues of $\mathcal{L}$, and $\SumInt$ indicates the summation and\slash or integration over its spectrum, leaving the corresponding spectral density implicit.
To proceed, we opt for a particularly convenient cutoff function $R_k$, namely the sharp cutoff\footnote{See ref. \cite{frank1} for a detailed discussion of the sharp cutoff. It is often used in quantum gravity since it allows for an easy closed-form evaluation of the threshold functions $\Phi^p_n$ and $\widetilde{\Phi}^p_n$ that frequently appear  in QEG beta-functions \cite{mr}.}:
\begin{align}
R_k(\Omega^2)= \lim_{\hat{R}\rightarrow\infty} \hat{R} \,\, \Theta(k^2-\Omega^2)
\label{eqn:cm_mot_048}
\end{align}
The limit $\hat{R}\rightarrow\infty$ in \eqref{eqn:cm_mot_048} is to be understood in the distributional sense.
It should be taken only after the integration over $\Omega^2$ has been performed.
If we formally use \eqref{eqn:cm_mot_048} in the equation \eqref{eqn:cm_mot_047} this leads us to 
$k\partial_k \cfunc_k= 2 \SumInt_{\Omega^2} \delta(1-\Omega^2\slash k^2)=2k^2 \Tr\left[\delta(k^2-\mathcal{L})\right]$, or equivalently,
\begin{align}
\frac{\md}{\md k^2} \cfunc_k = \Tr\left[\delta\left(k^2-\Gamma^{(2)}_k[0;\bar{\Phi}_k^{\scon}]\right)\right] \geq 0
\label{eqn:cm_mot_049}
\end{align}

The equation \eqref{eqn:cm_mot_049} is quite remarkable and sheds some light on the interpretation of $\cfunc_k$:
Its derivative equals exactly the spectral density of the Hessian operator evaluated at the $\scon$-background field configuration and for vanishing fluctuations, $\Gamma_k^{(2)}[0;\bar{\Phi}^{\scon}_k]$. 
It is a manifestly non-decreasing function of $k$ therefore.

Let us integrate \eqref{eqn:cm_mot_049} over $k^2$.
Provided the RG effects are weak and $\Gamma_k$ runs only very slowly, the $k$-dependence of the resulting field $\bar{\Phi}_k^{\scon}$  is weak, too, so that it may be a sensible approximation to neglect the $k$-dependence of $\Gamma_k^{(2)}[0;\bar{\Phi}_k^{\scon}]$ in the $\delta$-function of \eqref{eqn:cm_mot_049} relative to the explicit $k^2$.
Under these special circumstances, the integrated version of \eqref{eqn:cm_mot_049} reads:
\begin{align}
\cfunc_k=\Tr\left[\Theta\left(k^2-\Gamma_k^{(2)}[0;\bar{\Phi}_k^{\scon}]\right)\right]+\text{const}
\label{eqn:cm_mot_050}
\end{align}

Thus, our conclusion is that, at least under the conditions described, the function $\cfunc_k$ indeed counts field modes, in the almost literal sense of the word, namely the eigenfunctions of the Hessian operator which have eigenvalues not exceeding $k^2$.

Regardless of the present approximation we define in general
\begin{align}
\nBound_{k_1,k_2}\equiv \cfunc_{k_2}-\cfunc_{k_1}
\label{eqn:cm_mot_051}
\end{align}
Then, in the cases when the above assumptions apply and \eqref{eqn:cm_mot_050} is valid, $\nBound_{k_1,k_2}$ has a simple interpretation: it equals the number of eigenvalues between $k_1^2$ and $k_2^2>k_1^2$ of the Hessian operator $\EAA_k^{(2)}[0;\bar{\Phi}_k^{\scon}]$, when the spectrum is discrete.
When the assumptions leading to \eqref{eqn:cm_mot_050} are not satisfied, the interpretation of $\nBound_{k_1,k_2}$, and $\cfunc_k$ in the first place, is less intuitive, but these functions are well defined nevertheless.

As an aside let us also mention that the function \eqref{eqn:cm_mot_050} is closely related to the Chamseddine-Connes spectral action \cite{spect-act,cham-connes-surf,frank-spec-ac} in Noncommutative Geometry, where the squared Dirac operator plays the same role as the Hessian operator above.
\section{Asymptotically safe quantum gravity}
In this section we make the above ideas concrete and apply them to an appropriately truncated form of Quantum Einstein Gravity (QEG) which is asymptotically safe, that is, all physically relevant RG trajectories start out in the UV, for $k\,\text{`}\!=\!\!\text{'}\,\infty$, at a point infinitesimally close to a non-Gaussian fixed point (NGFP).
When $k$ is lowered they run towards the IR, always staying within the fixed point's UV critical manifold, and ultimately approach the (dimensionless) ordinary effective action.

\subsection{The single- and bi-metric Einstein-Hilbert truncations}
{\bf\noindent (A)}
In the following we study the $C$-function properties of $\cfunc_k$ in pure, metric-based quantum gravity in an arbitrary spacetime dimension.
We rely on results obtained with the so called single- and bi-metric Einstein-Hilbert truncations where the considered subspace of theory space is spanned by the invariants $\int\sqrt{g}$ and $\int \sqrt{g}\SR$ only, with $g_{\mu\nu}$- and $\bg_{\mu\nu}$-contributions disentangled in the bi-metric case.

In either case the ansatz for the EAA  in this subspace is given by
\begin{align}
\EAA_k[g,\Ghx,\GhAx,\bg]= \EAA^{\text{grav}}_k[g,\bg]+\EAA_k^{\text{gf}}[g,\bg]+ \EAA^{\text{gh}}_k[g,\Ghx,\GhAx,\bg].
\label{eqn:cm_qeg_ansatz}
\end{align}
 It consists of a purely gravitational part,
$\EAA_k^{\text{grav}}[g,\bg]$, and an essentially classical gauge sector%
\footnote{For a {\it single-metric} extension of the ghost sector, see refs. \cite{frank-ghost,astrid-ghost}.}
based on the coordinate condition $\big(\delta^{\beta}_{\mu}\bg^{\alpha\gamma} \bZ_{\gamma}-\varpi \bg^{\alpha\beta}\bZ_{\mu}\big)\,\flcb_{\mu\nu}=0$ from which the gauge fixing term  $\EAA_k^{\text{gf}}[g,\bg]$ and the corresponding ghost action $\EAA^{\text{gh}}_k[g,\Ghx,\GhAx,\bg]\propto \int \GhAx\, M(g,\bg) \Ghx$ are derived.
Here $\Ghx^{\mu}$ and $\GhAx_{\mu}$ denote the diffeomorphism ghosts, and $M(g,\bg)$ is the Faddeev-Popov operator \cite{mr}.

We will mostly focus in the following on the Einstein-Hilbert truncation in a bi-metric setting.
In this case, $\Gamma_k^{\text{grav}}$  comprises two separate Einstein-Hilbert terms built from the dynamical metric $g_{\mu\nu}$ and its background analog, $\bg_{\mu\nu}$, respectively:
\begin{align}
  \EAA_k^{\text{grav}}[g,\bg]&= - \frac{1}{16\pi \nkD } \int\md^d x \sqrt{g}\, \left(\SR(g) - 2 \KkbarD\right)\nonumber  \\
&\quad   - \frac{1}{16\pi \nkbB }  \int\md^d x \sqrt{\bg}\, \left(\SR(\bg)- 2 \KkbarB\right) \label{eqn:trA07}
\end{align}
The couplings, $\nkD$, $\KkbarD$ and $\nkbB$, $\KkbarB$ represent $k$-dependent  generalizations of the classical Newton or cosmological constant in the dynamical (`$\dyn$') and the background (`$\background$') sector, respectively. 
Expanding eq. \eqref{eqn:trA07} in terms of the fluctuation field $\flcb_{\mu\nu}=g_{\mu\nu}-\bg_{\mu\nu}$ yields the level-expansion of the EAA:
\begin{align}
  \EAA_k^{\text{grav}}[\flcb;\bg]&= - \frac{1}{16\pi G_k^{(0)} }  \int\md^d x \sqrt{\bg} \left(\SR(\bg) - 2\Kkbar_k^{(0)}\right)  \nonumber  \\
&\quad - \frac{1}{16\pi G_k^{(1)} }  \int\md^d x \sqrt{\bg}\, \Big[-\bar{G}^{\mu\nu}-\Kkbar_k^{(1)} \bg^{\mu\nu}\Big] \flcb_{\mu\nu} \nonumber \\
&\quad - \frac{1}{2}  \int\md^d x \sqrt{\bg}\ \,\flcb^{\mu\nu} \,\,{\EAA^{\text{grav}\,(2)}_k[\bg,\bg]_{\mu\nu}}^{\rho\sigma}\,\, \flcb_{\rho\sigma}
+ \Order{\flcb^3} \label{eqn:trA08}
\end{align}
In the level-description, the background  and  dynamical  couplings appear in certain combinations in front of  invariants that have a definite level, i.e. order in $\flcb_{\mu\nu}$.
The two sets of coupling constants are related by
\begin{subequations}
\begin{align}
 &\frac{1}{G_k^{(0)}}=\frac{1}{G_k^{\background}}+\frac{1}{G_k^{\dyn}}\,, &&
\frac{\Kkbar_k^{(0)}}{G_k^{(0)}}=\frac{\Kkbar_k^{\background}}{G_k^{\background}}+\frac{\Kkbar_k^{\dyn}}{G_k^{\dyn}}\, , \label{eqn:trA08BA}\\
&\frac{1}{G_k^{(p)}}=\frac{1}{G_k^{\dyn}}\, \text{ for } p\geq 1, && \frac{\Kkbar_k^{(p)}}{G_k^{(p)}}=\frac{\Kkbar_k^{\dyn}}{G_k^{\dyn}}\, \text{ for }p\geq 1. \label{eqn:trA08BB}
\end{align}
\label{eqn:trA08B}
\end{subequations}
Notice that the level-(0) couplings, $G_k^{(0)}$ and $\Kkbar_k^{(0)}$, multiply pure background invariants and thus do not contribute to the dynamical field equations. 
They are, however, relevant to the statistical mechanics of black holes, for instance \cite{daniel1}.

In the present ansatz all couplings of higher level, $p\geq1$, are identical and agree with the dynamical (`$\dyn$') ones. 
However, the level-($1$) Newton and cosmological constants, $G_k^{(1)}\equiv G_k^{\dyn}$ and $\Kkbar_k^{(1)}\equiv \Kkbar_k^{\dyn}$, which enter the effective field equations and the tadpole equation, differ  in general from the level-(0) couplings.

{\bf\noindent (B)}
When the distinction of the different levels is artificially suppressed in the truncation ansatz by hypothesizing perfect split-symmetry along the entire RG trajectory, i.e. if we set $G_k^{(0)}= G_k^{(p)}\equiv G_k^{\sm}$ and $\Kkbar_k^{(0)}= \Kkbar_k^{(p)}\equiv\Kkbar_k^{\sm}$ for all $p$ and $k$, then the gravitational action $\EAA_k^{\text{grav}}[g,\bg]$  reduces to a functional of a single metric:
\begin{align}
  \EAA_k^{\text{grav}}[g,\bg]&= - \frac{1}{16\pi G_k^{\sm} } \int\md^d x \sqrt{g}\, \Big(\SR(g) - 2 \Kkbar_k^{\sm}\Big) \label{eqn:trA0sm}
\end{align}
The second argument of the action functional, $\bg$, has actually disappeared from the RHS of \eqref{eqn:trA0sm}. 
This approximation, for obvious reasons, is referred to as the {single-metric} Einstein-Hilbert truncation. 
In general split-symmetry is violated during the RG evolution, and a detailed comparison with the more advanced bi-metric ansatz \eqref{eqn:trA08} has revealed that there are even qualitative differences in the respective flows, especially in the crossover regime \cite{daniel2}.

{\bf\noindent (C)}
In the sequel we will analyze the RG flows obtained in three different RG studies.
One is based upon the single-metric ansatz \eqref{eqn:trA0sm}, while the other two are bi-metric calculations which employ the same, more general 4-parameter ansatz \eqref{eqn:trA08}, but differ in various details of the computational setting, the gauge choice in particular.

All three calculations use a gauge fixing action of the form
\begin{align}
\EAA_k^{\text{gf}}[g,\bg]&= \frac{1}{32\pi \alpha \,G^{\dyn\slash \sm}_k} \int\md^d x \sqrt{\bg} \,\,\bg^{\mu\nu} \Big[\mathcal{F}_{\mu}^{\alpha\beta}[\bg] \left(g_{\alpha\beta}-\bg_{\alpha\beta}\right) \Big] \Big[\mathcal{F}_{\nu}^{\rho\sigma}[\bg] \left(g_{\rho\sigma}-\bg_{\rho\sigma}\right) \Big]		\label{eqn:trA04}
\end{align}
It depends on the gauge parameter $\alpha$ and the coefficient $\varpi$ occurring in the gauge condition $\mathcal{F}_{\mu}^{\alpha\beta}[\bg] \, \flcb_{\mu\nu}\equiv \big(\delta^{\beta}_{\mu}\bg^{\alpha\gamma} \bZ_{\gamma}-\varpi \bg^{\alpha\beta}\bZ_{\mu}\big)\,\flcb_{\mu\nu}$.
Both $\alpha$ and $\varpi$ are $k$-independent by assumption.
The single-metric results obtained in \cite{mr} are based on the choice $(\varpi=1\slash2,\,\alpha=1)$, whereas the bi-metric calculations performed in \cite{MRS2}, henceforth denoted [\Rmnum{1}], and \cite{daniel2}, in the following referred to as [\Rmnum{2}], use $(\varpi=1\slash d,\, \alpha\rightarrow0$), and $(\varpi=1\slash2,\, \alpha=1)$, respectively. 

In this paper, the investigation of $\cfunc_k$ will be based upon the beta-functions derived in refs. \cite{mr}, [\Rmnum{1}], and [\Rmnum{2}], respectively. 

{\bf\noindent (D)}
The beta-functions describing the flow on theory space pertain to the dimensionless couplings $\tg_k^{\cix}\equiv k^{d-2}G_k^{\cix}$ and $\Kk_k^{\cix}\equiv k^{-2}\Kkbar_k^{\cix}$ for $\cix \in\{\background,\dyn,\sm,(0),(1),\cdots\}$.
In the bi-metric case, the 4 independent RG differential equations are partially decoupled, displaying the hierarchical structure \cite{daniel2}:
\begin{align}
\left(\tg_k^{\dyn\slash (p)},\,\Kk_k^{\dyn\slash (p)}\right) \rightarrow \tg_k^{\background \slash (0)} \rightarrow \Kk_k^{\background \slash (0)}\,, \qquad \text{ for } p\geq1
\end{align}
In order to solve the system of differential equations one starts by finding solutions of the $\dyn$ or $p\geq1$-sector, and then substitutes them successively into the decoupled RG equations of the $\background$- or level-(0) couplings, depending on which `language' one uses.

In [\Rmnum{2}] it  was shown  that the  beta-functions obtained for the different gauge choices in [\Rmnum{1}] and [\Rmnum{2}] yield the same qualitative results, with  only minor numerical differences. 
In particular, a UV-fixed point was found in both cases with remarkably stable properties under the change of gauge.
Quite surprisingly, its `$\dyn$'-coordinates agree quite well with the results from the single-metric approximation; in fact the latter turned out to be unusually reliable within this regime.

\begin{figure}[h!]
\centering
 \psfrag{a}[tc]{${\scriptstyle\color[RGB]{46,103,255}\text{\bf Type \Rmnum{3}a} }$}
  \psfrag{b}{${\scriptstyle \color[RGB]{23,52,127}\text{\bf Type \Rmnum{1}a} }$}
 \psfrag{c}[tr]{${\scriptstyle\color[RGB]{123,159,255}\text{\bf Type \Rmnum{2}a} }$}
 \psfrag{g}{${ \tg }$}
  \psfrag{l}{${ \Kk }$}
\includegraphics[width=0.5\textwidth]{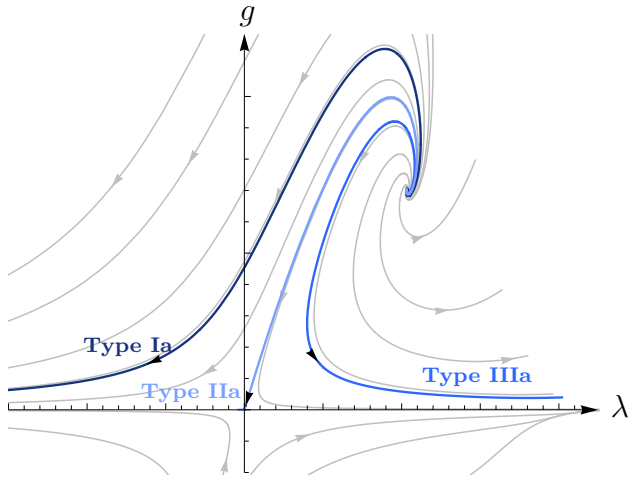}
 \caption{The schematic structure of the phase portrait on $\tg^{\sm}$-$\Kk^{\sm}$ or the projected $\tg^{\dyn}$-$\KkD$   plane  as predicted by all three  truncations considered. 
Here and in the following the arrows always point in the direction of decreasing $k$.
}\label{fig:typeStructure}
\end{figure}

Since the RG flows in the single- and bi-metric truncations are qualitatively similar their (projected) phase portraits in the $\tg^{\dyn}$-$\KkD$ and $\tg^{\sm}$-$\Kk^{\sm}$ plane, respectively, share the same overall structure, as depicted in Fig. \ref{fig:typeStructure}.
The integral curves in the upper half plane ($\tg^{\dyn\slash\sm}>0$) are classified as type \Rmnum{1}a, type \Rmnum{2}a, or type \Rmnum{3}a trajectories, depending on whether the cosmological constant $\Kk^{\dyn\slash\sm}$ approaches $-\infty$, $0$, or $+\infty$ in the IR,%
\footnote{%
The Einstein-Hilbert truncation is known to be inapplicable to type \Rmnum{3}a trajectories when $\Kk^{\dyn\slash{}\sm}$ approaches values of order unity.
In this paper we assume that their classical regime (having $\Kk^{\dyn\slash{}\sm}\ll 1$) represents their true $k\rightarrow0$ limit.
Even if ultimately this should turn out not to be the case, our treatment of the NGFP$\rightarrow$\crg{} crossover will remain valid.} respectively \cite{frank2}.
The type \Rmnum{2}a trajectory is a separatrix: it separates solutions with an ultimately positive  cosmological constant from those with a negative one at $k=0$. 
Likewise the trajectory $\tg^{\dyn\slash\sm}=0$ separates the upper and lower half plane, indicating that once the Newton coupling is chosen positive, it remains so on all scales.

The type \Rmnum{3}a trajectories display a generalized crossover of the kind mentioned in section \ref{subsec:2-06}, (B). 
It connects a fixed point in the UV to a classical regime in the IR.
The latter is located on its lower, almost horizontal branch where $\tg,\,\Kk\ll 1$ \cite{h3,entropy}.
\subsection{Gravitational instantons}

{\bf\noindent (A)} 
Let us now set up the tadpole equations  which result from the truncation ansatz $\Gamma_k=\Gamma_k^{\text{grav}}+\Gamma_k^{\text{gf}}+\Gamma_k^{\text{gh}}$.
To be consistent with the conventions in eq. \eqref{eqn:cm_mot_013} we must introduce background fields also for the ghosts, at least for a moment.
We decompose them as $\Ghx^{\mu}=\Xi^{\mu}+\eta^{\mu}$ and $\GhAx_{\mu}=\bar{\Xi}_{\mu}+\bar{\eta}_{\mu}$ where $(\Xi,\bar{\Xi})$ and $(\eta,\bar{\eta})$ denote their backgrounds and fluctuations, respectively.
Then the tadpole condition \eqref{eqn:cm_mot_013} amounts to the following three coupled equations for $\varphi\in \{\flcb,\,\eta,\,\bar{\eta}\}$: 
\begin{align}
0=\left.\frac{\delta \left(\Gamma_k^{\text{grav}}+\Gamma_k^{\text{gf}}+\Gamma_k^{\text{gh}}\right)}{\delta \varphi(x)}\right|_{\flcb=0,\eta=0,\bar{\eta}=0;\, \bg=\bg_k^{\scon},\, \Xi=\Xi^{\scon}_k,\, \bar{\Xi}=\bar{\Xi}^{\scon}_k}
\label{eqn:cm_inst_01}
\end{align}

If we begin by solving the equations involving $\delta \slash \delta \eta$ and $\delta \slash \delta \bar{\eta}$ and take advantage of the fact that $\Gamma_k^{\text{gh}}\propto \int (\bar{\Xi}+\bar{\eta})M(\Xi+\eta)$ is bilinear in the ghosts we conclude immediately that the only self-consistent background they admit for a non-degenerate $M$ is the trivial one,
$\Xi^{\scon}_k=0=\bar{\Xi}_k^{\scon}$.
As a consequence, the third equation, $\delta\Gamma_k\slash \delta \flcb\big|_{\cdots}=0$, receives no contribution from $\Gamma_k^{\text{gh}}$ since its $\flcb$-derivative vanishes upon inserting  the vanishing background ghosts and $\eta=0=\bar{\eta}$.
Furthermore, the gauge fixing action, too, does not contribute since $\Gamma_k^{\text{gf}}\propto \int (\mathcal{F}\flcb)^2$, being bilinear in $\flcb$, has a vanishing derivative at $\flcb=0$.

As a result, for every truncation ansatz of the above form, that is, for any choice of the `gravitational' piece $\Gamma_k^{\text{grav}}$, the tadpole condition for
$\bar{\Phi}^{\scon}_k\equiv \left(\bg_k^{\scon},\Xi_k^{\scon},\bar{\Xi}_k^{\scon}\right)=\left(\bg_k^{\scon},0,0\right)$
boils down to a single non-trivial equation, namely
\begin{align}
\left.\frac{\delta }{\delta \flcb_{\mu\nu}(x)}\Gamma_k^{\text{grav}}[\flcb;\bg]\right|_{\flcb=0; \bg=\bg_k^{\scon}}=0
\label{eqn:cm_inst_04}
\end{align}
This equation determines the self-consistent metrics which can `live' on a given spacetime manifold, $\mathcal{M}$, without being modified by the agitation of the quantum fluctuations.

For pure metric gravity, in truncations of the  type $\EAA_k=\EAA_k^{\rm grav}+\EAA_k^{\rm gf}+\EAA_k^{\rm gh}$, the $C$-function candidate $\cfunc_k\equiv \Gamma_k[\varphi=0;\bar{\Phi}_k^{\scon}]$ which we motivated above for a generic theory now becomes concretely
$\cfunc_k=\Gamma_k\big|_{\flcb=\eta=\bar{\eta}=\Xi=\bar{\Xi}=0, \bg=\bg_k^{\scon}}$.
It involves only the `grav'-part of the ansatz:
\begin{align}
\raalign{
\cfunc_k=\Gamma_k^{\text{grav}}[\flcb=0;\bg_k^{\scon}]}{}
\label{eqn:cm_instCandi}
\end{align}

{\bf\noindent (B)} 
In the special case of the  Einstein Hilbert truncation the tadpole equation \eqref{eqn:cm_inst_04} happens to have the same mathematical structure as the classical vacuum Einstein equation in presence of a cosmological constant.
 The $\flcb_{\mu\nu}$-derivative of the bi-metric ansatz \eqref{eqn:trA08} for $\EAA_k^{\text{grav}}$ yields at $\flcb_{\mu\nu}=0$:
\begin{align}
G_{\mu\nu}(\bg_k^{\scon})= - \Kkbar_k^{(1)} \, \bg^{\scon}_{k\,\mu\nu}\,, \quad 
\text{ or } \quad \Ric_{\mu\nu}(\bg_k^{\scon})=\tfrac{2}{d-2}\,\Kkbar_k^{(1)}\, \bg^{\scon}_{k\,\mu\nu}
\label{eqn:intro_eq01}
\end{align}
In the single-metric approximation the tadpole equation is the same, except that $\Kkbar^{(1)}_k$ is replaced with $\Kkbar_k^{\sm}$ then.
Thus $\bg_k^{\scon}$ is always an Einstein metric, and upon contraction we get from eq. \eqref{eqn:intro_eq01}:
\begin{align}
\SR(\bg_k^{\scon})=\frac{2\, d}{(d-2)}\,\Kkbar_k^{(1)}
\label{eqn:intro_eq02}
\end{align}
Inserting this expression for the curvature scalar into $\EAA_k^{\text{grav}}$ yields the following representation of the EAA, evaluated for $\flcb_{\mu\nu}=0$ and a self-consistent background geometry:
\begin{align}
\cfunc_k= \Gamma_k^{\text{grav}}[0;\bg_k^{\scon}]&= - \frac{1}{16\pi\,G_k^{(0)}}\int_{\MaFs}\md^d x\sqrt{\bg}\,\Big\{\SR(\bg)-2\Kkbar_k^{(0)}\Big\}\Big|_{\bg=\bg_k^{\scon}} \nonumber \\
&=-\frac{1}{8\pi\, G_k^{(0)}}\,\Big[ \left(\tfrac{d}{d-2}\right)\Kkbar_k^{(1)}-\Kkbar_k^{(0)}\Big]\, \vol(\MaFs,\bg_k^{\scon})
\label{eqn:intro_eq03}
\end{align}
Here $\vol(\MaFs,g)\equiv \int_{\MaFs}\md^d x \sqrt{g}$ denotes the Euclidean volume of the manifold $\MaFs$ measured with the metric written in the argument, $g_{\mu\nu}$.

Note that $\Gamma_k^{\text{grav}}$ evaluated at $(\flcb;\bg)=(0;\bg_k^{\scon})$ depends on both the level-(0) and the level-(1) couplings in a non-trivial way:
the former enter via the action $\EAA_k|_{\flcb=0}$ which has a level-(0) component only, the latter via the tadpole equation which is entirely `level-(1)'.

Assuming the running dimensionful couplings are regular at the scale $k$ considered, eq. \eqref{eqn:intro_eq03} shows that {$\cfunc_k$ is finite if, and only if, the spacetime manifold has finite volume.}
The self-consistent background being an Einstein metric, its curvature structure and other details play no role for the value of the action, it is only the volume that matters. 

{\bf\noindent (C)} 
Trying to find solutions to \eqref{eqn:intro_eq01} that exist for all scales from `$k=\infty$' down to $k=0$ the simplest situation arises when all metrics $\bg_k^{\scon}$, $k\in [0,\infty)$ can be put on {\it the same smooth manifold} $\MaFs$, leading in particular to the same spacetime topology at all scales, thus avoiding the delicate issue of a topological change.
This situation is realized, for example, if the level-(1) cosmological constant is positive on all scales, which is indeed the case along the type (\Rmnum{3}a) trajectories: $\Kkbar_k^{(1)}>0$, $k\in[0,\infty)$.

In the following we focus on precisely this situation.
The requirement of a finite  action is then met by a well studied class of Einstein spaces which exist for an arbitrary positive value of the cosmological constant, namely certain 4-dimensional gravitational instantons \cite{EGH, Besse, Page}, see Table \ref{tab:intro_tab01} for some examples. 
\begin{table}[h]
\begin{center}
\begin{tabular}{|l|c|c|}
\hline 
metric 			& $\MaFs$ 			& ${\cal V}$ 	\\ \hline 
Eucl. de Sitter & $S^4$				& $3\pi$ 		\\
Page 			& $P_2+\bar{P_2}$	& $1.8\pi$		\\
$S^2\times S^2$	& $S^2\times S^2$	& $ 2\pi$		\\
Fubini-Study	& $P_2(\noC)$		& $9\pi\slash 4$ \\
\hline
\end{tabular}
\caption{Various 4-dimensional gravitational instantons and the related normalized volumes ${\cal V}(\MaFs,\mathring{g})$. (See \cite{EGH} for a detailed account.)}
\label{tab:intro_tab01}
\end{center}
\end{table}

Let $\mathring{g}_{\mu\nu}$ be the metric of one such instanton, corresponding to a fixed reference value of the cosmological constant, $\mathring{\Kkbar}$, say.
Then the tadpole equation \eqref{eqn:intro_eq01}, at any $k$, is solved by the following rescaled metric \cite{jan1,jan2}:
\begin{align}
\bg^{\scon}_{k\,\mu\nu}=\frac{\mathring{\Kkbar}}{\Kkbar_k^{(1)}}\, \mathring{g}_{\mu\nu}
\label{eqn:intro_eq04}
\end{align}
As a result, the $k$-dependence of the total volume behaves as, for arbitrary $d$,
\begin{align}
\vol(\MaFs,\bg^{\scon}_k)=8\pi \big[\Kkbar_k^{(1)}\big]^{-d\slash 2}\,\cdot \, {\cal{V}}(\MaFs,\mathring{g})
\label{eqn:intro_eq05}
\end{align}

Here we introduced the dimensionless constant
\begin{align}
{\cal{V}}(\MaFs,\mathring{g})\equiv \frac{1}{8\pi}\, \mathring{\Kkbar}^{d\slash 2}\, \vol(\MaFs,\mathring{g})
\label{eqn:intro_eq06}
\end{align}
which is characteristic of the instanton under consideration.\footnote{It is closely related to the normalized volume $\tilde{v}(\MaFs,g)$ defined in the mathematical literature \cite{Besse, Carlip-V}.}
The number ${\cal{V}}$ is manifestly independent of $k$, and it is easy to see that it is also {\it independent of $\mathring{\Kkbar}$.}
The reason is that $\mathring{g}$ depends on $\mathring{\Kkbar}$ via the equation $\Ric_{\mu\nu}(\mathring{g})=\frac{2}{d-2}\,\mathring{\Kkbar}\,\mathring{g}_{\mu\nu}$. 
This implies that upon rescaling $\mathring{\Kkbar}$ by a constant factor, $\mathring{\Kkbar}\rightarrow c^2 \mathring{\Kkbar}$, the metric responds according to $\mathring{g}_{\mu\nu} \rightarrow c^{-2}\mathring{g}_{\mu\nu}$, and so the volume behaves as $\vol(\MaFs,\mathring{g})\rightarrow c^{-d}\,\vol(\MaFs,\mathring{g})$.
In the definition of ${\cal{V}}$, eq. \eqref{eqn:intro_eq06}, the factor $c^{-d}$ coming from the volume is therefore precisely canceled by a corresponding factor $c^{+d}$ which is produced by its prefactor, $\mathring{\Kkbar}^{d\slash 2} \rightarrow c^d\,\mathring{\Kkbar}^{d\slash 2}$.

Thus the value of ${\cal{V}}$ is a universal number which depends only on the type of the instanton considered\footnote{For a discussion of the topological properties of the normalized volume see \cite{Besse, Carlip-V}.}.
For the round metric on $S^d$ we find, for instance,
\begin{align}
{\cal{V}}(S^d) = \pi^{\frac{(d-1)}{2}} \,\frac{[(d-1)(d-2)]^{d\slash 2}}{2^{(d+4)\slash 2}\Gamma(\tfrac{d+1}{2})}
\label{eqn:intro_eq07}
\end{align}
Table \ref{tab:intro_tab01} contains the corresponding ${\cal{V}}$ values for some more examples in $d=4$.

{\bf\noindent (D)} 
Using \eqref{eqn:intro_eq05} in \eqref{eqn:intro_eq03} we obtain the following two equivalent representations of $\cfunc_k$:
\begin{align}
\cfunc_k &=-\frac{1}{G_k^{(0)}\, \big[\Kkbar_k^{(1)}\big]^{d\slash 2}} \, \left[\left(\tfrac{d}{d-2}\right)\Kkbar^{(1)}_k - \Kkbar_k^{(0)} \right] \, {\cal V}(\MaFs,\mathring{g}) \nonumber \\
&=- \frac{1}{g_k^{(0)}\left[\Kk_k^{(1)}\right]^{d\slash 2} } \left[\left(\tfrac{d}{d-2}\right)\Kk_k^{(1)}-\Kk_k^{(0)}\right]\, {\cal V}(\MaFs,\mathring{g}) 
\label{eqn:intro_eq08}
\end{align}
In the second line of \eqref{eqn:intro_eq08} we eliminated the dimensionful quantities $G_k^{(p)}$ and $\Kkbar_k^{(p)}$ in favor of their dimensionless analogs whereby all explicit factors of $k$ dropped out.
 
We observe that the result \eqref{eqn:intro_eq08} for the function $k\mapsto \cfunc_k$ has the general structure
\begin{align}
\cfunc_k\equiv \cfunc(\tg_k^{(0)},\Kk_k^{(0)},\Kk_k^{(1)})=\cF(g_k^{(0)}, \Kk_k^{(0)},\Kk_k^{(1)})\, {\cal V}(\MaFs,\mathring{g})
\label{eqn:intro_eq09}
\end{align}
Here $\cF(\,\cdot\,)\equiv \cfunc(\,\cdot\,)\slash {\cal{V}}$ stands for the following function over  theory space:
\begin{align}
\raalign{
\cF(g^{(0)}, \Kk^{(0)},\Kk^{(1)})= - \frac{\left[\left(\tfrac{d}{d-2}\right)\Kk^{(1)}-\Kk^{(0)}\right]}{g^{(0)}\, [\Kk^{(1)}]^{d\slash 2}}
}{}
\label{eqn:intro_eq10}
\end{align}
Equation \eqref{eqn:intro_eq10} represents our main result.
We shall study its properties below.
In 4 dimensions we have in particular 
\begin{align}
\cF(g^{(0)}, \Kk^{(0)},\Kk^{(1)})=- \frac{2\Kk^{(1)}-\Kk^{(0)}}{g^{(0)}\, (\Kk^{(1)})^2} \qquad \qquad \qquad \text{ $(d=4)$}
\label{eqn:intro_eq11}
\end{align}
Several comments are in order here.
\vspace{6px}

\noindent {\bf(1)}
The function $\cfunc$ depends on both the RG trajectory and on the solution to the running self-consistency condition, along this very trajectory, that has been picked.
In eq. \eqref{eqn:intro_eq09} those two  dependencies factorize: the former enters via the function $\cF$, the latter via the constant factor ${\cal V}(\MaFs,\mathring{g})$ that characterizes the gravitational instanton.

\noindent {\bf (2)}
The dependence on the RG trajectory, parametrized as $k\mapsto \big(g_k^{(0,1)},\Kk_k^{(0,1)}\big)$, is obtained by evaluating a {\it scalar function on theory space} along this curve, namely $\cF:\mathcal{T}\rightarrow \noR$, $\big(g^{(0,1)},\Kk^{(0,1)}\big)  \mapsto \cF(g^{(0)}, \Kk^{(0)},\Kk^{(1)})$.
It is defined at all points of $\mathcal{T}$ where $\tg^{(0)}\neq 0$ and $\Kk^{(1)}\neq 0$, and turns out to be actually independent of $\tg^{(1)}$.

\noindent {\bf (3)}
We shall refer to $\cF_k\equiv\cF(\tg_k^{(0)},\,\Kk_k^{(0)},\,\Kk_k^{(1)} )\equiv \cfunc_k\slash {\cal V}(\MaFs,\mathring{g})$ and $\cF(\,\cdot\,)\equiv \cfunc(\,\cdot\,)\slash {\cal V}(\MaFs,\mathring{g})$ as the {\it reduced $\cfunc_k$ and $\cfunc(\,\cdot\,)$ functions,} respectively.

\noindent {\bf (4)}
Denoting the collection of running couplings by $u(k)$, we may write the scale derivative of $\cfunc_k$ as 
$k\partial_k \cfunc_k= {\cal V}(\MaFs,\mathring{g}) \left(\beta_{\alpha}\frac{\partial}{\partial u_{\alpha}} \cF\right)(u(k))$
which involves the directional derivative $\vec{\beta}\cdot\vec{\nabla}$ acting upon scalar functions on theory space.
This motivates defining the subset $\mathcal{T}_+$ of $\mathcal{T}$ on which $\cF:\mathcal{T}\to \noR$ has a positive directional derivative in the direction of $\vec{\beta}$:
\begin{align}
\mathcal{T}_+\equiv\left\{u\in \mathcal{T}\, |\, \beta_{\alpha}(u)\tfrac{\partial}{\partial u_{\alpha}} \cF(u)> 0  \right\}\,.
\label{eqn:cm_inst_3plus}
\end{align}
The interpretation is that $\cfunc_k$ increases monotonically with $k$ along those (parts of) RG trajectories that lie  entirely inside $\mathcal{T}_+$, i.e. $k \partial_k \cfunc_k>0$ at all points of $\mathcal{T}_+$.

\noindent {\bf (5)}
Invoking the idealization of exact split-symmetry, i.e. assuming that the Newton and cosmological constants of different levels are all equal $(\tg_k^{(p)}\equiv \tg^{\sm}_k, \, \Kk_k^{(p)}\equiv \Kk^{\sm}_k,\, p=0,1,2,\cdots)$, we obtain $\cfunc_k$ in the single-metric approximation.
It reads $\cfunc_k=\cfunc(\tg_k^{\sm},\Kk_k^{\sm})=\cF^{\sm}(\tg_k^{\sm},\Kk_k^{\sm}){\cal V}(\MaFs,\mathring{g}) $ with
\begin{align}
\cF^{\sm}(\tg^{\sm},\Kk^{\sm})&= - \left(\frac{2}{d-2}\right) \, \frac{1}{\tg^{\sm}\, (\Kk^{\sm})^{d\slash 2 -1}}
\label{eqn:intro_eq12}
\end{align}
Note that it depends only on the dimensionless combination $G_k^{\sm}(\Kkbar_k^{\sm})^{d\slash2-1}=\tg^{\sm}_k\, (\Kk^{\sm}_k)^{d\slash 2 -1}$ whose robustness properties under changes of the cutoff and the gauge fixing has often been used to check the reliability of single-metric truncations \cite{oliver1, oliver2,prop, frank1}.

\noindent {\bf (6)}
In general, the beta-functions which govern the RG evolution of $\EAA_k$ may depend on the topology of $\MaFs$, see \cite{creh2} for an example.
Within the truncation considered here this is not the case, however, the reason being the universality of the heat-kernel asymptotics which is exploited in the computation of the beta-functions \cite{MRS2,daniel2}.

\noindent {\bf (7)}
Switching from the level language, which employs the couplings $\tg^{(p)}$ and $\Kk^{(p)}$, to the $\background$-$\dyn$ language, based upon the couplings $\{\tg^{\dyn},\,\KkD,\,\tg_k^{\background},\,\Kk^{\background}\}$, with
\begin{align}
\frac{1}{\tg^{(0)}}= \frac{1}{\tg^{\background}} + \frac{1}{\tg^{\dyn}}\,, \quad 
\frac{\Kk^{(0)}}{\tg^{(0)}}= \frac{\KkB}{\tg^{\background}} + \frac{\KkD}{\tg^{\dyn}} \quad
\text{ and } \tg^{(1)}=\tg^{\dyn},\quad \Kk^{(1)}=\KkD\,,
\nonumber 
\end{align}
the function $\cF:\mathcal{T}\rightarrow \noR$ assumes the following form, for $d=4$,
\begin{align}
\cF(\tg^{\dyn},\KkD,\tg_k^{\background},\Kk^{\background})=
- 
\frac{1}{\tg^{\dyn}\KkD} - \frac{1}{\tg^{\background}\KkD}\Big[2- \frac{\KkB}{\KkD}\Big]
\label{eqn:intro_eq13}
\end{align}
This representation is particularly convenient for part of the numerical analyses to which we turn in the next section.
\subsection{Numerical results}\label{sec:3-03}

Above we investigated the general properties of $\cfunc_k$ and we derived its explicit form for the Einstein-Hilbert truncation. 
In this subsection we will study the monotonicity properties of $\cfunc_k$ in 4 spacetime dimensions for solutions of the FRGE in both the single-metric \cite{mr} and the two bi-metric approaches [\Rmnum{1}], [\Rmnum{2}].
We will focus on type \Rmnum{3}a trajectories (see Fig.\ref{fig:typeStructure}) which exhibit the aforementioned NGFP$\rightarrow$\crg{} crossover.
Possibly those solutions are relevant to real Nature even \cite{h3,entropy}.
For the corresponding discussion of type \Rmnum{1}a and type \Rmnum{2}a trajectories we refer to appendix \ref{app:cmB}.


{\noindent\bf (i)}
Returning to $\cfunc_k$ for the bi-metric truncation given in eq. \eqref{eqn:intro_eq09}, the entire information on the RG trajectory is contained in the factor
\begin{align}
\cF_k\equiv \cF(g_k^{(0)}, \Kk_k^{(0)},\Kk_k^{(1)})= - \frac{2\Kk_k^{(1)}-\Kk_k^{(0)}}{g_k^{(0)}\, [\Kk_k^{(1)}]^{2}}
\label{eq:cm_nr01}
\end{align}
We are going to evaluate this function of $k$ for a number of RG trajectories on the {4-dimensional} theory space which we generate numerically.

{\noindent\bf (ii)}
Likewise we  compute the reduced $\cfunc_k$-function predicted by the single-metric truncation, that is, we evaluate
\begin{align}
\cF_k^{\sm}\equiv \cF^{\sm}(\tg_k^{\sm},\Kk_k^{\sm})= -\frac{1}{\tg_k^{\sm}\Kk^{\sm}_k}
\label{eq:cm_nr02}
\end{align}
for trajectories on the corresponding {2-dimensional} theory space.
We obtain them numerically by solving a system consisting of 2 differential equations only.

{\noindent\bf (iii)}
We are also going to perform a {\it hybrid calculation} which is intermediate between the 2- and 4-dimensional treatment, in the following sense.

As a rule, the single-metric approximation to a bi-metric truncation is a valid description of the flow if split-symmetry is only weakly broken, i.e. there is no significant difference between couplings at different levels: $u_{\alpha}^{(0)}=u_{\alpha}^{(1)}=u_{\alpha}^{(2)}=\cdots$.
If we make the corresponding identifications $\Kk_k^{(0)}=\Kk_k^{(1)}=\cdots\equiv \Kk_k$ and $\tg_k^{(0)}=\tg^{(1)}_k=\cdots\equiv \tg_k$ in \eqref{eq:cm_nr01} we obtain
\begin{align}
\cF^{\spsym}_k\equiv Y(\tg_k,\Kk_k,\Kk_k)= -\frac{1}{\tg_k\Kk_k}
\label{eq:cm_nr03}
\end{align}
Here $(\tg_k,\,\Kk_k)$ stands for $(\tg_k^{(0)},\,\Kk_k^{(0)})$ or, what should be the same, $(\tg_k^{(1)},\,\Kk_k^{(1)})\equiv (\tg_k^{\dyn},\,\Kk_k^{\dyn})$ as obtained from the bi-metric RG equations.
If the trajectory respects split-symmetry  
it does not matter from which level we take the couplings.
If split-symmetry is not perfect, it does however matter from which level they come, and  we obtain two, in general different functions:
\begin{align}
\cF_k^{\spsym, (0)}\equiv - \frac{1}{\tg_k^{(0)}\Kk_k^{(0)}}\,,\qquad  
\cF_k^{\spsym, (1)}\equiv - \frac{1}{\tg_k^{(1)}\Kk_k^{(1)}} 
\label{eqn:cm_nr0s}
\end{align}
It will be instructive to compare the two functions \eqref{eqn:cm_nr0s} for various representative trajectories on 4-dimensional theory space.
This will provide us  with some insights about what is more important in an approximate calculation of $\cfunc_k$: good control over the details of the underlying RG trajectory, or precise (analytic) knowledge about how $\EAA_k[0;\bar{\Phi}_k^{\scon}]$ depends on the couplings from the various levels when split-symmetry is broken.

Note that while \eqref{eq:cm_nr03} and \eqref{eqn:cm_nr0s} have the same {\it structure} as the single-metric result \eqref{eq:cm_nr02}, there is a crucial difference:
the former $\cF_k$-functions involve running couplings obtained from the 4-dimensional bi-metric system of RG equations, whereas the latter, $\cF_k^{\sm}$, has the solutions to the 2-dimensional single-metric flow equations as its input.

Note also that by virtue of \eqref{eqn:intro_eq13} the full-fledged bi-metric $\cF_k$ may be written  as 
\begin{align}
\cF_k=\cF_k^{\spsym,(1)}+\Delta \cF_k\quad \text{ with }\quad\Delta \cF_k\equiv - \frac{1}{\tg_k^{\background}\,\KkD_k} \left[2-\frac{\KkB_k}{\KkD_k}\right]
\label{eqn:cm_nrDelta}
\end{align}
The magnitude of the $\Delta\cF_k$-term is  a measure for the degree of split-symmetry violation as $\Delta \cF_k=0$ when the symmetry is exact%
\footnote{Of course, this can also be seen directly.
Reinstating dimensionful couplings, the two contributions to $\Delta\cF_k=-\frac{1}{G_k^{\background}\Kkbar_k^{\dyn}}\left[2-\frac{\Kkbar_k^{\background}}{\Kkbar_k^{\dyn}} \right]$ are proportional to $1\slash G_k^{\background}$ and $\Kkbar_k^{\background}\slash G_k^{\background}$, respectively.
Those quantities are the prefactors of the monomials responsible for the {\it extra} background dependence of $\EAA_k$, and so they must vanish to achieve split-symmetry.} 
and $\cF_k\equiv \cF_k^{\spsym,(p)}$ for all $p=0,1,2,\cdots$.

\subsubsection{Single-metric truncation}
We begin the computation of the reduced single-metric $\cfunc_k$-function by numerically calculating a number of type \Rmnum{3}a trajectories on the 2-dimensional $\tg^{\sm}$-$\Kk^{\sm}$ theory space, and then evaluate $\cF_k^{\sm}$ for them.
We find that the reduced $\cfunc_k$-functions thus obtained always have the same qualitative properties:
they become stationary (approach plateaus) for $k\rightarrow\infty$ and $k\rightarrow0$, but {\it they are not on all scales monotonically increasing with $k$.}
For all trajectories there exists a regime of scales where $\partial_k\cfunc_k<0$.
This negative derivative typically occurs while the trajectory crosses over from the NGFP to its turning point close to the GFP.
\begin{figure}[!ht]
\centering
\psfrag{b}[cm][0][1][90]{${\scriptscriptstyle k\partial_k 1\slash \cF^{\sm}_k}$}
\psfrag{a}{${\scriptscriptstyle k \slash m_{\text{Pl}}}$}
\psfrag{k}{${\scriptstyle k \slash m_{\text{Pl}}}$}
\psfrag{c}[r]{${\scriptstyle 1\slash \cF^{\sm}_k }$}             
\includegraphics[width=0.65\textwidth]{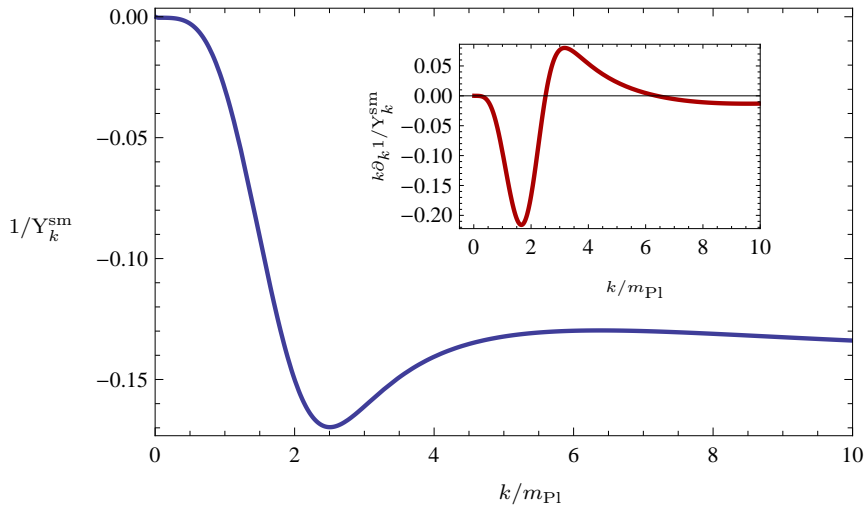}
\caption{The inverse of $\cF^{\sm}_k$ for a typical single-metric type \Rmnum{3}a trajectory. 
The inset shows its $k$-derivative whose positive values indicate a violation of monotonicity.} \label{fig:cfuncCorrectedSM}
\end{figure}

In Fig. \ref{fig:cfuncCorrectedSM} we display a representative single-metric example.
For reasons of a clearer presentation we plot here, and in all analogous diagrams that will follow, {\it the inverse} of the reduced $\cfunc_k$ function, along with its derivative.
Furthermore, here and in the following, the scale $k$ is always measured in units of the Planck mass defined by the classical regime, $m_{\text{Pl}}\equiv 1\slash \sqrt{G_{\crg}}$.
The example of Fig. \ref{fig:cfuncCorrectedSM} shows the `wrong sign' of the scale derivative ($\partial_k \cF_k^{\sm}<0$) for $k$ in an interval between about 3 and 5 Planck masses, which is the typical order of magnitude.

\subsubsection{Bi-metric Einstein-Hilbert truncation}
Turning now to the bi-metric Einstein-Hilbert truncation with its 4-dimensional theory space we employ the two sets of RG equations from [\Rmnum{1}] and [\Rmnum{2}], respectively, and compare the results they imply.

Furthermore, we must distinguish two fundamentally different cases with respect to the RG trajectories, namely trajectories which restore split-symmetry in the IR, and trajectories which do not.

In either case we begin by numerically computing a type \Rmnum{3}a trajectory of the decoupled $\tg^{\dyn}$-$\KkD$ subsystem.
Then, when we `lift' this 2D trajectory to a 4D one, we must pick initial conditions for $\tg^{\background}$ and $\KkB$, and it is at this point that we must decide about restoring, or not restoring the symmetry.
As we showed in detail in ref. \cite{daniel2}, the requirement of split-symmetry implies uniquely fixed values for the couplings $\tg_k^{\background}$ and $\KkB_k$ in the limit $k\searrow0$, namely precisely the coordinates $\left(\tg_{\attr}^{\background}(k),\,\KkB_{\attr}(k)\right)$ of the running UV-attractor.
We now discuss the cases with and without symmetry restoration in turn.

{\noindent\bf (A) Split-symmetry restoring trajectories.}
Opting for the symmetry restoring IR values of the $\background$-couplings, what remains free to vary is the underlying type \Rmnum{3}a trajectory in the $\dyn$ sector.
We find that the qualitative properties of the resulting functions $\cF_k$ are the same for all trajectories of this type, and that these properties do not depend on whether we use the RG equations from [\Rmnum{1}] or from [\Rmnum{2}].
\begin{figure}[!ht]
\centering
\psfrag{c}[cm][0][1][90]{${\scriptscriptstyle k\partial_k 1\slash \cF_k}$}
\psfrag{b}[cm][0][1][90]{${\scriptscriptstyle k\partial_k  1\slash \cF_k}$}
\psfrag{a}{${\scriptscriptstyle k \slash m_{\text{Pl}}}$}
\psfrag{k}{${\scriptstyle k \slash m_{\text{Pl}}}$}
\psfrag{x}[r]{${\scriptstyle 1\slash \cF_k }$}        
 \subfloat{
 \psfrag{c}[cm][0][1][90]{${\scriptstyle 1\slash \cF_k }$}  
 \includegraphics[width=0.450\textwidth]{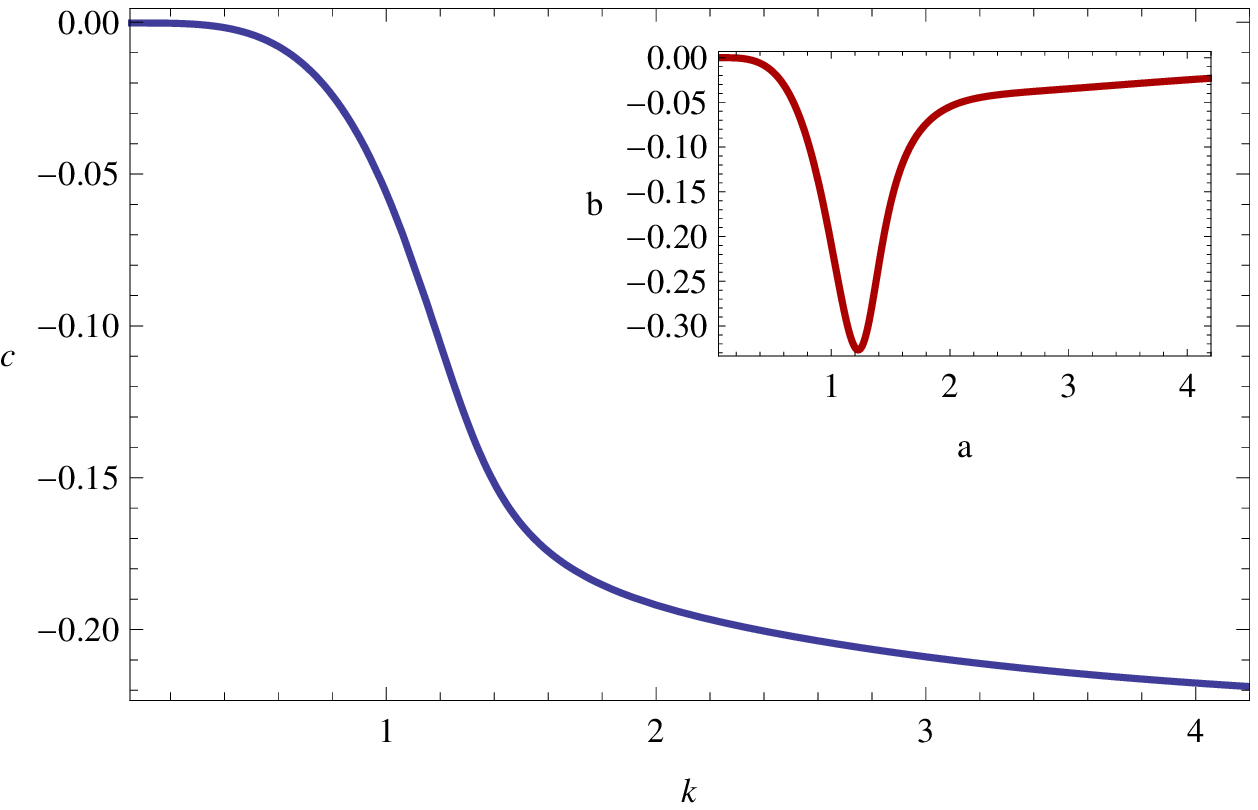}
 } 
 \hspace{0.04\textwidth} %
 \subfloat{%
\includegraphics[width=0.455\textwidth]{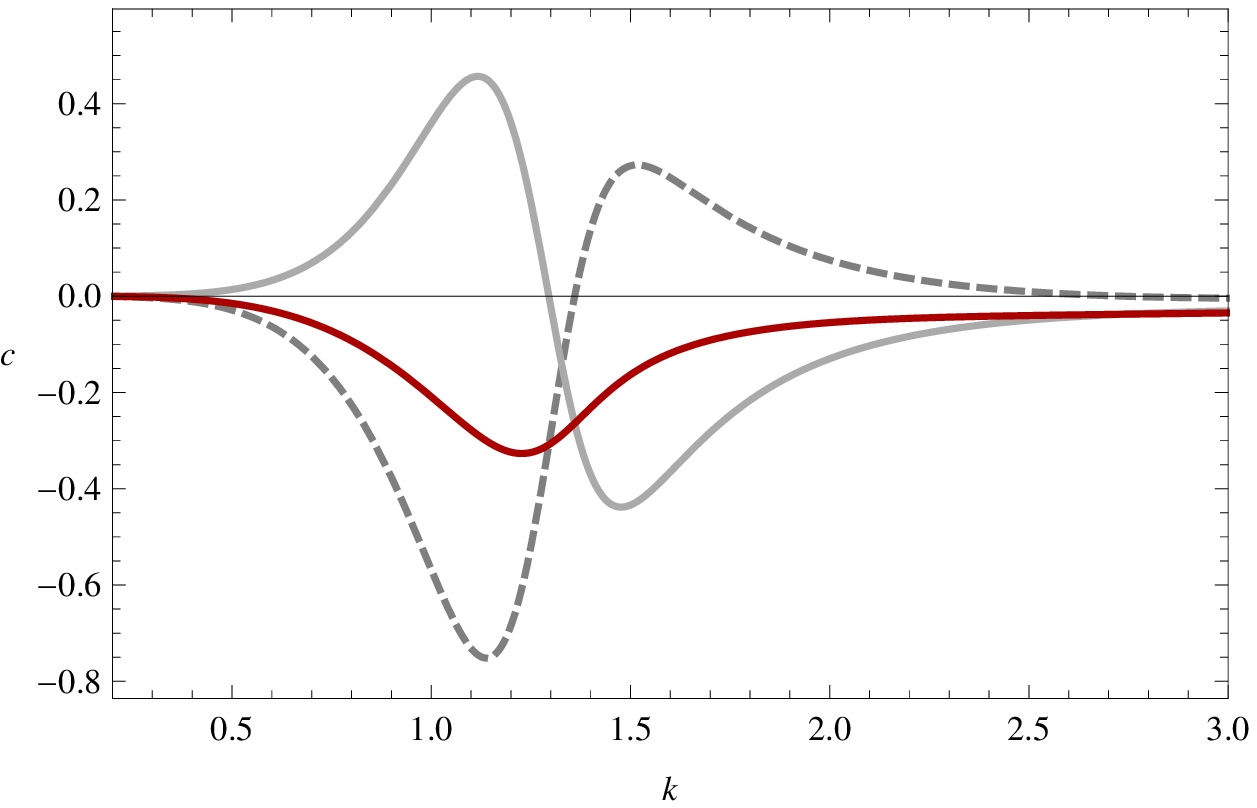}
}
\caption{The left plot shows $1\slash\cF_k$ and its scale derivative for a typical bi-metric type \Rmnum{3}a trajectory that restores split-symmetry in the  IR.
It is based on the RG equations of [\Rmnum{1}]. 
For these trajectories, $\cfunc_k$ is always found to be perfectly monotone.
The inset in the left plot shows $k\partial_k 1\slash \cF_k$, which is decomposed in the right plot into the derivative of the  split-symmetric component $1\slash \cF_k^{\spsym,(1)}$ (dashed, gray curve) and  of  $\Delta(1\slash \cF_k)$ (solid, gray curve). 
Neither  of the two contributions is negative definite separately, but their sum is (solid, dark red curve).} \label{fig:cfuncCorrectedBMMRS}
\end{figure}
\begin{figure}[!ht]
\centering
\psfrag{c}[cm][0][1][90]{${\scriptscriptstyle k\partial_k 1\slash \cF_k}$}
\psfrag{b}[cm][0][1][90]{${\scriptscriptstyle k\partial_k  1\slash \cF_k}$}
\psfrag{a}{${\scriptscriptstyle k \slash m_{\text{Pl}}}$}
\psfrag{k}{${\scriptstyle k \slash m_{\text{Pl}}}$}
\psfrag{x}[r]{${\scriptstyle 1\slash \cF_k }$}        
 \subfloat{
 \psfrag{c}[cm][0][1][90]{${\scriptstyle 1\slash \cF_k }$}  
 \includegraphics[width=0.450\textwidth]{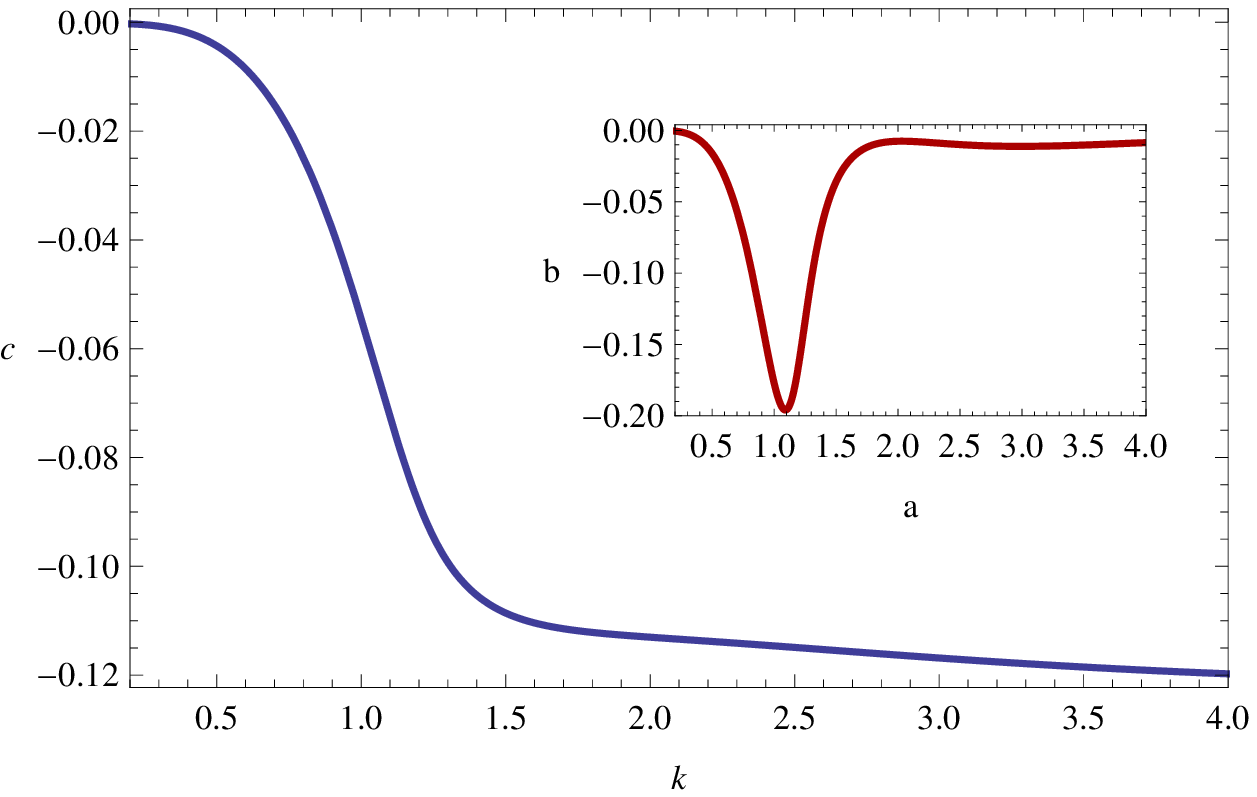}
 } 
 \hspace{0.04\textwidth} %
 \subfloat{%
\includegraphics[width=0.455\textwidth]{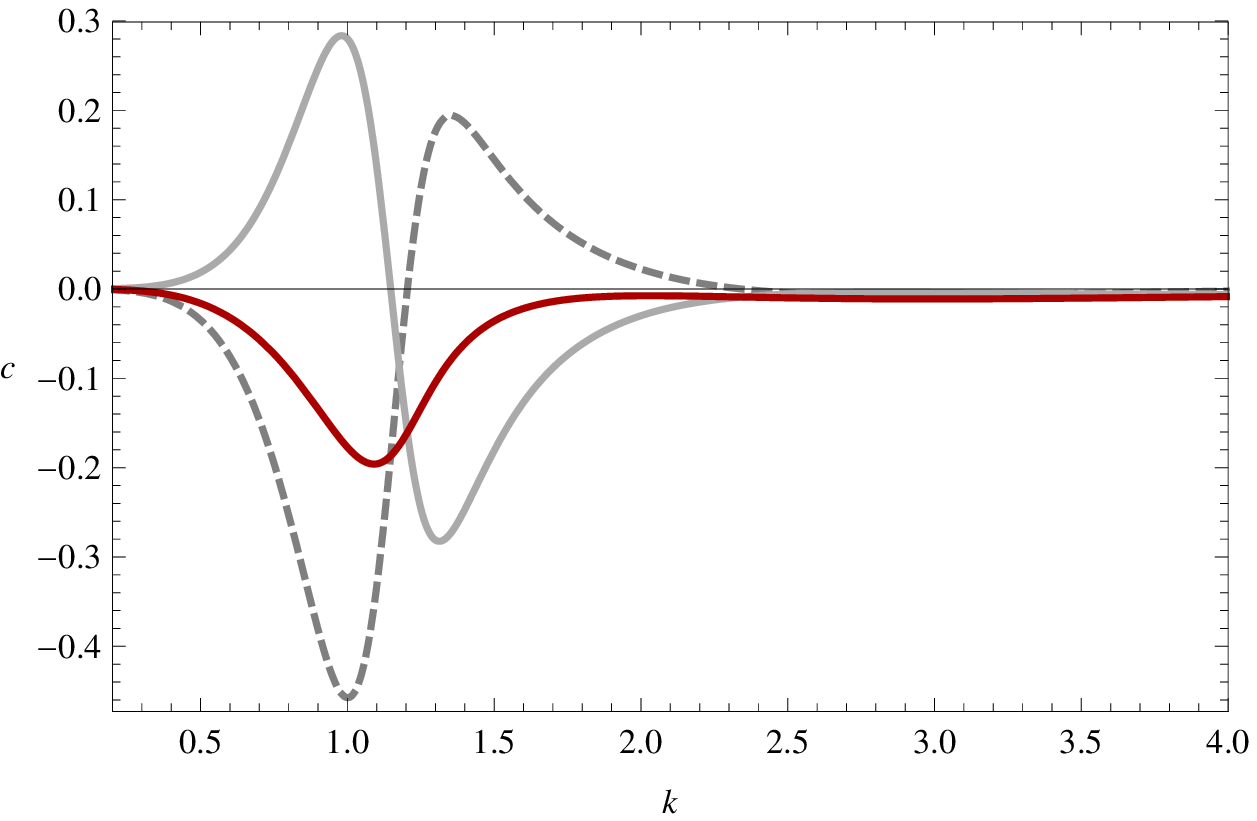}
}
\caption{The function $1\slash \cF_k$ as in Fig. \ref{fig:cfuncCorrectedBMMRS}, but now based on the RG equations of [\Rmnum{2}].} \label{fig:cfuncCorrectedBMD2}
\end{figure}
The picture is always the following:
The reduced $\cfunc_k$-function $\cF_k$ approaches plateau values in the fixed point and in the classical regime, i.e. it becomes stationary there, and {\it it is a strictly monotone function of $k$ on all scales in between,} $k\partial_k \cF_k>0$.

Clearly the latter property is in marked contrast with the single-metric results.
In Fig. \ref{fig:cfuncCorrectedBMMRS} we show $\cF_k$ for a representative \Rmnum{3}a-trajectory.
The plots were obtained with the RG equations derived in [\Rmnum{1}].
Their analogs based on the equations from [\Rmnum{2}] are displayed in Fig. \ref{fig:cfuncCorrectedBMD2}.
It is gratifying to see that there is hardly any difference between the results from the two calculational schemes.

In eq. \eqref{eqn:cm_nrDelta} we decomposed $\cF_k$ as $\cF_k=\cF_k^{\spsym,(1)}+\Delta\cF_k$ in order to make its split-symmetry violating part explicit.
For $1\slash \cF_k$ we have correspondingly $1\slash \cF_k=1\slash \cF_k^{\spsym,(1)}+\Delta(1\slash \cF_k)$ with $\Delta(1\slash\cF_k)=- \Delta\cF_k\slash (\cF_k^{\spsym,(1)} \cF_k)$ and exact split-symmetry ($\Delta \cF_k=0$) amounts to $\Delta(1\slash \cF_k)=0$, of course.

In the Figs. \ref{fig:cfuncCorrectedBMMRS} and \ref{fig:cfuncCorrectedBMD2} we show how the  scale derivative of $1\slash \cF_k$ decomposes into the derivative of $1\slash \cF_k^{\spsym,(1)}$ and of the symmetry violation term $\Delta (1\slash\cF_k)$.
For all trajectories of the class considered, and with the RG equations from both [\Rmnum{1}] and [\Rmnum{2}], we always find that the scale derivative of neither $1\slash \cF_k^{\spsym,(1)}$, nor of $\Delta(1\slash \cF_k)$ is negative definite separately, but their sum is!

When split-symmetry is intact, $\Delta \cF_k=0$, (violation of) monotonicity for $1\slash \cF_k^{\spsym,(1)}$ is equivalent to a (non-) monotone function $1\slash \cF_k$.
Now, for {\it generic} RG trajectories from the bi-metric calculations [\Rmnum{1}] and [\Rmnum{2}] this condition is known to be approximately satisfied  only for $k\rightarrow\infty$, i.e. in the vicinity of the NGFP.
The trajectories considered in the present paragraph are fine-tuned to fulfill the requirement of  split-symmetry restoration in the IR, so $\Delta \cF_k$ vanishes also there.
But on all intermediate scales split-symmetry is broken, the monotonicity of $\cfunc_k$ is not guaranteed by any general argument, and in general $\cF_k\neq \cF_k^{\spsym,(1)}$.
And indeed Figs. \ref{fig:cfuncCorrectedBMMRS} and \ref{fig:cfuncCorrectedBMD2} show a strong violation of this equality. 
In fact, the part $\cF_k^{\spsym,(1)}$ is seen to be non-monotone exactly in the regime where the split-symmetry of the RG trajectory is known to be significantly broken.

The status of split-symmetry violation displayed by an RG trajectory is thus also reflected by the deviation of the pertinent $\cF_k$  from its `split-symmetry enforced' version $\cF_k^{\spsym,(1)}$.
Quite remarkably, for the present class of trajectories {\it a perfect compensation of the split-symmetry violation the RG trajectories suffer from, and a nonzero correction term $\Delta \cF_k$ takes place.} 
Miraculously,  the term $\Delta\cF_k$ modifies the non-monotone $\cF_k^{\spsym,(1)}$ in precisely such a way that the total $\cF_k$ is monotone.

This perfect compensation {\it for all eligible trajectories} strongly supports our hope that the candidate $\cfunc_k$-function really qualifies as a `$C$-function' since the structure of $\Delta \cF_k$, i.e. the way how it depends on the couplings, is a direct consequence of having set $\cfunc_k=\EAA_k[0;\bar{\Phi}_k^{\scon}]$.
It is indeed surprising to see that a function as simples as the $\Delta\cF_k$ of eq. \eqref{eqn:cm_nrDelta} can do the job of rendering $\cfunc_k$ monotone for all physically relevant trajectories at once.

{\noindent\bf (B) Split-symmetry violating trajectories.}
We continue to use the bi-metric RG equations from [\Rmnum{1}] and [\Rmnum{2}], but now we deliberately break split-symmetry by selecting a generic trajectory in the $\tg^{\background}$-$\KkB$-subspace, one that would {\it not} hit the running UV-attractor for $k\searrow0$.
After having generated solutions $k\mapsto \left(\tg_k^{\dyn},\,\KkD_k\right)$ of the two decoupled $\dyn$-equations, again corresponding to a type \Rmnum{3}a trajectory, we solve the resulting $\background$-equations with initial values for $(\tg_{k}^{\background},\,\KkB_{k})$ that explicitly break split-symmetry even in the IR.

\begin{figure}[!ht]
\centering
\psfrag{c}[cm][0][1][90]{${\scriptscriptstyle k\partial_k 1\slash \cF_k}$}
\psfrag{b}[cm][0][1][90]{${\scriptscriptstyle k\partial_k  1\slash \cF_k}$}
\psfrag{a}{${\scriptscriptstyle k \slash m_{\text{Pl}}}$}
\psfrag{k}{${\scriptstyle k \slash m_{\text{Pl}}}$}
\psfrag{x}[r]{${\scriptstyle 1\slash \cF_k }$}        
 \subfloat{
 \psfrag{c}[cm][0][1][90]{${\scriptstyle 1\slash \cF_k }$}  
 \includegraphics[width=0.450\textwidth]{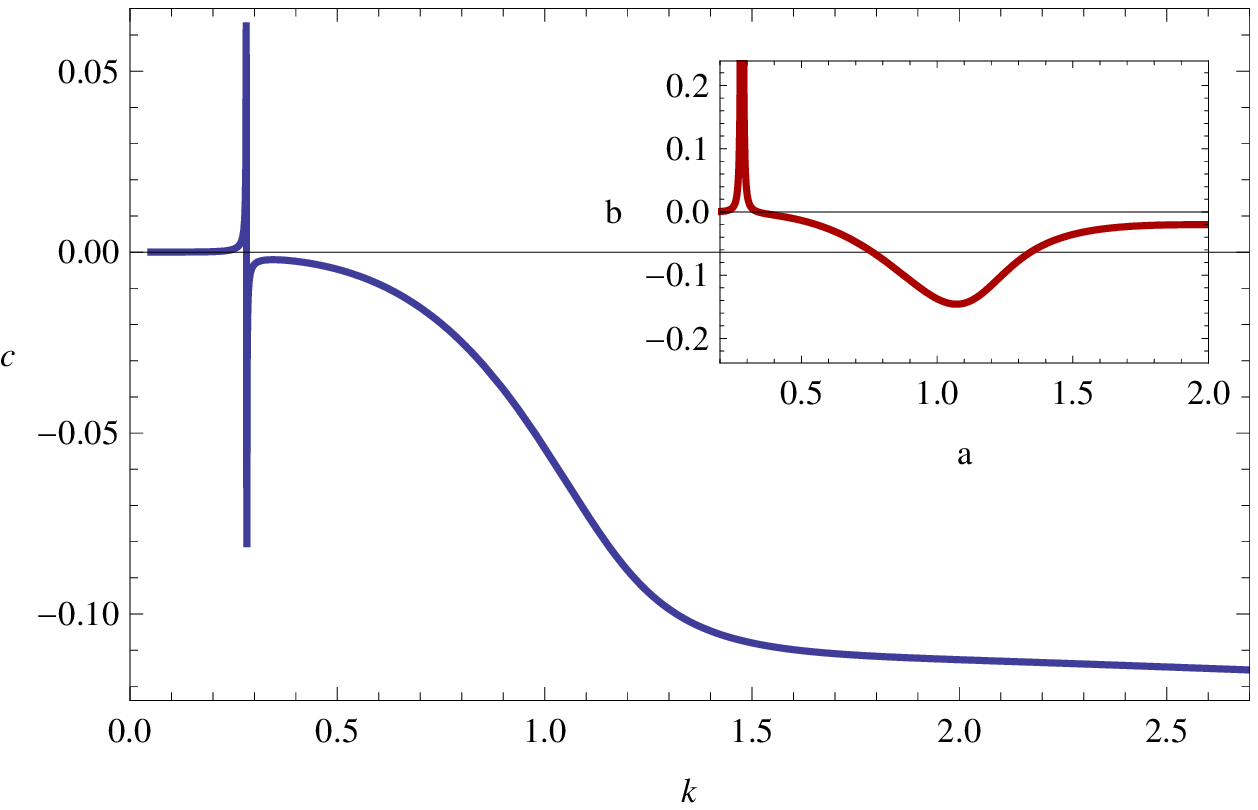}
 } 
 \hspace{0.04\textwidth} %
 \subfloat{%
\includegraphics[width=0.455\textwidth]{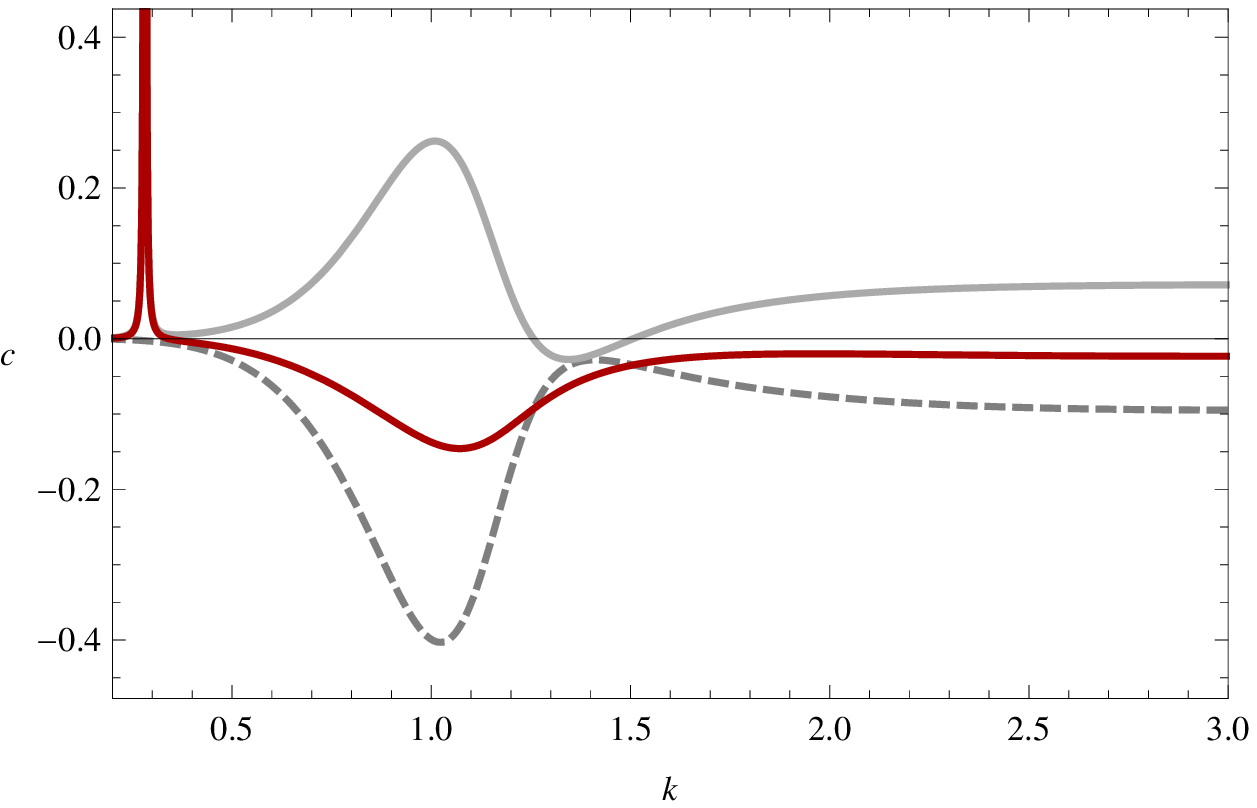}
}
\caption{The left plot shows the function $1\slash \cF_k$ for a bi-metric trajectory of type \Rmnum{3}a that does {\it not} restore split-symmetry in the IR.
It is based on the RG equations of [\Rmnum{1}].
We observe a sign-change of $k\partial_k (1\slash \cF_k)$ at moderate values of $k$, indicating a violation of monotonicity.
In the decomposed form of $k\partial_k( 1\slash \cF_k)$, shown in the right plot, we see that the contribution $1\slash \cF^{\spsym,(1)}_k$ is in fact monotone, but the correction term $\Delta (1\slash \cF_k)$ is not, and neither is their sum. 
Not restoring split-symmetry in the IR results in a violation of the monotonicity of $\cfunc_k$ along the trajectory considered.} \label{fig:cfuncCorrectedBMMRS2vio}
\end{figure}

\begin{figure}[!ht]
\centering
\psfrag{c}[cm][0][1][90]{${\scriptscriptstyle k\partial_k 1\slash \cF_k}$}
\psfrag{b}[cm][0][1][90]{${\scriptscriptstyle k\partial_k  1\slash \cF_k}$}
\psfrag{a}{${\scriptscriptstyle k \slash m_{\text{Pl}}}$}
\psfrag{k}{${\scriptstyle k \slash m_{\text{Pl}}}$}
\psfrag{x}[r]{${\scriptstyle 1\slash \cF_k }$}        
 \subfloat{
 \psfrag{c}[cm][0][1][90]{${\scriptstyle 1\slash \cF_k }$}  
 \includegraphics[width=0.450\textwidth]{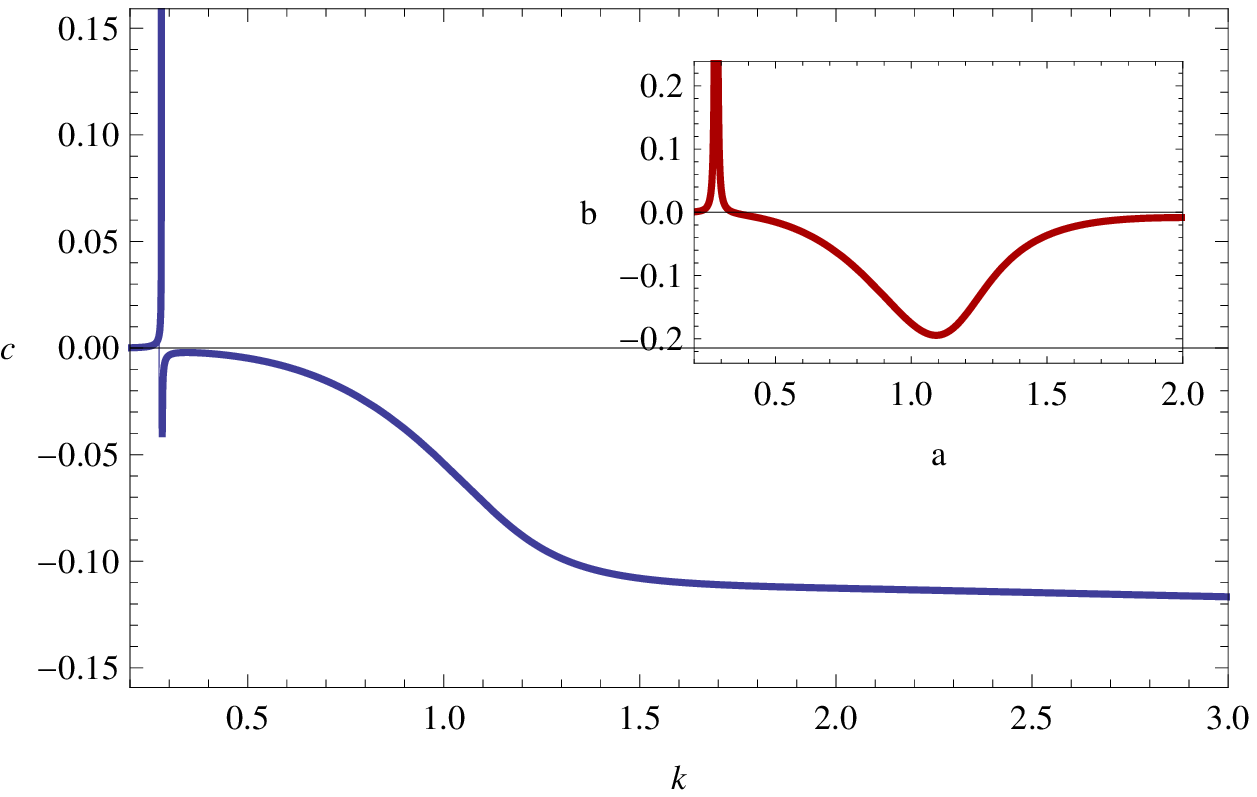}
 } 
 \hspace{0.04\textwidth} %
 \subfloat{%
\includegraphics[width=0.455\textwidth]{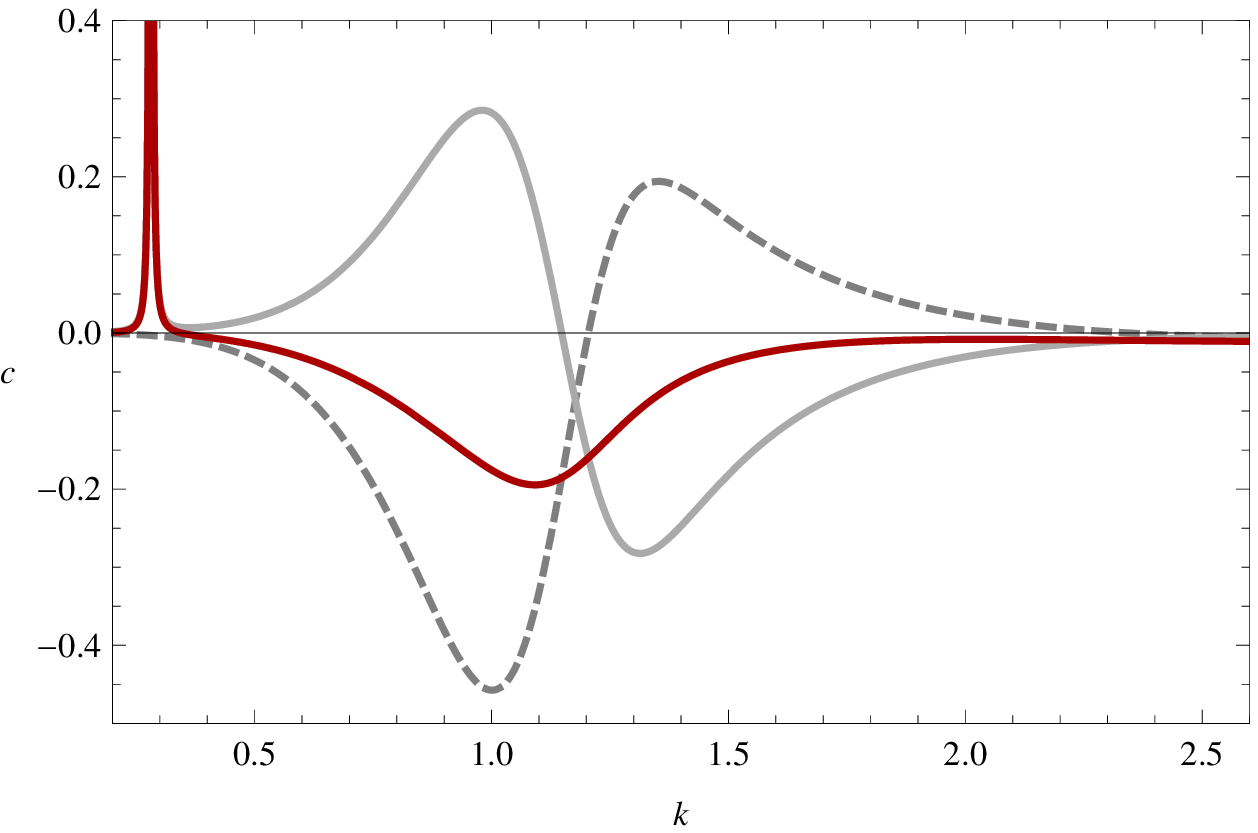}
}
\caption{The function $1\slash\cF_k$ as in Fig. \ref{fig:cfuncCorrectedBMMRS2vio}, but now based on the RG equations of [\Rmnum{2}].} \label{fig:cfuncCorrectedBMvio}
\end{figure}

In Figs. \ref{fig:cfuncCorrectedBMMRS2vio} and \ref{fig:cfuncCorrectedBMvio} the numerical results for $1\slash \cF_k$ are displayed for the RG equations of [\Rmnum{1}] and [\Rmnum{2}], respectively.
They show the same qualitative behavior: 
While the function $\cF_k$ becomes stationary towards the NGFP-regime in the UV, the second  plateau in the IR, which we had found for trajectories restoring split-symmetry,  is now destroyed by the appearance of extrema in the function $\cF_k$, rendering it non-monotone.
In the right panels of Figs. \ref{fig:cfuncCorrectedBMMRS2vio} and \ref{fig:cfuncCorrectedBMvio} this is reflected by the  changing sign of the derivative $k\partial_k \cF_k$ plotted there.

In order to visualize how sign flips of $k\partial_k \cF_k$ can come about it is helpful to define, and to determine numerically, the following subset of the $\tg^{\background}$-$\KkB$-plane:
\begin{align}
\mathcal{T}_+^{\background}(k)\equiv\left\{\left(\tg^{\background},\,\KkB\right)\in \noR^2\,|\, \left(\tg_k^{\dyn},\,\KkD_k,\, \tg^{\background},\,\KkB\right) \in \mathcal{T}_+\subset \mathcal{T} \right\}
\end{align}
The RG time-dependent set $\mathcal{T}_+^{\background}(k)$ consists of all those points of the 4D theory space at which the directional derivative is positive, $\beta_{\alpha} \partial \cF \slash \partial u_{\alpha}>0$, and which have $\left(\tg^{\dyn},\,\KkD\right)$-coordinates that agree with the current position of the selected `$\dyn$' trajectory at time $k$, i.e. $\left(\tg_k^{\dyn},\,\KkD_k\right)$.
\begin{figure}[pt!]
\centering
\psfrag{B}[tc]{${\scriptscriptstyle  }$}
\psfrag{C}[tc]{${\scriptscriptstyle  }$}
\psfrag{g}[tc]{${\scriptscriptstyle \tg^{\background} }$}
\psfrag{l}[c]{${\scriptscriptstyle \KkB }$}
 \subfloat{%
 \psfrag{B}[tc]{${\scriptscriptstyle P_1 }$}
 \psfrag{C}[tc]{${\scriptscriptstyle P_2 }$}
 \psfrag{l}[c]{${\scriptscriptstyle  }$}
\label{fig:snaps3aBA}\includegraphics[width=0.4\textwidth]{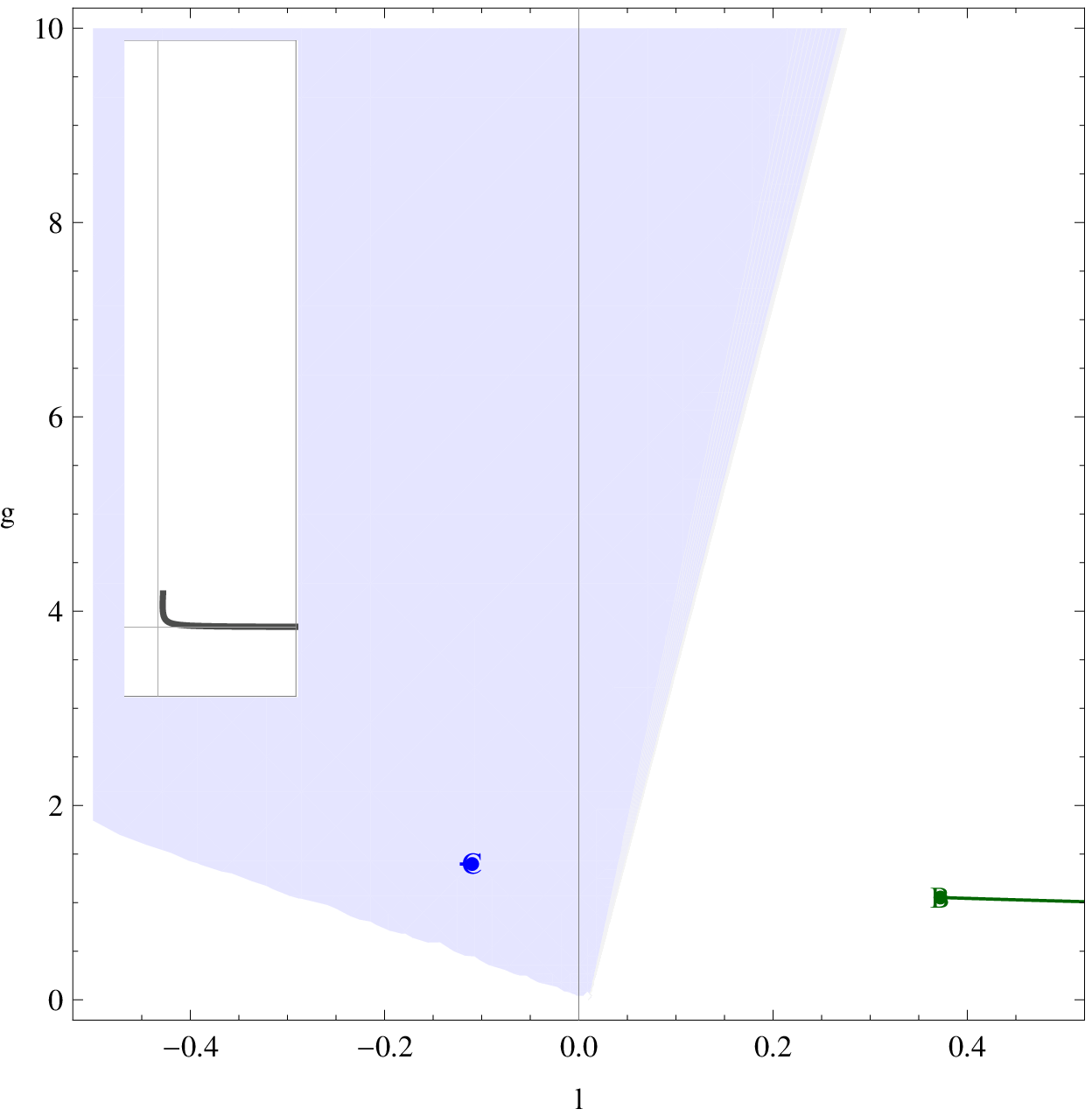}}
\hspace{0.05\textwidth}
%
%
 \subfloat{%
 \psfrag{l}[c]{${\scriptscriptstyle }$}      
\label{fig:snaps3aBC}\includegraphics[width=0.4\textwidth]{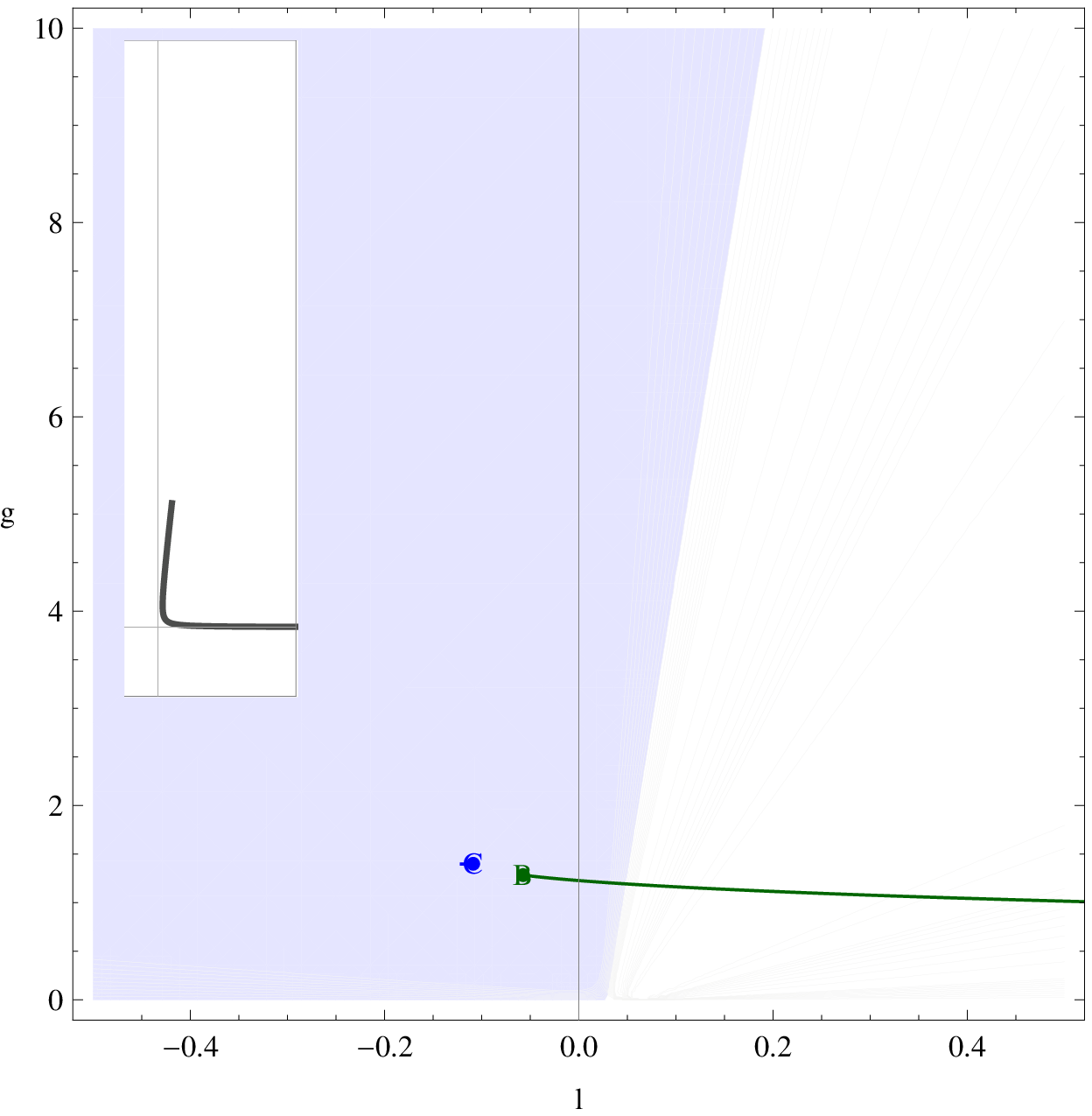}}
\hspace{0.05\textwidth}
 \subfloat{%
\label{fig:snaps3aBD}\includegraphics[width=0.4\textwidth]{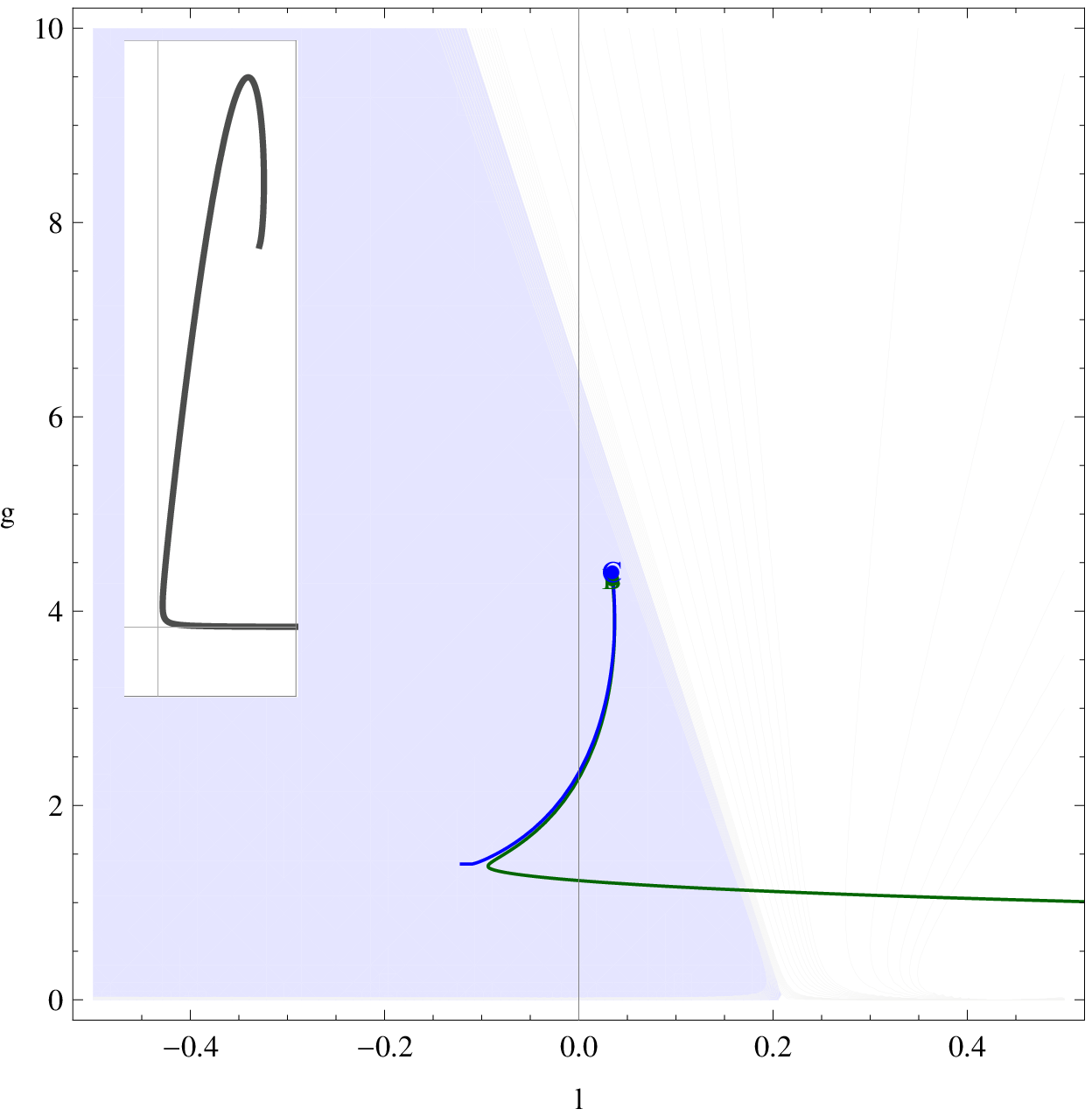}}
\hspace{0.05\textwidth}
%
%
 \subfloat{%
\label{fig:snaps3aBF}\includegraphics[width=0.4\textwidth]{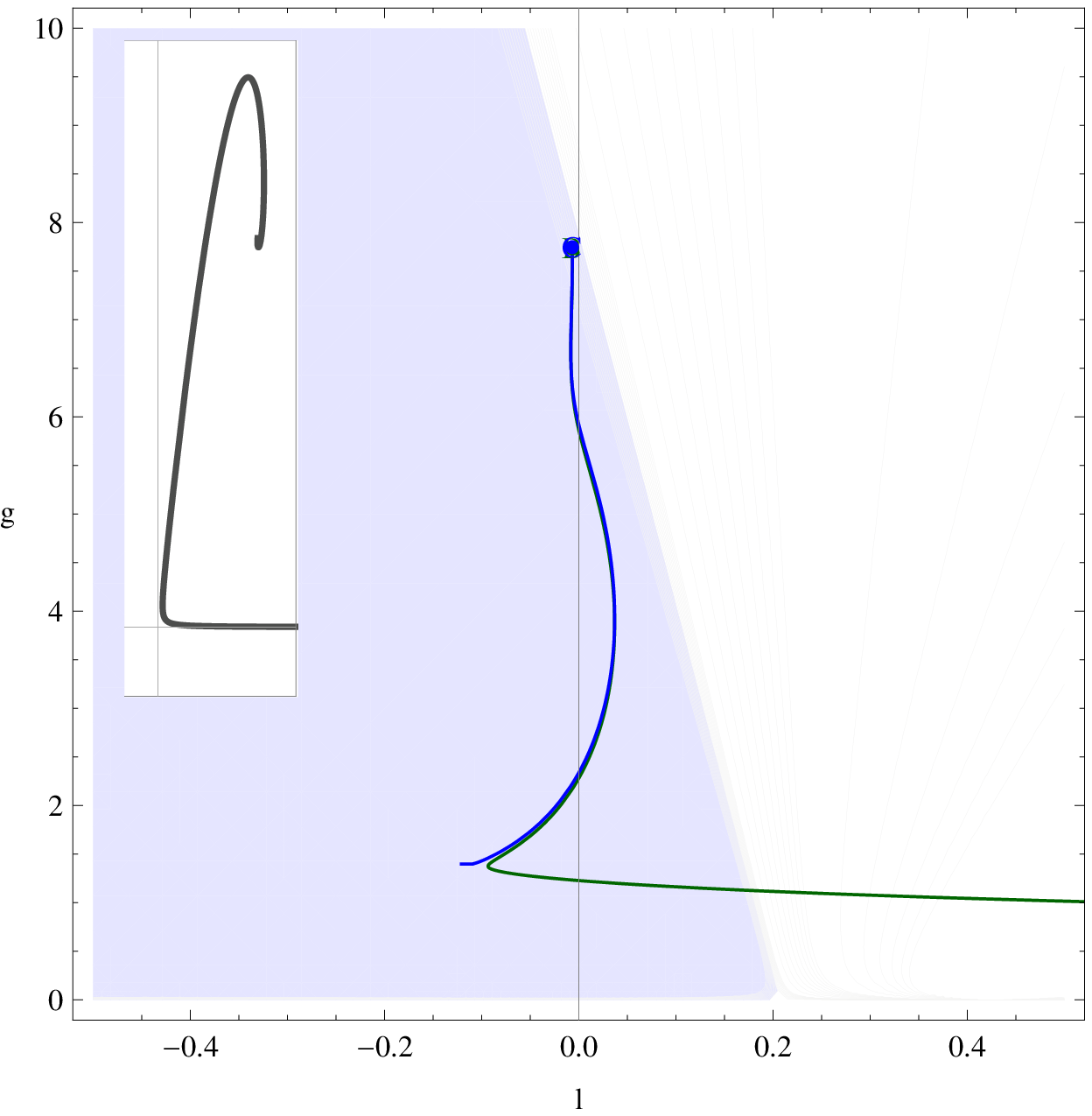}}
 \caption{This series of snapshots represents the $\tg^{\background}$-$\KkB$-plane at  four RG times which increase from the upper left to the lower right diagram.
 They are given by the maximum $k$-value of the incomplete dynamical trajectory $k\mapsto (\tg^{\dyn}_k,\,\KkD_k)$ shown in the respective inset.
 The shaded regions correspond to $\mathcal{T}_+^{\background}(k)$ at that particular time; hence every trajectory in the shaded (white) region will give rise to a  positive (negative) value of $k\partial_k \cfunc_k$ at the instant of time $k$.
 Furthermore, two different $\background$-trajectories that are evolved upward (towards increasing scales $k$) are shown at the corresponding moments.
 The one passing the point $P_1$ ($P_2$) is split-symmetry violating (restoring).
 The symmetry restoring trajectory starts its upward evolution close to $P_2$, the position of the running UV attractor \cite{daniel2}; 
 we see that this trajectory never leaves the shaded area, and thus its $\cfunc_k$-function is strictly monotone.
This is different for the   trajectory through $P_1$: 
Attracted by the running UV-attractor, it is pulled into the shaded region, thus unavoidably crossing the boundary of $\mathcal{T}_+^{\background}(k)$, which causes a sign flip of $\partial_k \cfunc_k$,  rendering $\cfunc_k$ non-monotone.
  }\label{fig:snaps3aB}
\end{figure}

Using the same typical \Rmnum{3}a trajectory in the dynamical sector as above in the split-symmetry restoring case, we now study the RG evolution in the $\tg^{\background}$-$\KkB$-plane, see Fig. \ref{fig:snaps3aB}.
To this end, we subdivide this plane into $\mathcal{T}_+^{\background}(k)$, the shaded regions in the diagrams of Fig.  \ref{fig:snaps3aB},  and its complement, the white regions in Fig.  \ref{fig:snaps3aB}.
This subdivision is different at each instant of RG time.
We are particularly interested in those 4D trajectories $k\mapsto \left(\tg_k^{\dyn},\,\KkD_k,\,\tg_k^{\background},\,\KkB_k\right)$ that give rise to a monotonically increasing $\cfunc_k$-function, or in other words, in trajectories whose projection on the $\background$-plane is such that $\left(\tg_k^{\background},\,\KkB_k\right)\in \mathcal{T}_+^{\background}(k)$ holds true for all $k$.

From Fig. \ref{fig:snaps3aB} it is now clear why the trajectory restoring split-symmetry in the IR, starting at the UV-attractor located at $P_2$, is so special:
As the scale $k$ changes, so does the region defined by $\mathcal{T}_+^{\background}(k)$.
In the IR, the domain $\mathcal{T}_+^{\background}(k)$ defines a narrow band around the  running UV-attractor ($P_2$), and this results in a monotonically increasing $\cfunc_k$.
While the distinguished split-symmetry restoring trajectory is safely within this band, most of the split-symmetry breaking trajectories lie well outside $\mathcal{T}_+^{\background}(k)$ at low $k$.
Increasing $k$, we move towards the UV, and $\mathcal{T}_+^{\background}(k)$ extends especially to regions with negative $\KkB$, while its boundary approaches, and ultimately touches the asymptotic position of the NGFP.

The crucial fact to notice is the following.
At all scales, the symmetry restoring trajectory is seen to stay within $\mathcal{T}_+^{\background}(k)$, and this is in agreement with the results obtained in paragraph (A).
Since when $k$ is increased sooner or later all trajectories converge to this particular one\footnote{See ref. \cite{daniel2} for a detailed demonstration of this behavior.}, they are necessarily all pulled towards a regime $\mathcal{T}_+^{\background}(k)$, if they are not yet inside already. 
This can  be observed in Fig. \ref{fig:snaps3aB} by following  the trajectory that passes through the  point $P_1$ at some low scale.
As $P_1$ lies outside $\mathcal{T}_+^{\background}(k)$ this implies that $\partial_k \cfunc_k<0$ in the IR.
Increasing $k$ the trajectory is  pulled towards the running UV-attractor and between the first and second snapshot of Fig. \ref{fig:snaps3aB} it crosses the boundary of $\mathcal{T}_+^{\background}(k)$.
At this moment the derivative of $\cfunc_k$ crosses zero and from this point onward we have $\partial_k \cfunc_k>0$.
In the third snapshot the trajectory is already well inside $\mathcal{T}_+^{\background}$ and it approaches the symmetry-restoring one.
Once close to this `guiding trajectory' it remains in its vicinity and together, for $k\rightarrow\infty$, they approach the boundary of $\mathcal{T}_+^{\background}(k)$ from its interior.
This is as it should be since we know that $k\partial_k \cfunc_k=0$ at the NGFP.

{\noindent\bf (C) The hybrid calculation}
\begin{figure}[h!]
\centering
\psfrag{k}{${\scriptstyle   k \, \slash m_{\text{Pl}}}$}
\psfrag{m}[bc][bc][1][90]{$\scriptstyle   1\slash \cF_k^{\spsym}  $}
\psfrag{a}[ r]{${\scriptstyle  \text{level-(1)} }$}
\psfrag{b}{${\scriptstyle   \text{level-(0)} }$}
\psfrag{c}[r]{${\scriptstyle  \sm }$}
 \subfloat{%
\label{fig:res4Dpp4T3aBA}\includegraphics[width=0.5\textwidth]{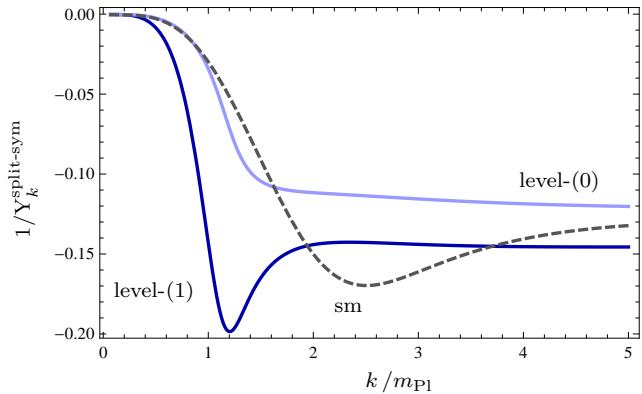}}
 \caption{The inverse of $\cF^{\spsym,(0)}_k$ and $\cF_k^{\spsym,(1)}$ is shown for a generic bi-metric type \Rmnum{3}a trajectory that restores split-symmetry in the IR.
 On intermediate scales, the  RG trajectory is not split-symmetric, as is evident from the different graphs of the level-(1) (dark, solid) and the level-(0) (light, solid) variants of $\cF_k^{\spsym}$. 
For comparison, the single-metric result $\cF^{\sm}_k$ is also included  (dashed curve).
}\label{fig:res4DcpT3aB}
\end{figure}
In the hybrid calculation, we retain only the $\cF_k^{\spsym}$-part of $\cF_k$, omitting the correction term $\Delta\cF_k$.
In Fig.  \ref{fig:res4DcpT3aB} we show (the inverse of) its two variants $\cF_k^{\spsym,(0)}$ and $\cF_k^{\spsym,(1)}$ which are obtained by extracting the couplings from, respectively, the $0^{\rm th}$ and the $1^{\rm st}$ level of a bi-metric type \Rmnum{3}a trajectory.
This particular trajectory restores split-symmetry in the IR.
Fig.  \ref{fig:res4DcpT3aB} shows that the graphs of the resulting functions $\cF_k^{\spsym,(0)}$ and $\cF_k^{\spsym,(1)}$ are quite different, the former function is monotone, the latter is not.

This observation once more tells us that the correction term $\Delta\cF_k$ is needed in order to compensate for the split-symmetry violation that goes into $\cfunc_k$ via the trajectories.
In fact, at intermediate scales, {\it all} trajectories suffer from this disease, both the symmetry restoring and the non-restoring ones; the unmistakable symptoms are the substantial differences among the levels.

This confirms our earlier findings:
The correction term $\Delta\cF_k$ is indispensable. It is needed in order to protect the sum $\cF_k^{\spsym}+\Delta\cF_k$ against the otherwise unavoidable infection with the symmetry violation the trajectories must live with.
This protection is successful, i.e. $\cfunc_k=(\cF_k^{\spsym}+\Delta \cF_k){\cal V}$ has a monotone dependence on $k$, provided we do not break split-symmetry by hand, that is, by selecting inappropriate initial conditions for the background couplings.

{\bf\noindent(D) Testing pointwise monotonicity.}
We have seen in eq. \eqref{eqn:cm_mot_043}
that for {\it exact} RG trajectories $k\mapsto \Gamma_k$ the only source of obstructions for $\cfunc_k$ to become a monotone function is the second term on its RHS, which measures to what extent $\Gamma_k$ breaks split-symmetry.
In the case of exact RG solutions, we know that the first term on the RHS is positive, $\left(\partial_k \Gamma_k\right)[0;\bar{\Phi}_k^{\scon}]\geq 0$, since this is a special case of the pointwise monotonicity, $\left(\partial_k \Gamma_k\right)[\varphi;\bar{\Phi}]\geq0$ $\forall\, (\varphi,\bar{\Phi})$, $\forall\,k$.
However, the latter property might not always be true for approximate solutions to the flow equation, those obtained by using truncations, for instance. 
Testing pointwise monotonicity, $\left(\partial_k \Gamma_k\right)[\varphi;\bar{\Phi}]\geq0$, may therefore serve as a device to judge the validity of a truncation. 
We will come back to this method for arbitrary arguments $(\varphi,\bar{\Phi})$ in ref. \cite{wip}.
We focus here only on $\left(\partial_k \Gamma_k\right)$ evaluated at the special arguments  $(\varphi,\bar{\Phi})=(0,\bar{\Phi}_k^{\scon})$.
\begin{figure}[h!]
\centering
\psfrag{k}[t]{${\scriptstyle   k \, \slash m_{\text{Pl}}}$}
\psfrag{g}{${\scriptstyle \left(\partial_k \Gamma_k\right)[0;\bar{\Phi}_k^{\scon}] }$}   
\psfrag{a}{${\scriptstyle \left(\partial_k \Gamma_k\right)[0;\bar{\Phi}_k^{\scon}] }$}   
 \subfloat{%
 \includegraphics[width=0.4\textwidth]{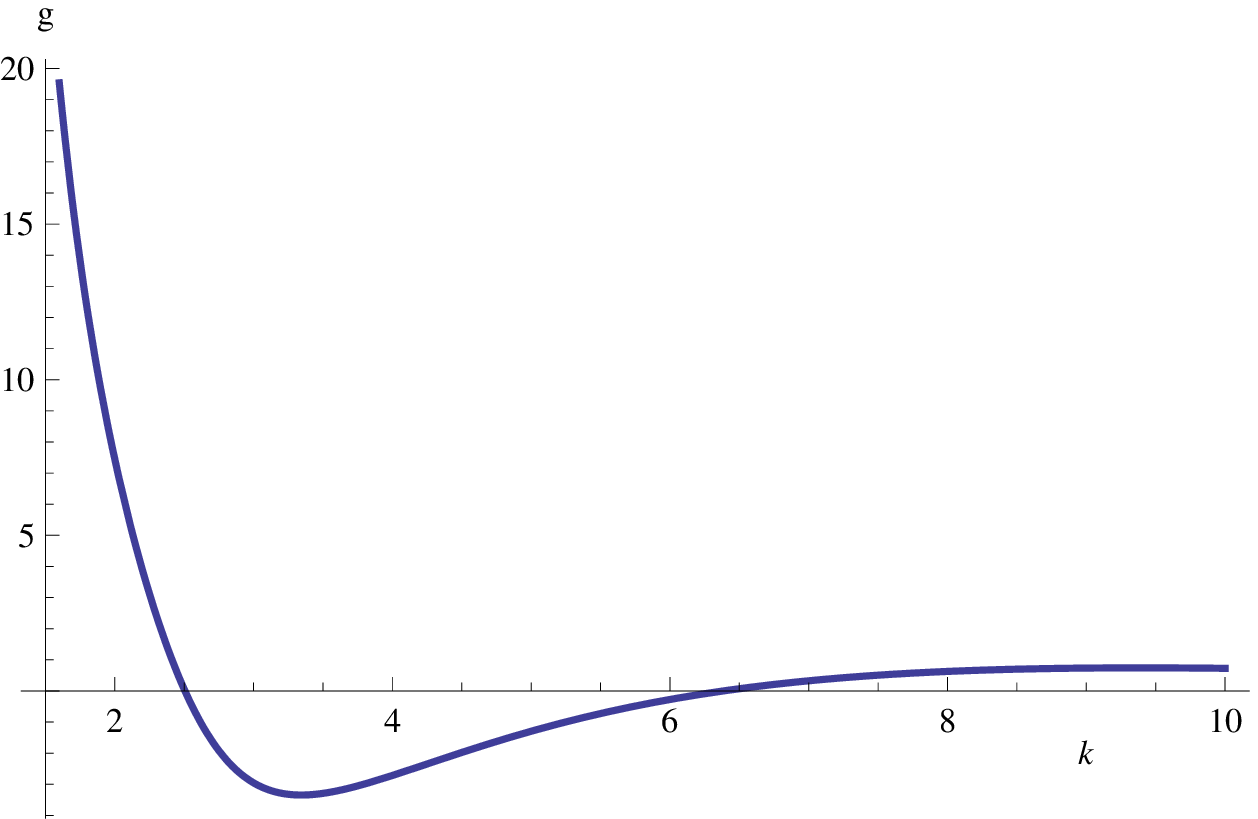}}
 \subfloat{%
\includegraphics[width=0.4\textwidth]{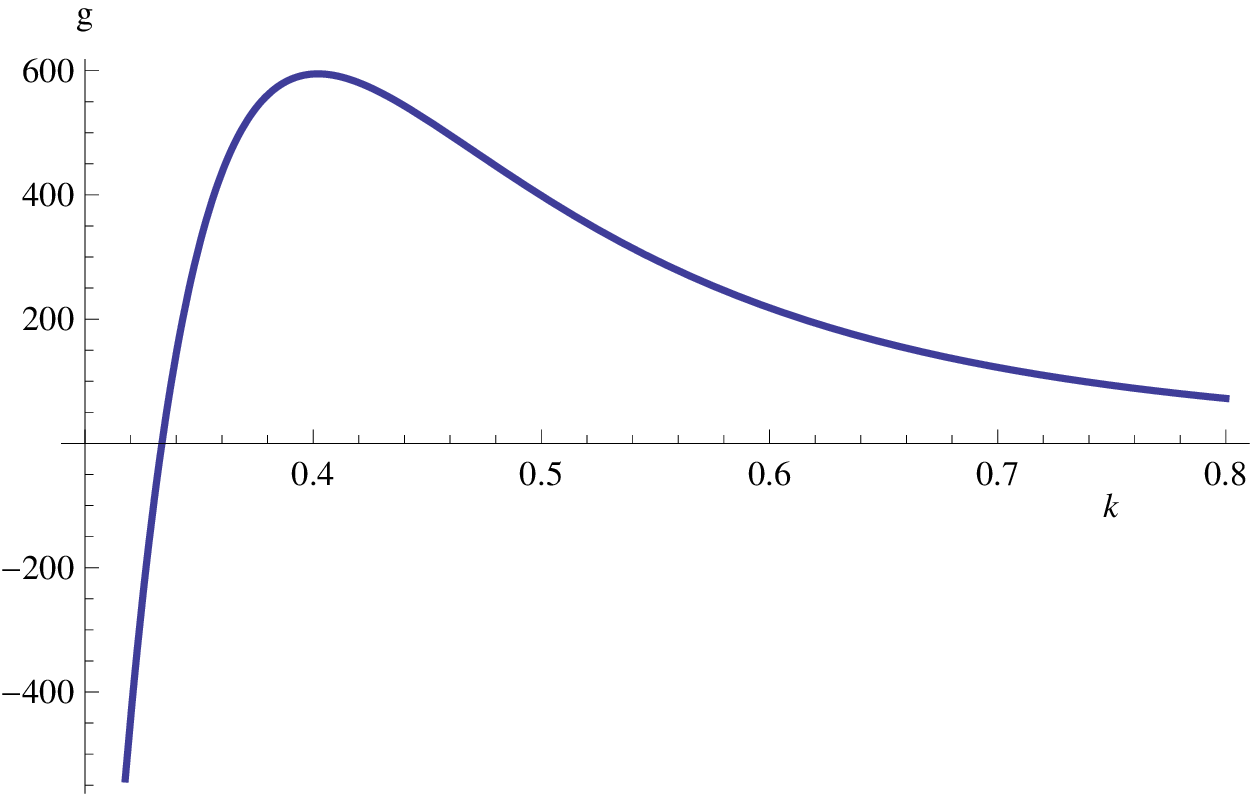}
}
\caption{The first term on the RHS of eq. \eqref{eqn:cm_mot_043}, i.e. $\left(\partial_k \Gamma_k\right)[0;\bar{\Phi}_k^{\scon}]0$ is evaluated for a typical single-metric (left) and split-symmetry violating bi-metric (right) trajectory.
In both cases, it is seen to be negative for certain scales.
This indicates a severe failure of the underlying approximation since, at the exact level, $\left(\partial_k \Gamma_k\right)$ is known to be positive at all field arguments and for any $k$.}
\label{fig:pwmfail}
\end{figure}
\begin{figure}[h!]
\centering
\psfrag{k}{${\scriptstyle   k \, \slash m_{\text{Pl}}}$}
\psfrag{g}{${\scriptstyle \left(\partial_k \Gamma_k\right)[0;\bar{\Phi}_k^{\scon}] }$}   
\psfrag{a}{${\scriptstyle \left(\partial_k \Gamma_k\right)[0;\bar{\Phi}_k^{\scon}] }$}   
\includegraphics[width=0.6\textwidth]{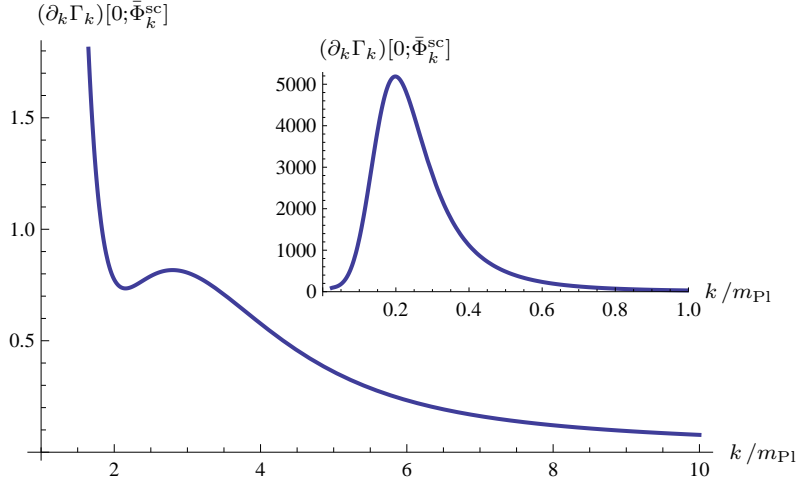}
\caption{The quantity $\left(\partial_k \Gamma_k\right)[0;\bar{\Phi}_k^{\scon}]$ is now evaluated for a split-symmetry restoring bi-metric trajectory.
It always stays non-negative, even in those regimes where the single-metric or the split-symmetry violating bi-metric (see inset) trajectories fail the pointwise monotonicity test.}
\label{fig:pwmsucc}
\end{figure}

It turns out that the single-metric and the `unphysical' bi-metric RG trajectories (those without split-symmetry restoration) actually fail this pointwise monotonicity test.
As shown in Fig. \ref{fig:pwmfail}, there are $k$-intervals on which $\left(\partial_k \Gamma_k\right)[0;\bar{\Phi}_k^{\scon}]$ is {\it negative}.
On the other hand, for the split-symmetry restoring bi-metric trajectories this quantity is positive throughout, as it is at the exact level, see Fig. \ref{fig:pwmsucc}.
These findings make it very clear that the non-monotonicity displayed by our $\cfunc_k$-function candidate, when applied to single-metric and symmetry violating bi-metric truncations, is {\it not due to a defect of the proposed form of $\cfunc_k$} but rather originates in insufficient approximations.
Only the symmetry-restoring, bi-metric trajectories are close enough to the exact ones to render both $\left(\partial_k \Gamma_k\right)[0;\bar{\Phi}_k^{\scon}]$ and the full $\partial_k \cfunc_k$ positive.

\subsubsection{Summary and Conclusion}
To sum it up we can say that the expected monotonicity of $\cfunc_k$ arises under the following conditions:
First, the bi-metric version of the Einstein-Hilbert truncation is used, and second, the underlying RG trajectory is split-symmetry restoring.
Violating either of these conditions may destroy the monotonically increasing behavior of $\cfunc_k$.
We saw that there are cases where the split-symmetry violation of $\EAA_k$ is sufficiently small to leave the monotonicity of $\cfunc_k$ intact, the main example being the bi-metric type \Rmnum{3}a trajectories approaching the UV attractor for $k\rightarrow0$.

We have seen that down-grading the bi-metric truncation ansatz to the level of a single-metric approximation is paid by loosing the monotonicity property of $\cfunc_k$.
As for its dependence on the running couplings, the reduced $\cfunc_k$-function in the bi-metric truncation, $\cF_k$, differs from its single-metric counterpart $\cF_k^{\sm}$ by the correction term $\Delta\cF_k$, which vanishes when $\EAA_k^{\text{grav}}$ is exactly split-symmetric.
(In the single-metric approximation, this is always the case, by decree.)
We found that there is a numerically highly non-trivially conspiracy and compensation between $\Delta\cF_k$ and those properties of the RG trajectories which stem from the split-symmetry violation in the flow equation, and which could easily destroy the monotonicity of $\cfunc_k$.
The fact that this does not happen for any of the physically relevant trajectories is directly linked to the specific properties of our candidate function, $\cfunc_k=\EAA_k[0;\bar{\Phi}_k^{\scon}]$, its scale dependent argument in particular, since it determines the structure of $\Delta\cF_k$.

Taken together these findings strongly support the following conjecture:
{\it In the full theory, QEG in 4 dimensions, or in a sufficiently general truncation thereof, the proposed candidate for a generalized $C$-function is a monotonically increasing function of $k$ along all RG trajectories that restore split-symmetry in the IR and thus comply with the fundamental requirement of Background Independence.}

If the conjecture can be established we will have a particularly easy to apply diagnostic tool for testing the reliability of  truncations.
Since then solutions to the untruncated flow equation for sure have a monotone $\cfunc_k$, any truncation that violates the monotonicity misses qualitatively important features of  the RG flow and would therefore be judged an insufficient approximation to the full flow.
In this light we provisionally conclude that the single-metric approximation is not fully reliable, while the bi-metric Einstein-Hilbert truncation is superior as it  keeps the  monotonicity of $\cfunc_k$ intact at least.
Of course this conclusion is fully consistent with all other results available on the bi-metric Einstein-Hilbert truncation \cite{MRS2,daniel2,daniel-eta}.

As split-symmetry is essential in this context, it might be helpful to recall its physical contents.
Split-symmetry and the corresponding Ward identity (WISS) are the technical device by means of which Background Independence in the physical sector (`on-shell') is imposed on the effective action and similar `off-shell' quantities\footnote{For them, `Background Independence' is not naively `independence of $\bg_{\mu\nu}$'.}.
The prototypical example of a Background Independent theory is classical General Relativity \cite{ARR}. Now, even though we describe it by an effective action $\Gamma[g,\bg]$, we would like QEG to enjoy Background Independence exactly at the same level as General Relativity.

Let us contrast QEG with {\it genuine bi-metric theories}, in the original sense of the word, that is, extensions of General Relativity employing two {\it physically distinct} metrics, $g_{\mu\nu}^{(1)}$ and $g_{\mu\nu}^{(2)}$, say.
Depending on the structure of their action $S[g_{\mu\nu}^{(1)},g_{\mu\nu}^{(2)},\cdots]$ they could differ, for instance, in their coupling to matter, or their propagation properties.
In an appropriate limit, matter particles of a certain species could, for example, follow the geodesics of $g_{\mu\nu}^{(1)}$, or of $g_{\mu\nu}^{(2)}$; but it also can happen that trajectories are no geodesics at all and have no geometric interpretation.
In a genuine bi-metric theory, these different cases are {\it experimentally distinguishable}.
The metrics have equal status in that both of them, independently, can make their way into observables.
In canonical quantization both $g_{\mu\nu}^{(1)}$ and $g_{\mu\nu}^{(2)}$ are turned into operators.
This is fundamentally different when one applies the background field technique to the quantization of a system with a bare action $S[\hat{g}]$ depending on one metric only, and introduces $\bg_{\mu\nu}$ only as a technical convenience, for coarse-graining and gauge-fixing purposes in particular.
Setting $\hat{g}_{\mu\nu}=\bg_{\mu\nu}+\hat{\flcb}_{\mu\nu}$ we transfer the physical degrees of freedom entirely from $\hat{g}_{\mu\nu}$ to $\hat{\flcb}_{\mu\nu}$ which is made the new dynamical quantum field by replacing $\int\mathcal{D}\hat{g}_{\mu\nu}$ with $\int\mathcal{D}\hat{\flcb}_{\mu\nu}$.
From the functional perspective, $\bg_{\mu\nu}$ is merely an arbitrary shift on which no observable consequence of the theory may depend.
In canonical quantization, $\hat{g}_{\mu\nu}$ and $\hat{\flcb}_{\mu\nu}$ are operators, while $\bg_{\mu\nu}$ continues to be a classical $c$-number field.
The logical dissimilarity between dynamical and background metric gets slightly obscured at the level of the expectation values $\flcb_{\mu\nu}\equiv\langle \hat{\flcb}_{\mu\nu}\rangle$ and $g_{\mu\nu}\equiv\langle \hat{g}_{\mu\nu}\rangle = \bg_{\mu\nu}+\flcb_{\mu\nu}$ since $\Gamma_k[\flcb;\bg]\equiv \Gamma_k[g,\bg]$ depends on two independent fields, two metrics in fact, if one uses the EAA in the `comma notation', $\Gamma_k[g,\bg]$.
Now, the role of the split-symmetry as encoded in the Ward identity (WISS) of eq. \eqref{eqn:cm_mot_020}, is to express the requirement that {\it there is only one physical metric} and that no observable quantity may depend on how $\bg_{\mu\nu}$ was chosen.
Setting for example $k=0$ and ignoring gauge fixing issues for a moment, invariance of the bare action under $\{\delta \hat{\flcb}_{\mu\nu}=\epsilon_{\mu\nu},\, \delta \bg_{\mu\nu}=-\epsilon_{\mu\nu}\}\, \iff\,\{\delta \hat{g}_{\mu\nu}=0,\, \delta \bg_{\mu\nu}=-\epsilon_{\mu\nu}\}$ implies that $\Gamma_0[\flcb;\bg]$ can depend on the sum $\bg+\flcb\equiv g$ only, while $\Gamma_0[g,\bg]\equiv \Gamma_0[g]$ simply does not depend on its second argument.
In reality, because we use a `background-type' gauge fixing condition, $\Gamma_0[g,\bg]$ does have a certain $\bg$-dependence, again dictated by the WISS, but it disappears upon going on-shell and cannot be seen in any experiment therefore.

\subsection{Crossover trajectories and their mode count}
{\bf\noindent (A)}
In subsection \ref{subsec:2-05} we proved that the exact $\cfunc_k$ is stationary  at fixed points as well as in classical regimes.
The explicit $\cfunc_k$ functions obtained from both the single- and the bi-metric truncation indeed display this  behavior. 
Looking at the two alternative formulas for $\cfunc_k$ in \eqref{eqn:intro_eq08} it is indeed obvious that $\cfunc_k$ becomes stationary when the {\it dimensionless} couplings are at a fixed point of the flow, and when the {\it dimensionful} ones become scale independent; this is the case in a classical regime (`$\crg$') where by definition no physical RG effects occur.
If $\Kkbar_{\crg}^{\cix}$ and $G_{\crg}^{\cix}$ are the constant values of the cosmological and Newton constants there, this regime amounts to the trivial canonical scaling $\Kk_k^{\cix}=k^{-2}\Kkbar_{\crg}^{\cix}$ and $\tg_k^{\cix}=k^{d-2}G_{\crg}^{\cix}$.

\begin{figure}[h!]
\centering
 \psfrag{T}[tc]{${\scriptstyle \mathcal{T} }$}
  \psfrag{A}{${\scriptstyle \text{FP}_2 }$}
 \psfrag{B}[tc]{${\scriptstyle \text{FP}_1 }$}
 \psfrag{C}[tc]{${\scriptstyle \text{CR} }$}
 
 \subfloat[Crossover: FP $\rightarrow$ FP.]{%
\label{fig:crossoverA}\includegraphics[width=0.4\textwidth]{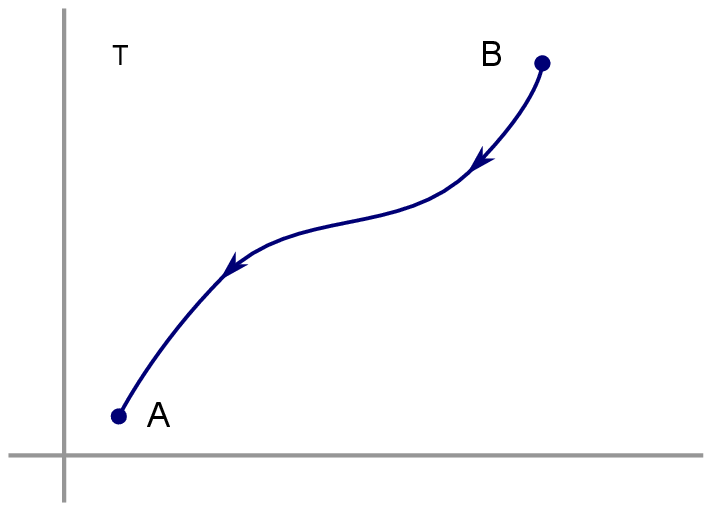}}
\hspace{0.05\textwidth}
 \subfloat[Crossover: FP $\rightarrow$ \crg.]{%
   \psfrag{A}{${\scriptstyle \text{FP}}$}
\label{fig:crossoverB}\includegraphics[width=0.4\textwidth]{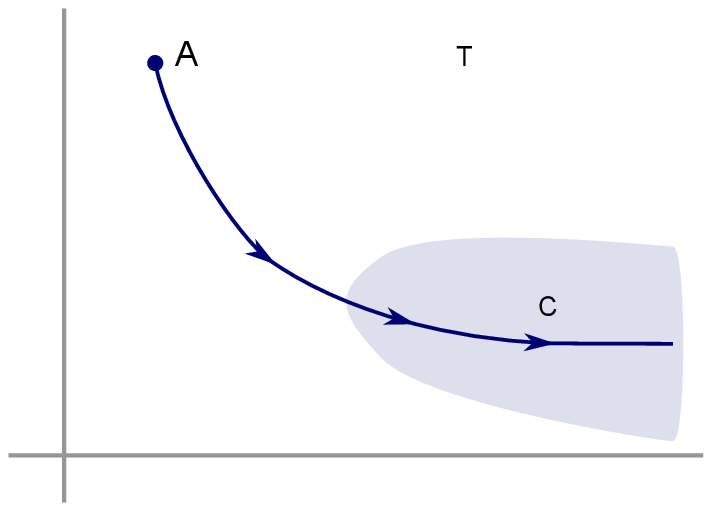}}
 \caption{Crossover trajectories in theory space: from one fixed point to another, (a), and from a UV fixed point to a classical regime, (b).}\label{fig:cm_crossover}
\end{figure}

{\bf\noindent (B)}
In subsection \ref{subsec:2-05} we  mentioned already the possibility of generalized crossover transitions, not only in the standard way from one fixed point to another, but rather from a fixed point to a classical regime or vice versa.
Thereby $\cfunc_k$ will always approach well defined stationary values $\cfunc_*$ and $\cfunc_{\crg}$ in the respective fixed point or classical regime.
(See Fig. \ref{fig:cm_crossover} for a schematic sketch.)

In the case of an asymptotically safe RG trajectory, the initial point in the UV is a non-Gaussian fixed point, by definition.
For the corresponding limit $\cfunc^{\UV}\equiv \lim_{k\rightarrow\infty}\cfunc_k$ the bi-metric calculation yields $\cfunc^{\UV}=\cfunc_*$, with
\begin{align}
\raalign{\cfunc_*=-\frac{\left(\frac{d}{d-2}\right) \Kk_*^{(1)}-\Kk_*^{(0)} }{\tg_*^{(0)}\left[\Kk_*^{(1)}\right]^{d\slash 2} }\, {\cal V}(\MaFs,\mathring{g}) }{}
\label{eqn:cm_cross_01}
\end{align}
This result simplifies to
\begin{align}
\cfunc_*^{\sm}&=-\left(\frac{2}{d-2}\right)\frac{ {\cal V}(\MaFs,\mathring{g}) }{\tg_*^{\sm}\left[\Kk_*^{\sm}\right]^{d\slash 2-1} } 
\label{eqn:cm_cross_02}
\end{align}
in the single-metric approximation.
Note that $\cfunc_*$ diverges at the trivial (`Gaussian') fixed point at which all dimensionless couplings vanish.

If the trajectory ends in an IR fixed point the corresponding limit $\cfunc^{\IR}\equiv \lim_{k\rightarrow0}\cfunc_k$, if it exists, is again given by the formula \eqref{eqn:cm_cross_01}, $\cfunc^{\IR}\equiv \cfunc_*$, but for different fixed point coordinates.
If the trajectory is instead destined to enter  a classical regime and to approach $k=0$ by an infinitely `long', and boring, since purely canonical running on $\mathcal{T}$, then the value at the end point equals $\cfunc^{\IR}=\cfunc_{\crg}$ where
\begin{align}
\raalign{\cfunc_{\crg} =-\left(\frac{2}{d-2}\right)\frac{ {\cal V}(\MaFs,\mathring{g}) }{G_{\crg}\left[\Kkbar_{\crg}\right]^{d\slash 2-1} } }{}
\label{eqn:cm_cross_03}
\end{align}
In writing down eq. \eqref{eqn:cm_cross_03} we assumed that the same values of $G_{\crg}$ and $\Kkbar_{\crg}$ apply at all levels, as required by split-symmetry.

Thus, for any of the above crossover types we expect a finite value of 
\begin{align}
\nBound\equiv \nBound_{0,\infty}\equiv \cfunc^{\UV}-\cfunc^{\IR}
\label{eqn:cm_cross_04}
\end{align}

{\bf\noindent (C)}
Beside monotonicity and stationarity, $\cfunc_k$ has another essential property in common with a $C$-function: 
The limiting value $\cfunc_*$ has a genuine inherent interpretation at the fixed point itself.
It is a number characteristic of the NGFP which does not depend on the direction it is approached, and in this role it is analogous to the central charge.
The interpretation of $\cfunc_*$ is best known for $d=4$ and the single-metric approximation where, apart from inessential constants, it is precisely the inverse of the dimensionless combination
$\tg_*\Kk_*=G_{k\rightarrow\infty}\Kkbar_{k\rightarrow\infty}$.
Its physical interpretation is that of an `intrinsic' measure for the size of the cosmological constant at the fixed point, namely the limit of the running cosmological constant in units of the running Planck mass ($G_k^{-1\slash2}$).
In numerous single-metric studies the product $\tg_*\Kk_*$ has been investigated, and it was always found that {\it $\tg_* \Kk_*$ is a universal quantity}, i.e. it is independent of the cutoff scheme and the gauge fixing, within the accuracy permitted by the approximation.
In fact, typically the universality properties of $\tg_* \Kk_*$ were even much better than those of the critical exponents.
Completely analogous remarks apply to $d\neq 4$, and to the quantity \eqref{eqn:cm_cross_01} in the bi-metric generalization.

We interpret this number, with all due care, as a measure for the total `number of field modes' integrated out along the entire trajectory.
Clearly $\cfunc^{\UV}$ and $\cfunc^{\IR}$ play a role analogous to the central charges of the 2D conformal field theories sitting at the end points of the trajectory in the case of Zamolodchikov's  theorem.

Within the Einstein-Hilbert truncation, our numerical results for $\cfunc_*$ at the NGFP in $d=4$ are as follows for the three calculations we compared:
\begin{align}
\cfunc_*=-{\cal V}(\MaFs,\mathring{g})\times \begin{cases}
7.3 & \text{ single-metric}\\
4.3 & \text{ bi-metric \Rmnum{1}}\\
8.1 & \text{ bi-metric \Rmnum{2}}
\end{cases}
\label{eqn:cm_cross_05}
\end{align}
Obviously, in all calculations $\cfunc_*$ is a negative number of order unity.

Let us now focus on the case depicted in Fig. \ref{fig:crossoverB}, which can be seen as a simple caricature of the real Universe.
We consider a family of de Sitter spaces along a type \Rmnum{3}a trajectory which is known to possess a classical regime with $\Kkbar_{\crg}>0$.
Assuming that this regime represents the true final state of the evolution, we obtain
\begin{align}
\cfunc^{\IR}=-\frac{3\pi}{G_{\crg} \Kkbar_{\crg}}
\label{eqn:cm_cross_06}
\end{align}
Note that this $\cfunc^{\IR}$ is negative, too, and that $-\cfunc^{\IR}$ equals precisely the well known semi-classical Bekenstein-Hawking entropy of de Sitter space \cite{BH-Therm}.

Thus, combining \eqref{eqn:cm_cross_05} and \eqref{eqn:cm_cross_06} for $\cfunc^{\UV}$ and $\cfunc^{\IR}$, respectively, we arrive at the following important conclusion: 
In an asymptotically safe theory of quantum gravity which is built upon a generalized crossover trajectory from criticality (the NGFP) to classicality {\it the total number of modes integrated out, $\nBound=\cfunc^{\UV}-\cfunc^{\IR}$, is  finite} according to the natural counting device provided by the EAA itself.

In the special situation when $G_{\crg} \Kkbar_{\crg}\ll 1$, like in the real world, we have $|\cfunc^{\IR}|\gg 1$, while $|\cfunc^{\UV}|=\Order{1}$ according to the NGFP data of \eqref{eqn:cm_cross_05}.
As a consequence, the number $\nBound$ is completely dominated by the IR part of the trajectory, $\nBound= \cfunc^{\UV}-\cfunc^{\IR}\approx - \cfunc^{\IR}$, and so we obtain
\begin{align}
\nBound\approx + \frac{3\pi}{G_{\crg} \Kkbar_{\crg}}\gg 1
\label{eqn:cm_cross_07}
\end{align}

It is tempting to identify $\Kkbar_{\crg}$ and $G_{\crg}$ with the corresponding values measured in the real Universe.
Because of their extremely tiny product $G_{\crg} \Kkbar_{\crg}$ we find a tremendous number of modes then: $\nBound\approx 10^{120}$.
Nevertheless, in sharp contradistinction to what standard perturbative field theory would predict, this number is {\it finite}.

{\bf\noindent (D)}
Concerning the finiteness of $\nBound$, the situation changes if we try to define the function $k\mapsto \cfunc_k$ along trajectories of the type \Rmnum{1}a, those heading for  a negative cosmological constant $\KkD$ after leaving the NGFP regime, and of type \Rmnum{2}a, the single trajectory which crosses over from the NGFP to the Gaussian fixed point (GFP) at which all 4 couplings vanish.

In the {\bf type \Rmnum{2}a} case, eq. \eqref{eqn:cm_cross_01} yields `$\cfunc_*=-\infty$' at the IR end point of the trajectory so that the total number of modes diverges, `$\nBound=+\infty$'.
Clearly this behavior can be seen as the limit $\Kkbar_{\crg}\rightarrow 0$ of eq. \eqref{eqn:cm_cross_07} since the GFP has a vanishing cosmological constant.
The divergent value of $\nBound$ is the signal of a `topology change' that occurs at $\Kkbar^{(1)}=0$:
While the self-consistent backgrounds (of maximal symmetry, say) are spheres $S^d$ for $\Kkbar^{(1)}>0$, it is flat space ($R^d$) if $\Kkbar^{(1)}=0$.
The Euclidean volume of the former is always finite, but that of $R^d$ is infinite.

While along the type \Rmnum{2}a trajectory the divergence of $\cfunc_k$ occurs only at the very end of the RG evolution, i.e. in the limit $k\rightarrow 0$, for {\bf type \Rmnum{1}a} trajectories $\cfunc_k$ becomes singular already at a {\it finite} scale $k=k_{\text{sing}}>0$.
All trajectories of this type cross the hyperplane  $\KkD=0$ at a nonzero scale, $k_{\text{sing}}$.
However, as eq. \eqref{eqn:intro_eq13} shows, $\cF(\,\cdot\,)$ and $\cfunc(\,\cdot\,)$ are singular on this plane\footnote{One might be worried about the hyperplanes $\tg^{\dyn}=0$ and $\tg^{\background}=0$ on which $\cF(\,\cdot\,)$ is singular too, see eq. \eqref{eqn:intro_eq13}. 
However, within the truncation considered there exist no trajectories that would ever cross or touch those planes.},
so that $\cfunc_k$ diverges in the limit $k\searrow k_{\text{sing}}$.
The number of modes, $\nBound_{k_{\text{sing}},\infty}$ is infinite then, which however by no means implies that  {\it all} modes have been integrated out already.
In fact, there is a non-trivial RG evolution also between $k_{\text{sing}}$ and $k=0$.

Along a type \Rmnum{1}a trajectory, the tadpole equation has qualitatively different solutions for $k>k_{\text{sing}}$, $k=k_{\text{sing}}$, and $k<k_{\text{sing}}$, namely spherical, flat, and hyperbolic spaces, respectively ($S^d$, $R^d$, and $H^d$, say).
This topology change prevents us from smoothly continuing the mode count across the $\KkD=0$ plane.
This is the reason why in this  paper we mostly focused on type \Rmnum{3}a trajectories.
Some further details for the type \Rmnum{1}a and \Rmnum{2}a cases can be found in appendix \ref{app:cmB}, however.
\section{Discussion and outlook}
{\bf\noindent (A)}
The effective average action is a variant of the standard effective action which has an IR cutoff built-in at a sliding scale $k$.
As such, it possesses a natural `mode counting' and monotonicity property which is strongly reminiscent of Zamolodchikov's $C$-function in 2 dimensions, at least at a heuristic level.
For a broad class of systems, this property (`pointwise monotonicity') is easy to demonstrate, the essential input being  that in every system with a well-defined RG flow the action $\Gamma_k+\Delta S_k$ is a strictly convex functional on all scales, that is, the Hessian operator satisfies the positivity constraint $\EAA_k^{(2)}+\mathcal{R}_k >0$, $k\in(0,\infty)$.
Motivated by this observation, and taking advantage of the structures and tools that are naturally provided by the manifestly non-perturbative EAA framework, we tried to find a map from the functional $\EAA_k[\Phi,\bar{\Phi}]$ to a single real valued function $\cfunc_k$ that shares two main properties with the $C$-function in 2 dimensions, namely monotonicity along RG trajectories and stationarity at RG fixed points.

We do not expect such a map to exist in full generality.
In fact, an essential part of the research program we are proposing consists in finding suitable restrictions on, or specializations of the {\it admissible trajectories} (restoring split-, or other symmetries, etc.), the {\it theory space} (with respect to field contents and symmetries), the underlying {\it space of fields} (boundary conditions, regularity requirements, etc.), and the {\it coarse graining methodology} (choice of cutoff, treatment of gauge modes, etc.) that will guarantee its existence.

In the present paper we motivated and analyzed a specific candidate for a map of this kind, namely $\cfunc_k=\EAA_k[\bar{\Phi}^{\scon}_k,\bar{\Phi}^{\scon}_k]$ where $\bar{\Phi}^{\scon}_k$ is a running self-consistent background, a solution to the tadpole equation implied by $\EAA_k$.
We showed that the function $\cfunc_k$ is stationary at fixed points, and a non-decreasing function of $k$ when the breaking of the split-symmetry which relates fluctuation fields and backgrounds is sufficiently weak.
Thus, for a concrete system the task is to identify the precise conditions under which the split-symmetry violation does not destroy the monotonicity property of $\cfunc_k$, and to give a corresponding proof then. 

It would be interesting to work out the properties of this $\cfunc_k$-function for further concrete examples, but also to explore and test structurally different maps from $\Gamma_k$ to $\cfunc_k$.
A new kind of map should in particular be devised if one wants to count the modes of fermionic fields with the same weight as those of bosons.
Because of the sign factors produced by the super-trace in the flow equation, $\cfunc_k=\Gamma_k[\bar{\Phi}_k^{\scon},\bar{\Phi}_k^{\scon}]$ rather counts bosons and fermions with opposite signs, and so $\nBound$ equals the total number of bosonic modes integrated out minus the number of fermionic ones.
(The different count for bosons and fermion is reminiscent of, but not precisely identical to the functional integral representing the Witten index instead of the partition function. 
There, fermionic states of fermion number $F$ contribute with the weight $(-1)^F$.)
While the `boson minus fermion' counting is at variance with the standard $c$-theorem, the properties and potential applications of the present $\cfunc_k$ with fermions should be explored in more detail before dismissing it prematurely.
As we stressed already, rather then reproducing known results, our main goal consists in finding maps $\Gamma_k\rightarrow \cfunc_k$ that are simple and `geometrically natural' in the theory space and functional RG context.

{\bf\noindent (B)}
By means of a particularly relevant example, QEG in $d>2$ dimensions, we demonstrated that our approach is viable in principle and can indeed lead to interesting candidates for `$C$-functions' under conditions which are not covered by the known $c$- and $a$-theorems.
As exact proofs are not within reach for the time being, the practical problem is of course the same as in all non-perturbative functional RG studies, namely the necessity to truncate the theory space.
Here, for asymptotically safe quantum gravity we computed  $\cfunc_k$ directly from the RG trajectories obtained with both the single- and bi-metric Einstein-Hilbert truncation, respectively.

It is one of our main results that in the bi-metric truncation the function $\cfunc_k$ has the desired properties of monotonicity and stationarity, while the single-metric truncation is too poor an approximation to correctly reproduce the monotonicity which we expect at the exact (un-truncated) level.
We demonstrated explicitly that the monotonicity property obtains only for the RG trajectories which are physically meaningful, that is, those which lead to a restoration of split-symmetry once all field modes are integrated out.

We studied generalized crossover trajectories from a fixed point in the UV to a classical regime in the IR, in which by definition the dimensionful cosmological and Newton constants loose their $k$-dependence.
In many ways they are analogous to a standard (fixed point $\rightarrow$ fixed point) crossover.
In quantum gravity they are of special importance as one of the main challenges consists in explaining the emergence of a classical spacetime from the quantum regime.

For the trajectories with positive cosmological constant (`type \Rmnum{3}a') the self-consistent background configurations needed are gravitational instantons.
The resulting $\cfunc_k$ depends on the instanton type via the normalized volume, a quantity of topological significance.
For the example of Euclidean de Sitter space, the sphere $S^4$, for instance, 
we obtained the `integrated $\cfunc$-theorem' 
\begin{align}
\nBound=\cfunc^{\UV}-\cfunc^{\IR}\approx 3\pi\slash G_{\crg} \Kkbar_{\crg}
\label{eqn:cm_con01}
\end{align}
This result is intriguing for several reasons.
First of all, $\nBound$, and also the values of $\cfunc^{\UV}$ and $\cfunc^{\IR}$ separately, are well defined finite numbers, in marked contrast to expectations based on the counting in perturbative field theory.
The quantity $\nBound$ can be interpreted as a measure for the `number of modes' which are integrated out while the cutoff is decreased from $k\,\text{`}\!=\!\!\text{'}\infty$ to $k=0$.

Hereby the notion of `counting' and the precise meaning of a `number of field modes' is defined by the EAA itself, namely via the identification $\cfunc_k=\EAA_k[\bar{\Phi}^{\scon}_k,\bar{\Phi}^{\scon}_k]$.
Under special conditions it reduces to a literal counting of the $\EAA_k^{(2)}$-eigenvalues in a given interval.
Generically we are dealing with a non-trivial generalization thereof which, strictly speaking, amounts to a {\it definition} of `counting'.
As such it is the most natural one from the EAA perspective, however.

The approximate equality $\nBound\approx 3\pi \slash G_{\crg}\Lambda_{\crg}$  is valid if $G_{\crg}\Lambda_{\crg}\ll 1$ in the classical regime.
In this limiting case, $\nBound\approx |\cfunc^{\IR}|$ equals exactly the Bekenstein-Hawking entropy of de Sitter space, and the contribution from the UV fixed point is negligible, $|\cfunc^{\UV}|\ll|\cfunc^{\IR}|$.
Asymptotic Safety is crucial for this result, making $\cfunc^{\UV}$ finite.

So we are led to the following interpretation of the entropy of de Sitter space: it equals the number $\nBound$ of  metric and ghost fluctuation modes that are integrated out between the NGFP in the UV and the classical regime in the IR.
Asymptotic Safety is `taming' the ultraviolet and renders this number perfectly finite. 
(In the real Universe, $\nBound\approx 10^{120}$.)
\vspace{6px}

{\bf\noindent (C)}
Future work along these lines will be in various different directions.
Clearly one of the goals will be to corroborate our result based on the bi-metric Einstein-Hilbert truncation on larger theory spaces.
This should also lead to a better understanding and to a physics interpretation of the stationary values $\cfunc_*$ and $\cfunc_{\crg}$ which replace the central charge of  2D conformal field theory.
We found that $\cfunc_{\crg}$ coincides with the familiar Bekenstein-Hawking entropy\footnote{It is nevertheless intriguing that the Bekenstein-Hawking entropy appears here in a role analogous to the central charge in conformal field theory. In fact the thermodynamics of 2-dimensional black holes, or correspondingly dimensionally reduced ones, is closely related tot the Virasoro algebra and its central charge \cite{mr-Virasoro}.}, but this is likely to change beyond the Einstein-Hilbert truncation.

Ultimately one could hope to establish, perhaps even at some level of rigor, the existence of a $C$-function in 4D asymptotically safe Quantum Einstein Gravity by using a combination of WISS and FRGE in order to derive bounds for the `disturbing' term in the equation \eqref{eqn:cm_mot_045} for $\partial_k\cfunc_k$.

Besides its obvious relevance to the global structure of the RG flow, this will also allow us to use the monotonicity of $\cfunc_k$ as a powerful criterion and easy to apply practical test for assessing the reliability of truncations or other approximations.

Furthermore, it would be interesting to find further examples, based on other theory spaces which admit a simple map from $\EAA_k$ to some $\cfunc_k$.
It also remains to clarify the precise relationship between the existing $c$- and $a$-theorems on the one side, and the present framework on the  other.
We shall come back to this question elsewhere \cite{wip}.

In the existing work on the known (generalized) $c$-theorems in 2, 3, and 4 dimensions, usually no reference is made to a (bare) action and fields it depends on; only the existence of a local energy momentum tensor is assumed.
In the present approach the emphasis is instead on (effective) action functionals depending on a set of fields that is fixed from the start.
However, recalling the discussion (of the `reconstruction problem') in ref. \cite{elisa1} it becomes clear that this clash is much less profound than it seems:
In the EAA approach to Asymptotic Safety, $\Gamma_k[\,\cdot\,]$ should primarily be seen as a generating function (or functional) for a set of $n$-point functions.
Hereby the field arguments of the EAA serve a purely technical purpose, and in general {\it the relationship between those field arguments and the fundamental physical degrees of freedom (dof) whose quantization would result in a given RG trajectory is at best a highly indirect one.}

The main reason is that in the case of an asymptotically safe theory $\Gamma_{k\rightarrow \infty}$, i.e. the fixed point action is extremely complicated, highly non-linear, contains higher derivatives, and is nonlocal probably.
Furthermore, $\Gamma_{k\rightarrow\infty}$ is a gauge-fixed action, but it will {\it not} be of the familiar form $S+S_{\text{gauge-fixing}}+S_{\text{ghost}}$ in general, with some invariant action $S$ and a quadratic, second derivative ghost action $S_{\text{ghost}}$, as it is the case when one applies the Faddeev-Popov trick.
This is another issue that complicates the identification of the physical contents of the fixed point theory.
In perturbation theory, $\Gamma_{k\rightarrow\infty}$ which is essentially the same as the bare action $S$, contains only a few relevant field monomials and only second derivative terms.
As a result, there is basically a one-to-one relation between degrees of freedom and fields.
In Asymptotic Safety, rather than an ad hoc input, the theory's bare action, or what comes closest to it, $\Gamma_{k\rightarrow\infty}$, is the result of a complicated nonperturbative evaluation of the fixed point condition.
As the structure of propagating modes is crucially affected by higher derivative and nonlocal terms, it is the fixed point condition that decides about the nature of the underlying dof's.
To identify them, a phase-space functional integral of the form $\int\mathcal{D}x\int \mathcal{D}\pi\,\exp\left(i\int \pi_j \dot{x}_j-H[\pi,x]\right)$ must be found which reproduces the RG trajectory obtained from the FRGE.
We can then read off canonically conjugate pairs $x_j$, $\pi_j$ and the (local) Hamiltonian $H[\pi,x]$ which governs their bare dynamics. 
To bring the original functional integral \cite{elisa1} $\int \mathcal{D}\hat{\Phi}\,e^{-\Gamma_{k\rightarrow\infty}}$ to this form, field redefinitions and the introduction of further, or different fields will be necessary in order to remove nonlocal and higher-derivative terms.
It is quite conceivable that there is more than one set $\{\pi_j,\,x_j\}$ and Hamiltonian $H$ that reproduces a given RG trajectory.
In this case we would say that the quantum theory defined by the latter has `dual' descriptions employing different (bare) actions and fields.
As yet, not much work has been devoted to this `reconstruction problem', see however ref. \cite{elisa1} for a first step.

{\bf\noindent (D)}
To close with, we mention another intriguing aspect of the integrated $\cfunc$-theorem \eqref{eqn:cm_con01} which deserves being investigated further, namely its connection to the hypothesis of the {\bf `$N$-bound'} which is due to Banks \cite{Banks} and, in a stronger form, to Bousso \cite{Bousso}.
In Bousso's formulation, the claim is that in any universe with a positive cosmological constant, containing arbitrary matter that even may dominate at all times, the observable entropy $S_{\rm obs}$ is bounded by $S_{\rm obs}\leq 3\pi \slash G \Lambda\equiv N$.

Here $S_{\rm obs}$ includes both matter and horizon entropy, but excludes entropy that cannot be observed in a causal experiment.
As for the notion of an `observable entropy', it is identified \cite{Bousso} with the entropy contained in the causal diamond of an observer, i.e. the spacetime region which can be both influenced and seen by the observer.
It is bounded by the past and future light cones based at the endpoints of the observer's world line.

Remarkably, while the number $N$ equals the Bekenstein-Hawking entropy of {\it empty} de Sitter space, the bound is believed to apply in presence of arbitrary matter, and for arbitrary spacetimes with $\Lambda>0$, which not even asymptotically need to be de Sitter.\footnote{In \cite{Bousso} the original requirement \cite{Banks} of spacetimes that are asymptotically de Sitter has been dropped.}

Given the methods developed in the present paper the intriguing possibility arises to check whether the $N$-bound holds in asymptotically safe field theories and to tentatively identify $N$ with $\nBound\equiv \cfunc^{\UV}-\cfunc^{\IR}$.
In principle we have all tools available for a fully non-perturbative test that treats gravity at a level well beyond the semi-classical approximation.
We would have to add matter fields to the truncation ansatz \cite{vacca,matterPerc} and include for all types of fields the corresponding (Gibbons-Hawking, etc.) surface terms that are needed on spacetimes with a non-empty boundary \cite{daniel1,Jac-Satz}.

Originally the $N$-bound grew out of string theory based arguments which hinted at the possibility of a {\bf `$\Lambda$-$N$-connection'} \cite{Banks,Bousso}.
It would be such that all universes with a {\it positive} cosmological constant are described by a fundamental quantum theory which has only a {\it finite} number of degrees of freedom, and that this number is determined by $\Lambda$.%
\footnote{For a similar discussion in Loop Quantum Gravity see ref. \cite{Smolin}.}

Is there a corresponding `$\Lambda$-$\nBound$-connection' in asymptotically safe field theory?
For pure gravity we can answer this question in the affirmative already now: 
The fundamental quantum field theory is defined by the Asymptotic Safety construction with an RG trajectory of the type \Rmnum{3}a, we get the required positive cosmological constant in the IR, $\Lambda_{\crg}$, which in turn fixes the number of degrees of freedom, here to be interpreted as $\cfunc^{\UV}-\cfunc^{\IR}$, by $\nBound=3\pi\slash G_{\crg}\Lambda_{\crg}<\infty$.
Recalling our discussion of the \Rmnum{1}a and \Rmnum{2}a trajectories in section \ref{sec:3-03} and appendix \ref{app:cmB} we can now easily understand what is special about a strictly {\it positive} $\Lambda$, and why the connection fails for a negative or vanishing classical cosmological constant:
in the latter cases, we found that  $\nBound$ is {\it not} finite.

%


\subsection*{Acknowledgment}
\noindent We are grateful to A.~Codello, G.~D'Odorico, C.~Pagani, R.~Percacci, and F.~Saueressig for helpful discussions.

\clearpage
 \begin{appendix}
 \section*{Appendix}
  \section{The special status of Faddeev-Popov ghosts}\label{app:cmA}

In subsection \ref{subsec:2-03} we argued that Faddeev-Popov ghosts, even though they contribute with a negative sign to the supertrace on the RHS of the flow equation, do not destroy the pointwise monotonicity of the EAA when they are the only fields present with odd Grassmann parity.
The reason was that the ghosts are merely a way of representing the Faddeev-Popov determinant, $\det(\mathcal{M})$, the functional integral actually being $Z=\int \mathcal{D}g \, \det(\mathcal{M})\, e^{-S_{\text{gf}}}e^{-S}$, wherein the gauge fixing term and the determinant effectively restrict the integration over all metrics to an integral over the gauge orbit space of metrics modulo diffeomorphisms.
If we had parametrized the latter directly we were dealing with a purely Grassmann-even integral \cite{Mottola}, which when modified by an IR cutoff, obviously leads to a pointwise monotone EAA as the gauge orbit space is independent of $k$.
In this appendix, we briefly indicate how this general argument can be made concrete.

We start out from the functional integral that has been gauge-fixed \`{a} la Faddeev-Popov, but without IR cutoff yet.
Then, after the usual background split, we perform a partial\footnote{It is `partial' in that the vector field $V_{\mu}$ is not decomposed further here as this is usually done, setting $V_{\mu}=V_{\mu}^T+\bZ_{\mu}\sigma$ with $\bg^{\mu\nu}\bZ_{\mu}V_{\nu}^T=0$.} 
TT-decomposition of the fluctuation field,
\begin{align}
\flcb_{\mu\nu}=\flcb_{\mu\nu}^T+ \left(\bZ_{\mu}V_{\nu}+\bZ_{\nu}V_{\mu}-\frac{2}{d}\bg_{\mu\nu} \bZ^{\alpha}V_{\alpha}\right)+\frac{1}{d}\bg_{\mu\nu}\flcb
\label{eqn:cm_ap0A_01}
\end{align}
with $\bZ^{\mu}\flcb^T_{\mu\nu}=0$, $\bg^{\mu\nu}\flcb_{\mu\nu}^T=0$, $\flcb\equiv \bg^{\mu\nu}\flcb_{\mu\nu}$.
Henceforth we interpret $\mathcal{D}g_{\mu\nu}$ as $\mathcal{D}\flcb_{\mu\nu}^T\mathcal{D}\flcb \mathcal{D}V_{\mu}$.
Furthermore, we write the Faddeev-Popov determinant as $\det\left(\mathcal{M}[g,\bg]\right)\equiv \det\left(M\right)e^{-S_1}$ with $M\equiv\mathcal{M}[\bg,\bg]$.
Diagrammatically speaking the action $S_1$ contains the ghost-antighost-graviton vertices and $M^{-1}$ is the `free' ghost propagator.
Thus 
\begin{align}
Z[\bg]=\int\mathcal{D}\flcb_{\mu\nu}^T\mathcal{D}\flcb \mathcal{D}V_{\mu}\, \det(M)e^{-S_{\text{gf}}}e^{-\widetilde{S}}
\end{align}
with $\widetilde{S}\equiv S+S_1$ and the representation $\det(M)=\int \mathcal{D}\Ghx\mathcal{D}\GhAx\, \exp \int \GhAx M \Ghx$.

Next, consider the family of gauge fixing functions
\begin{align}
F_{\mu}&= \bZ^{\nu}\flcb_{\mu\nu}-\varpi \bZ_{\mu}\flcb\nonumber\\
&= \bZ^2 V_{\mu}+ \left(1-\frac{2}{d}\right)\bZ_{\mu}\bZ^{\nu}V_{\nu}+\SRb_{\mu}^{\nu} V_{\nu}+ \left(\frac{1}{d}-\varpi\right)\bZ_{\mu}\flcb
\label{eqn:cm_ap0A_02}
\end{align}
As for the parameter $\varpi$, we choose the `un-harmonic' gauge \cite{anharmRoberto,frankmach}, setting $\varpi=1\slash d$. (The harmonic gauge has $\varpi=1\slash2$ instead.)
This gauge leads to a remarkable conspiracy of the gauge fixing action, $S_{\text{gf}}=\frac{1}{2\alpha} \int \md^d x\sqrt{\bg}\, \bg^{\mu\nu} F_{\mu}F_{\nu}$, and the ghost action at $\flcb_{\mu\nu}=0$, $S_{\text{gh}}=\int \md^d x\sqrt{\bg}\, \GhAx_{\mu} \, {M^{\mu}}_{\nu}\Ghx^{\nu}$.
One finds that $S_{\text{gf}}$ is a bilinear form in $V_{\mu}$  (and only $V_{\mu}$!) whose kernel is precisely the {\it square of the inverse ghost propagator\footnote{This `magic' property has been discovered by F. Saueressig et al. \cite{frank-square, RG-machine}.}}:
\begin{align}
S_{\text{gf}}=\frac{1}{2\alpha}\int \md^d x\sqrt{\bg}\, V_{\mu} \, {M^{\mu}}_{\alpha} {M^{\alpha}}_{\nu}\, V^{\nu}
\label{eqn:cm_ap0A_03}
\end{align}
In this gauge, ${M^{\mu}}_{\nu}\equiv {\mathcal{M}[\bg,\bg]^{\mu}}_{\nu}=\bZ^2{\delta^{\mu}}_{\nu}-(1-2\slash d) \bZ^{\mu}\bZ_{\nu}-{\SRb^{\mu}\,\!}_{\nu}$.
As a consequence, the integral over the ghosts, producing the determinant $\det(M)$, when combined with $e^{-S_{\text{gf}}}$, yields a Dirac $\delta$-functional in the limit $\alpha\rightarrow0$:
\begin{align}
\det(M)e^{-\frac{1}{2\alpha}\int V M^2 V } \rightarrow \delta[V]
\label{eqn:cm_ap0A_04}
\end{align}
It satisfies $\int \mathcal{D}V\,\delta[V] \rightarrow 1$.
That $\delta[V]$ is indeed correctly normalized, up to a constant, follows from $\det(M)\int \mathcal{D}V\, e^{-\frac{1}{2\alpha}\int V M^2 V}=\det(M) \det^{-1\slash 2}(M^2)=\det(M)\det^{-1}(M)=1$.

As a result, the limit $\alpha\rightarrow0$ simplifies the integral for $Z$ quite considerably:
after integrating over $V_{\mu}$ and using \eqref{eqn:cm_ap0A_04} we are left with 
\begin{align}
Z=\int \mathcal{D}\flcb^{T}_{\mu\nu}\,\mathcal{D}\flcb \, \exp\left(-\widetilde{S}[\flcb_{\mu\nu}^T+d^{-1}\bg_{\mu\nu}\flcb;\bg_{\mu\nu}]\right)
\label{eqn:cm_ap0A_05}
\end{align}
This functional integral is manifestly over fields of even Grassmann parity only.
So when we go through the usual procedure and define the associated EAA, the derivation of $\partial_k \EAA_k\geq0$ in the main part of this paper applies to it, provided the above exact compensation of the ghost and $V_{\mu}$ contributions persists in presence of an IR cutoff.
While this is not the case for a generic cutoff, it has been shown \cite{frank-square} that if the cutoff operators $\mathcal{R}_k$ of the ghost and metric fluctuations, respectively, are appropriately related, which always can be achieved, the compensation does indeed persist.
For further details the reader is referred to \cite{frank-square}.

Thus we have shown that (at the very least) when the ghosts are the only Grassmann-odd fields it is in principle always possible to set up the gauge fixing and ghost sector of the EAA and its FRGE in such a way that $\partial_k \EAA_k\geq 0$ holds true  pointwise.

The various sets of beta-functions studied in this paper were {\it not} obtained using this very special set-up for the gauge-fixing and ghost sector.
However, as the truncations considered here anyhow neglect all RG effects in this sector we have the freedom to use any gauge at this level of accuracy since this should not lead to an extra error.
A similar remark applies to the choice of the cutoff operators.
  \section{The trajectory types \Rmnum{1}a and \Rmnum{2}a} \label{app:cmB}
In this appendix we evaluate $\cfunc_k=\EAA_k^{\text{grav}}[0;\bg_k^{\scon}]$ along type \Rmnum{1}a and \Rmnum{2}a trajectories in $d=4$ for the bi-metric Einstein-Hilbert truncation.
The analysis parallels to some extent the one in subsection \ref{sec:3-03} for the type \Rmnum{3}a case.
Since in the \Rmnum{1}a and \Rmnum{2}a cases the cosmological constant $\Kk^{(1)}$ turns zero or even negative at some scale the corresponding self-consistent background undergoes a topological change, from  $S^4$ to  flat Euclidean space $R^4$ and to $H^4$, respectively.
While for the separatrix, the type \Rmnum{2}a trajectory,  the change from $S^4$ to $R^4$  happens in the limit $k\rightarrow0$ only, the type \Rmnum{1}a solutions have a negative $\Kk^{(1)}$ at finite scales $k<k_{\text{sing}}$ already.
At the transition  point $k=k_{\text{sing}}$ equation \eqref{eq:cm_nr01} is no longer valid and $\cF_k$ diverges: $\lim_{k\searrow k_{\text{sing}}}\cF_k=\infty$.
For $k<k_{\text{sing}}$ a different formula for $\cfunc_k$, employing a new background configuration, could be derived.
We shall not do this here and rather restrict our attention to the subspace of $\mathcal{T}$ with $\Kk^{(1)}>0$.

\begin{figure}[!ht]
\centering
\psfrag{c}[cm][0][1][90]{${\scriptscriptstyle k\partial_k 1\slash \cF_k}$}
\psfrag{b}[cm][0][1][90]{${\scriptscriptstyle k\partial_k  1\slash \cF_k}$}
\psfrag{a}{${\scriptscriptstyle k \slash m_{\text{Pl}}}$}
\psfrag{k}{${\scriptstyle k \slash m_{\text{Pl}}}$}
\psfrag{x}[r]{${\scriptstyle 1\slash \cF_k }$}        
 \subfloat{
 \psfrag{c}[cm][0][1][90]{${\scriptstyle 1\slash \cF_k }$}  
 \includegraphics[width=0.450\textwidth]{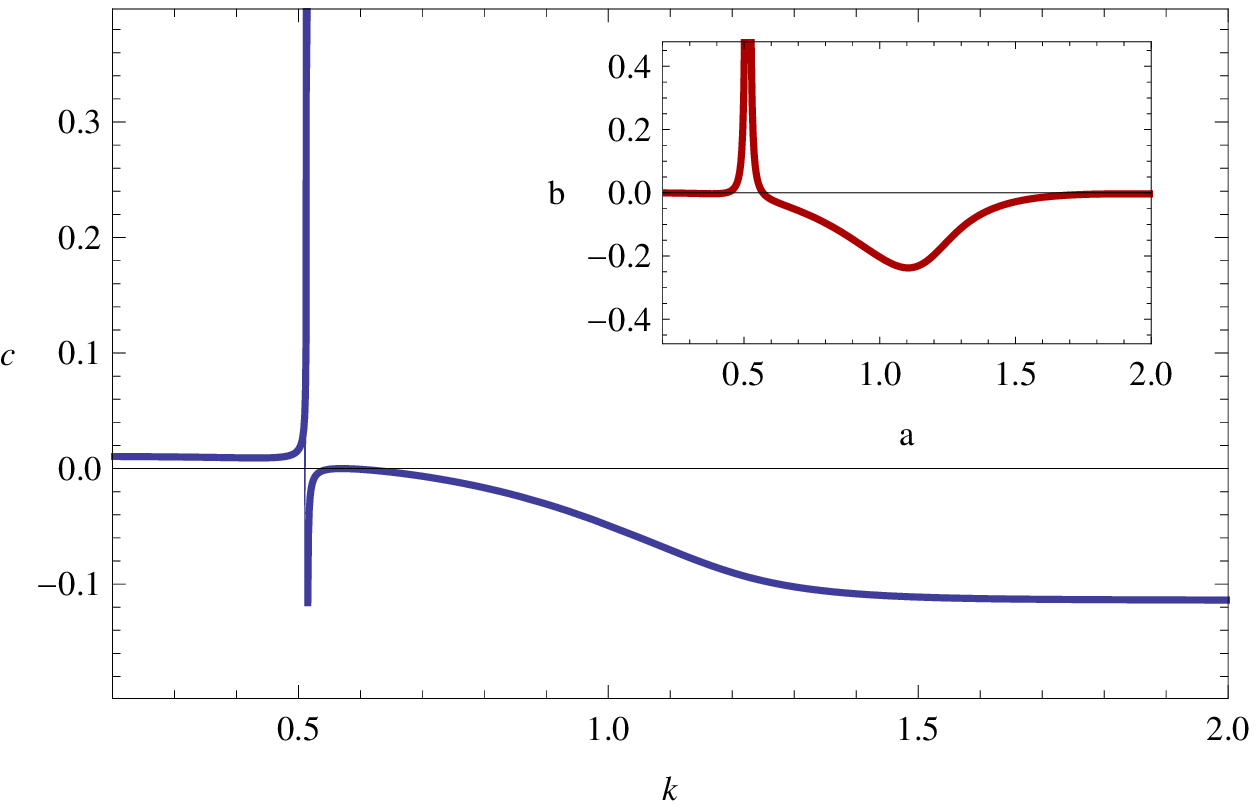}
 } 
 \hspace{0.04\textwidth} %
 \subfloat{%
\includegraphics[width=0.455\textwidth]{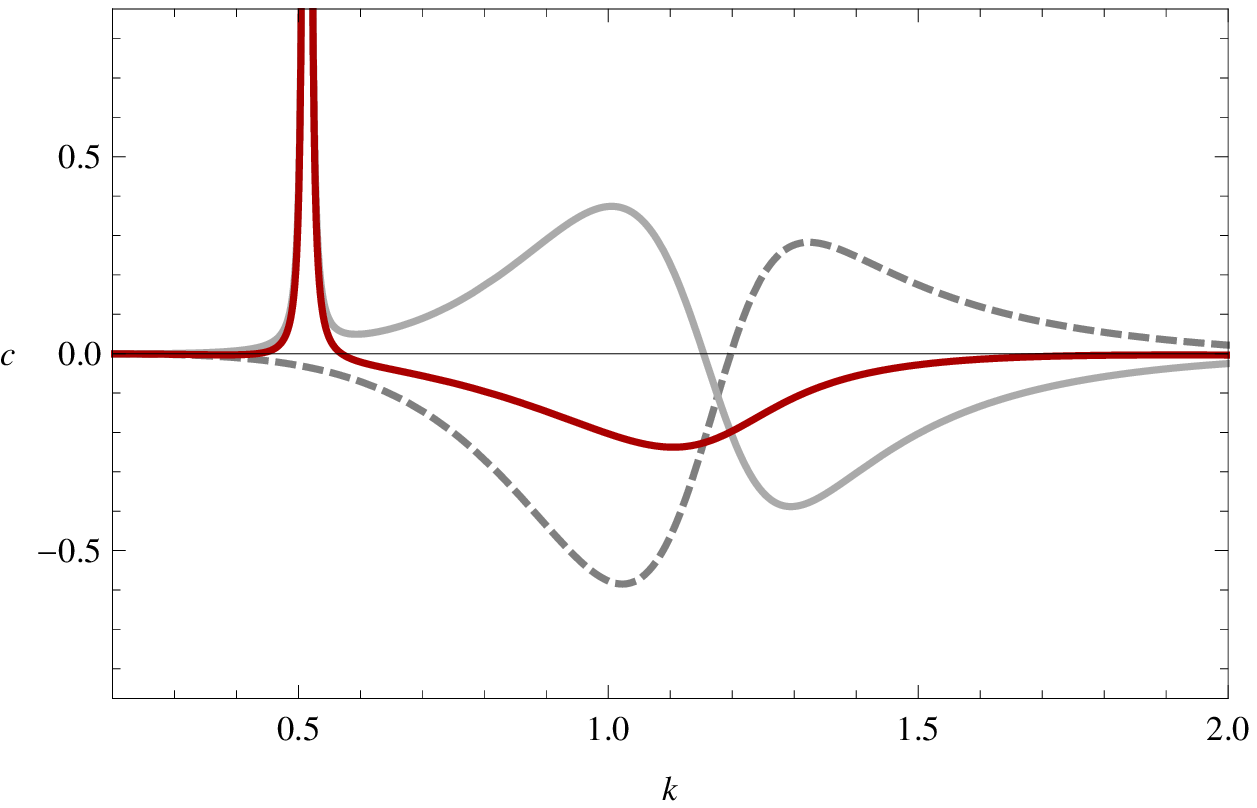}
}
\caption{The function $1\slash\cF_k$ and its derivative are shown for a typical bi-metric type \Rmnum{1}a trajectory which restores split-symmetry in the IR. 
Due to the sign flip of $\Kk_k^{(1)}$ near $k_{\text{sing}}\approx 0.5 m_{\text{Pl}}$, the function $1\slash \cF_k$ has a zero there and $\cfunc_k$ diverges.
As long as \eqref{eq:cm_nr01} is still valid, to $k>k_{\text{sing}}$, the function $\cF_k$ is seen to be monotone.
} \label{fig:cfuncCorrectedBMIa}
\end{figure}

\begin{figure}[!ht]
\centering
\psfrag{c}[cm][0][1][90]{${\scriptscriptstyle k\partial_k 1\slash \cF_k}$}
\psfrag{b}[cm][0][1][90]{${\scriptscriptstyle k\partial_k  1\slash \cF_k}$}
\psfrag{a}{${\scriptscriptstyle k \slash m_{\text{Pl}}}$}
\psfrag{k}{${\scriptstyle k \slash m_{\text{Pl}}}$}
\psfrag{x}[r]{${\scriptstyle 1\slash \cF_k }$}        
 \subfloat{
 \psfrag{c}[cm][0][1][90]{${\scriptstyle 1\slash \cF_k }$}  
 \includegraphics[width=0.450\textwidth]{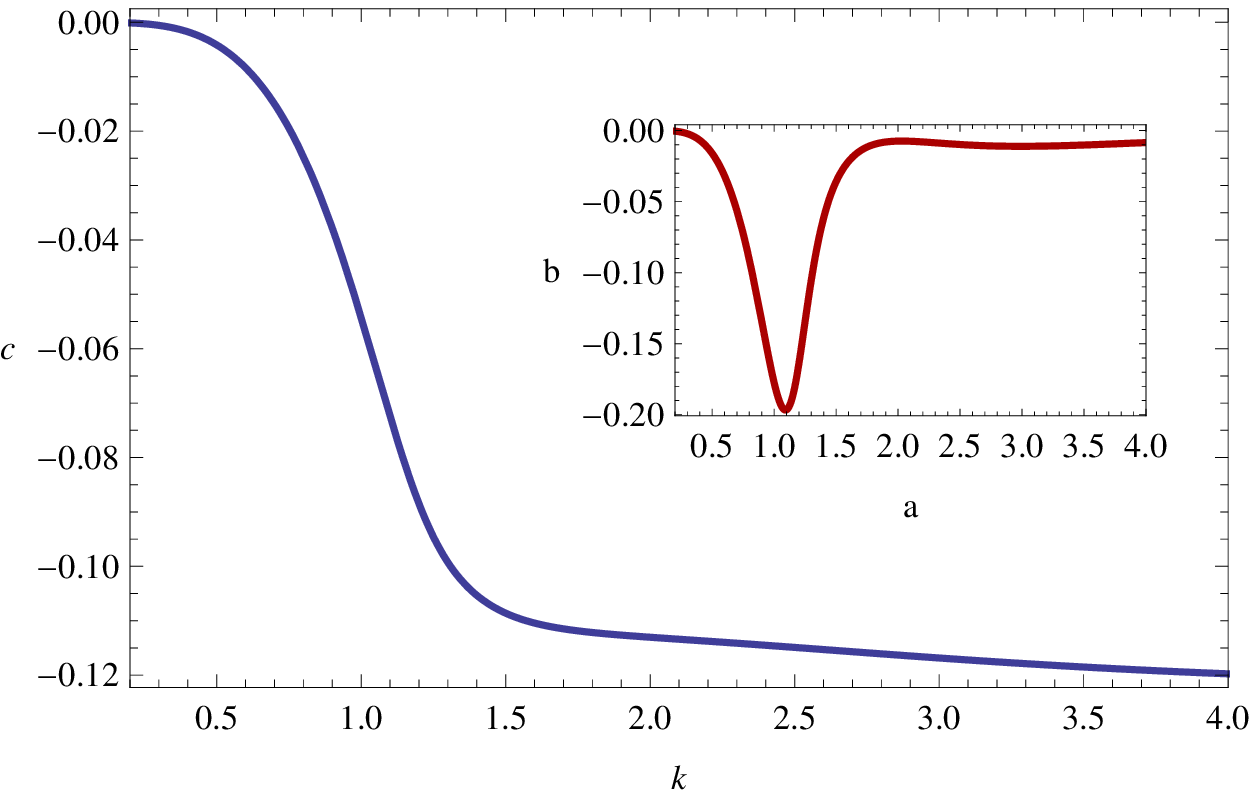}
 } 
 \hspace{0.04\textwidth} %
 \subfloat{%
\includegraphics[width=0.455\textwidth]{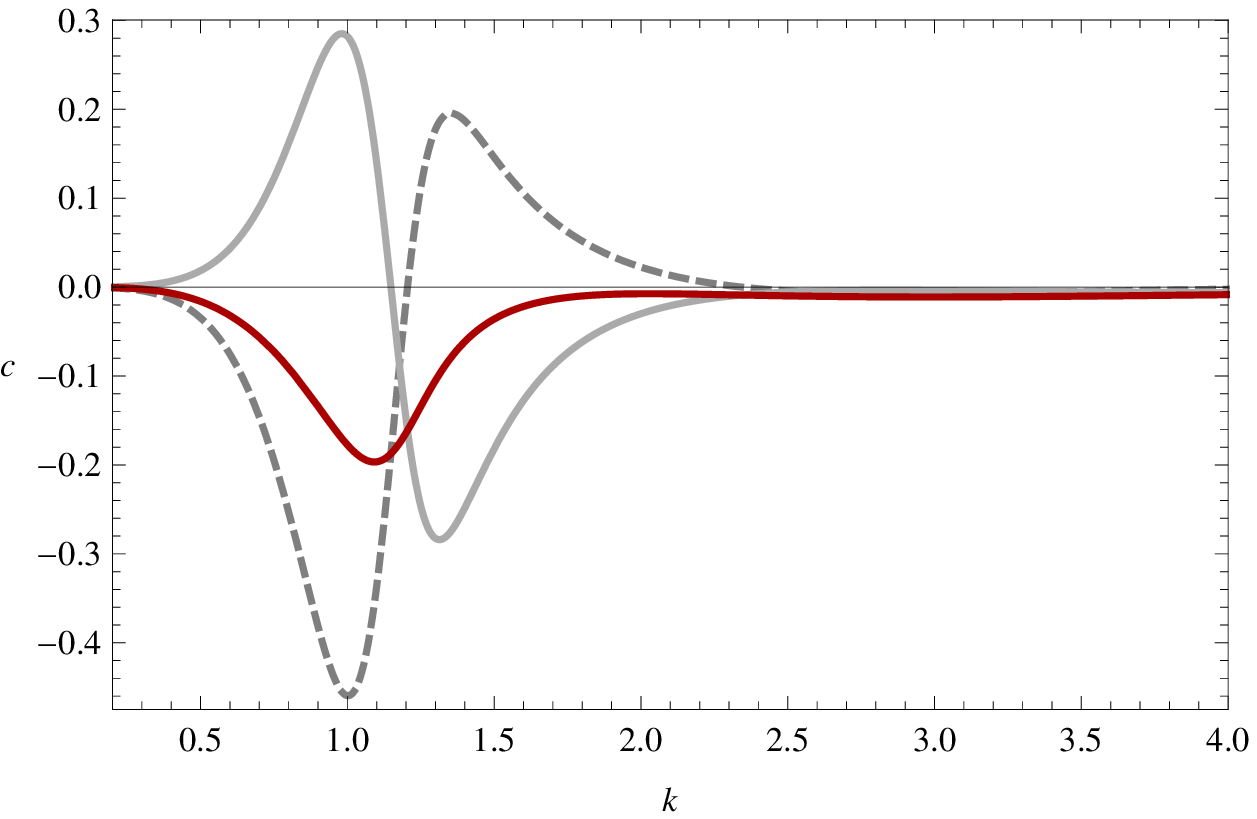}
}
\caption{
The function $1\slash \cF_k$ for the bi-metric type \Rmnum{1}a trajectory (separatrix) and its derivative.
Its properties are similar to the type \Rmnum{3}a results obtained in subsection \ref{sec:3-03}.
In particular $\cF_k$ is seen to be monotone.
The asymptotic topology change of the self-consistent background in the limit $k\rightarrow0$ is not directly visible in these plots.
It can be checked though that $\cF_k$ diverges for $k\rightarrow 0$ (and that it does not in the \Rmnum{3}a case).
} \label{fig:cfuncCorrectedBMIIa}
\end{figure}

The Figs. \ref{fig:cfuncCorrectedBMIa} and \ref{fig:cfuncCorrectedBMIIa} depict the explicit $k$-dependence of $1\slash \cF_k$ for a representative type \Rmnum{1}a trajectory and the unique \Rmnum{2}a trajectory, respectively, both for the case of restored split-symmetry in the IR.
The plots are based on the RG equations of [\Rmnum{2}]; the results with those from [\Rmnum{1}] are quite similar.
For the type \Rmnum{1}a trajectory in Fig. \ref{fig:cfuncCorrectedBMIa} the singularity of $\cF_k$ occurs at about $k_{\text{sing}}\approx 0.5 m_{\text{Pl}}$, and $\cF_k$ is seen to be perfectly monotone above this scale.

Along this trajectory, we integrated out $\nBound_{k_{\text{sing}},\infty}=\infty$ modes already before the end of the trajectory.
But as there is a nontrivial RG evolution also below $k_{\text{sing}}$, there are still further modes left to be integrated out.
Using a hyperbolic background we could count how many there are in some interval $[k_1,k_2]$ with $k_1<k_2<k_{\text{sing}}$.
But clearly there is no meaningful way of associating a finite number $\nBound_{0,\infty}$ to the complete trajectory as this was possible in the \Rmnum{3}a case.

For the symmetry-restoring bi-metric separatrix, the plot of $1\slash \cF_k$ is very similar to the case of the \Rmnum{3}a-trajectories discussed in the main part of the paper, see Fig. \ref{fig:cfuncCorrectedBMIIa}.
The only difference is that $1\slash \cF_k$ vanishes exactly at $k=0$, while $1\slash \cF_k$ was always nonzero for the type \Rmnum{3}a solutions.
So the separatrix is the marginal case where $\nBound=\infty$ is reached precisely at the IR-end point of the trajectory.


\begin{figure}[pt!]
\centering
\psfrag{B}[tc]{${\scriptscriptstyle  }$}
\psfrag{C}[tc]{${\scriptscriptstyle  }$}
\psfrag{g}[tc]{${\scriptscriptstyle \tg^{\background} }$}
\psfrag{l}[c]{${\scriptscriptstyle \KkB }$}
 \subfloat{%
  \psfrag{B}[tc]{${\scriptscriptstyle P_1 }$}
%
\label{fig:snapsIaA}\includegraphics[width=0.35\textwidth]{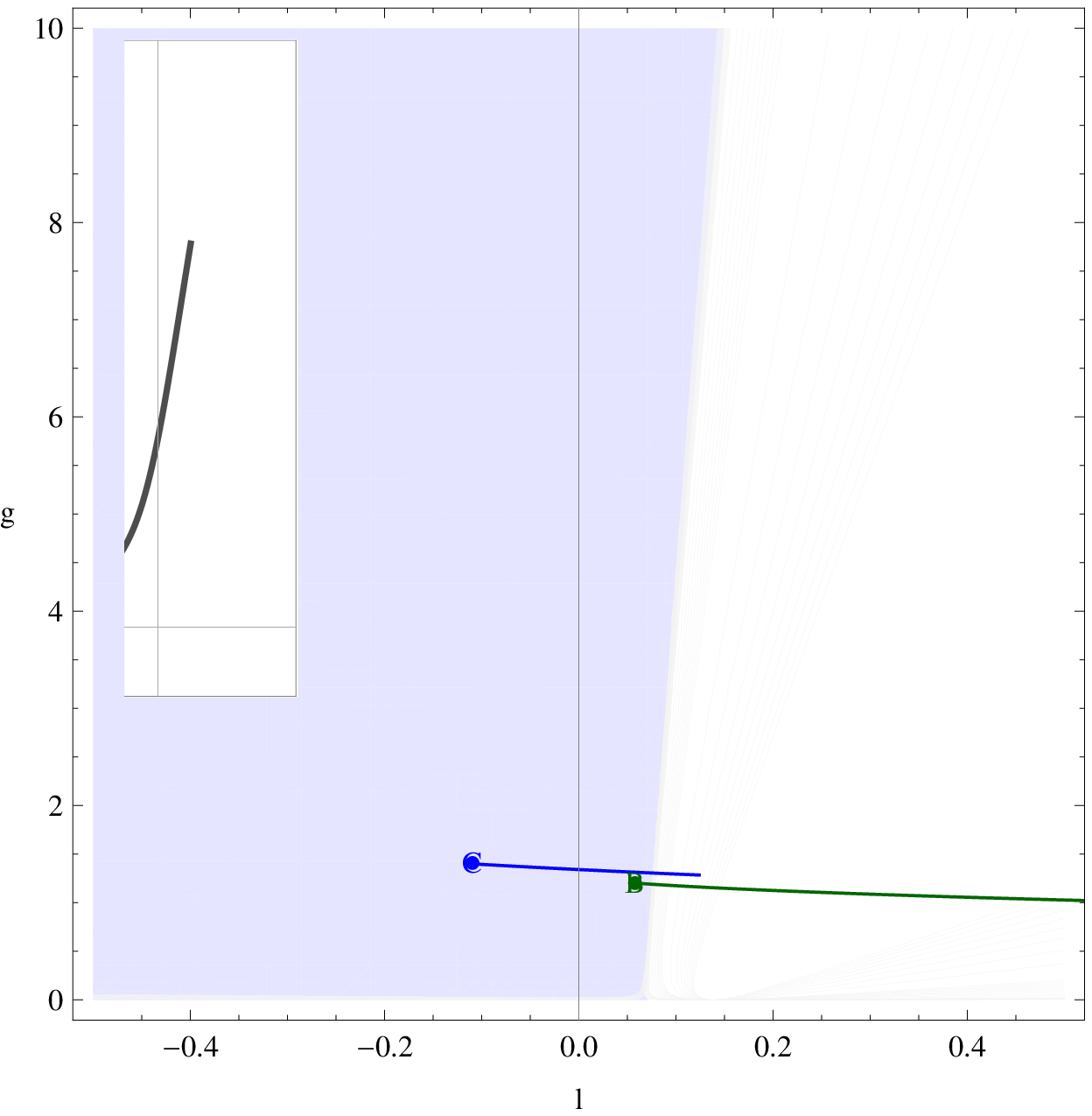}}
\hspace{0.05\textwidth}
 \subfloat{%
\label{fig:snapsIaB}\includegraphics[width=0.35\textwidth]{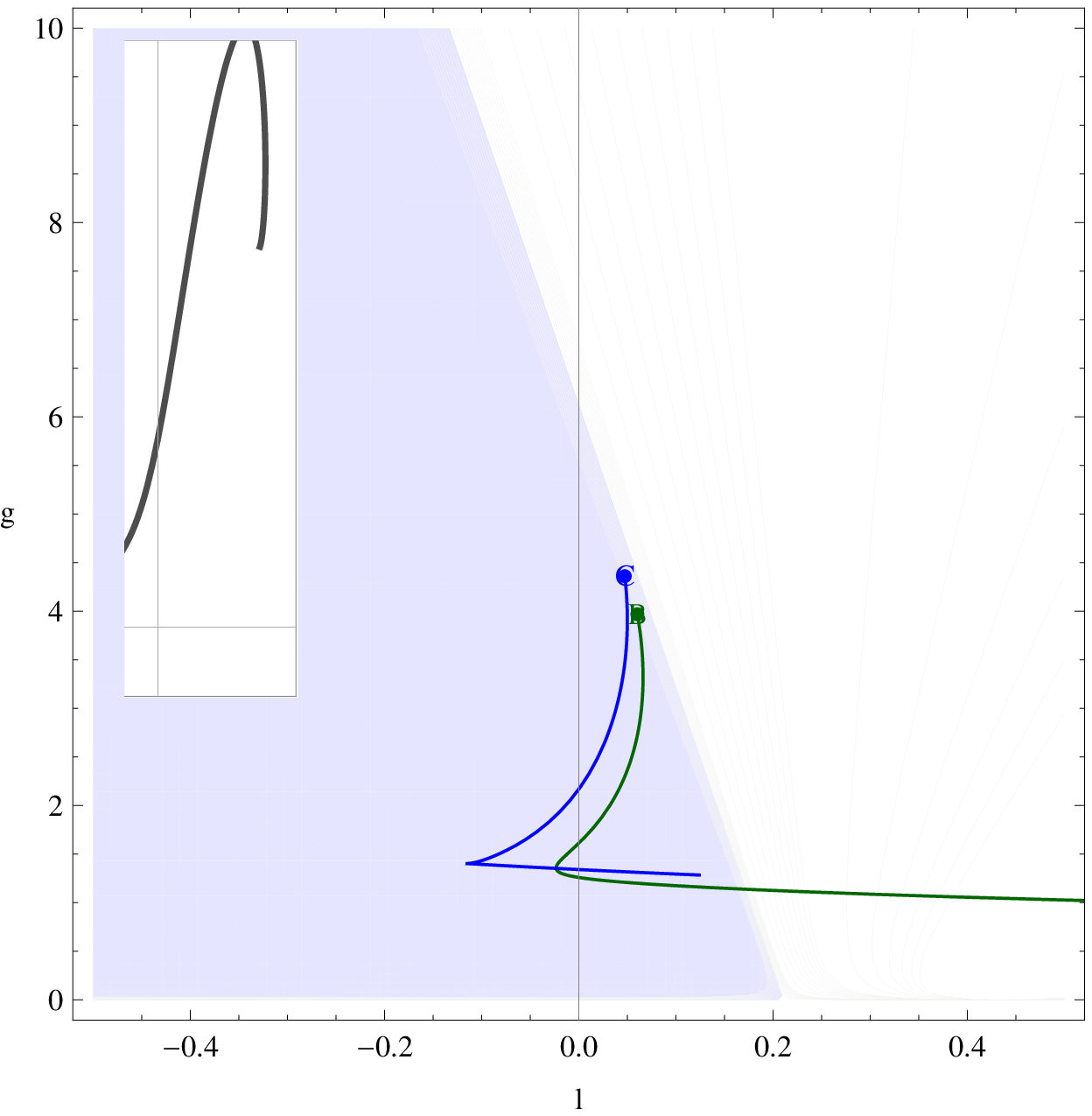}}
\hspace{0.05\textwidth}
 \subfloat{%
\label{fig:snapsIaC}\includegraphics[width=0.35\textwidth]{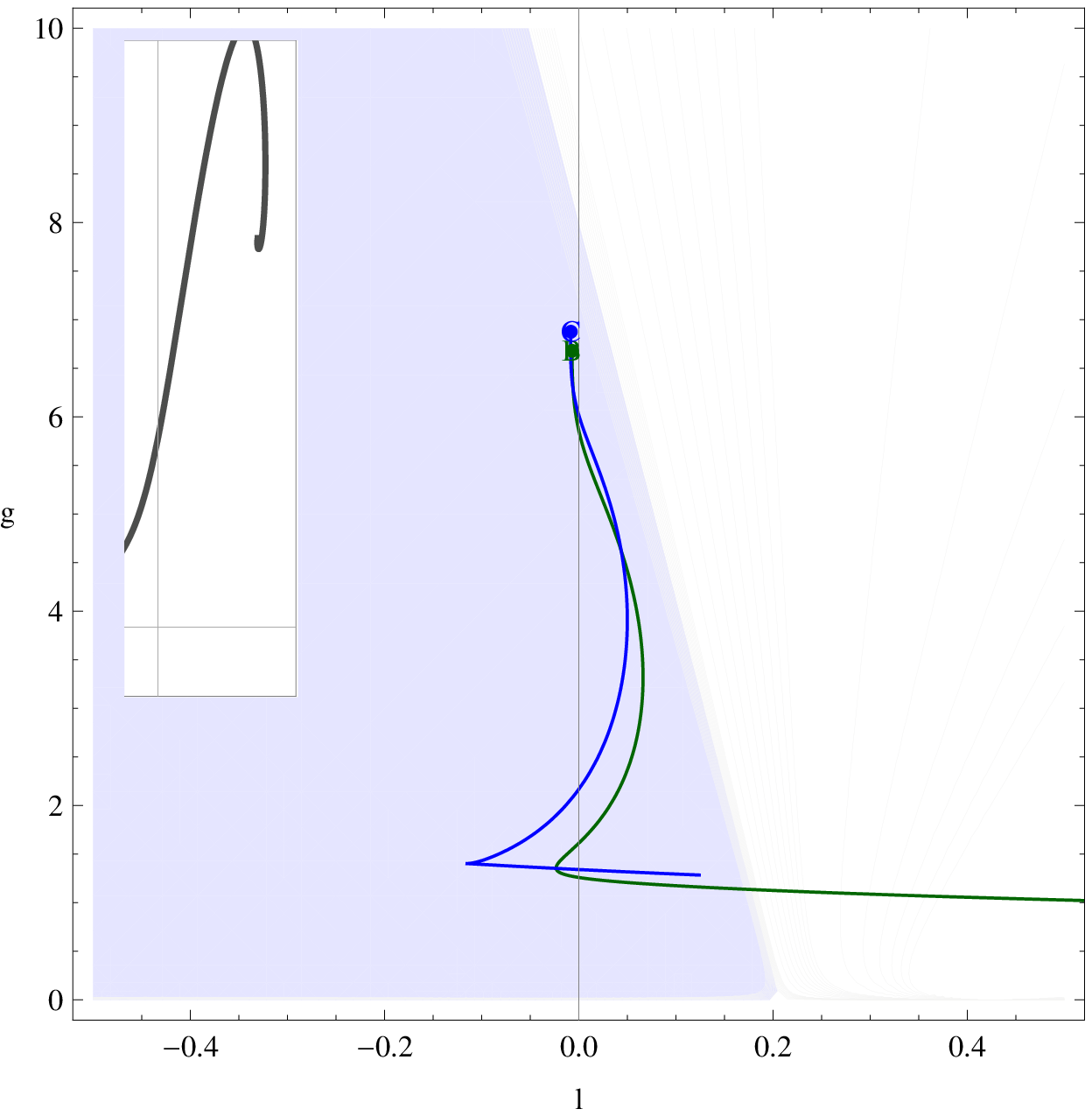}}
\hspace{0.05\textwidth}
 \subfloat{%
 \label{fig:snapsIaD}\includegraphics[width=0.35\textwidth]{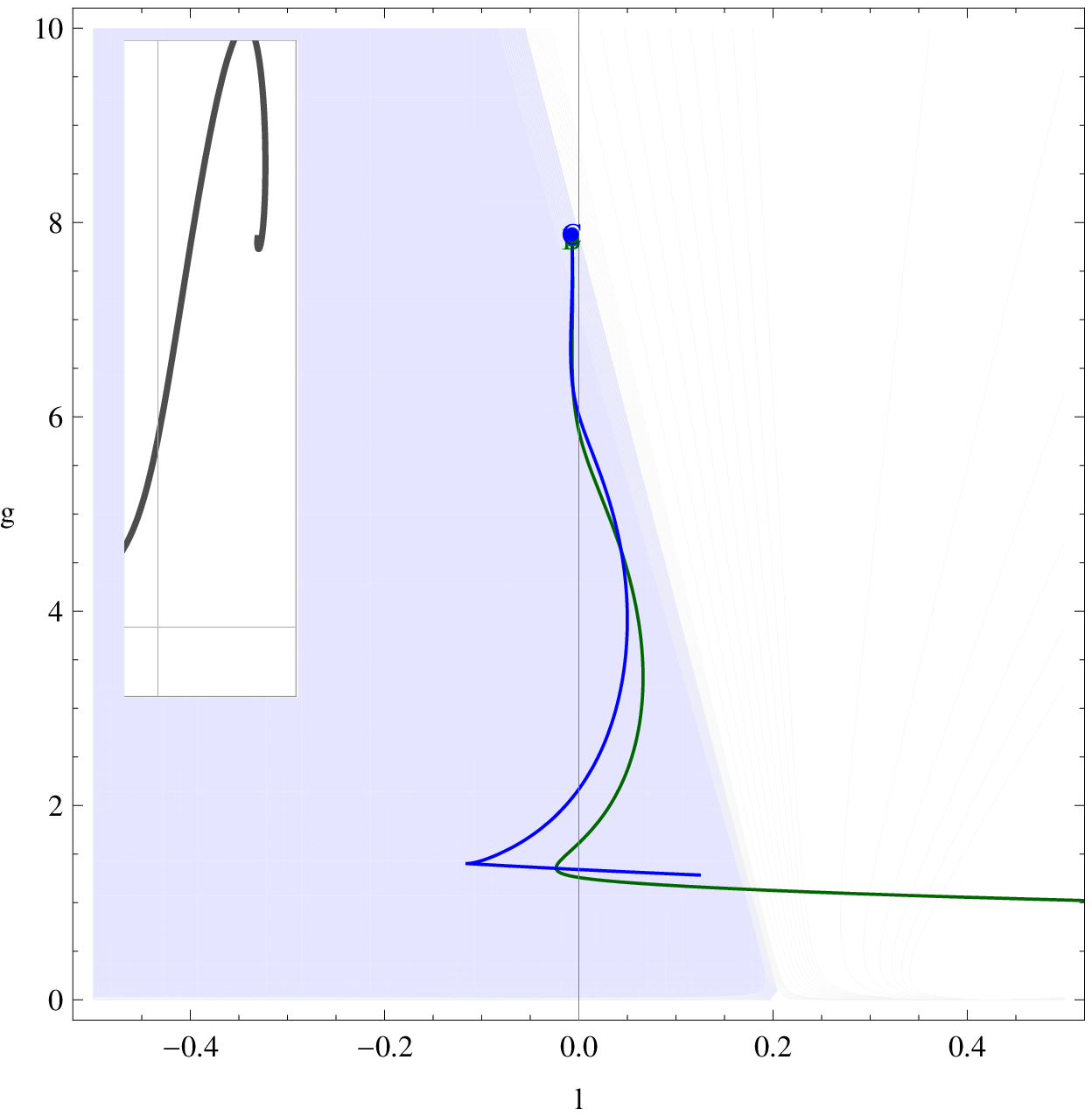}}
 \caption{The $\tg^{\background}$-$\KkB$-plane is shown in a series of subsequent `snapshots' at different RG-times which increase from the upper left to the lower right diagram.
  They are given by the maximum $k$-value of the incomplete dynamical {\it type \Rmnum{1}a} trajectory $k\mapsto (\tg^{\dyn}_k,\,\KkD_k)$ shown in the respective inset.
  The shaded regions corresponds to $\mathcal{T}_+^{\background}(k)$ at that particular time, so that every trajectory in the shaded (white) region will give rise to a  positive (negative) value of $k\partial_k \cfunc_k$ at the instant of time $k$.
  Furthermore, two different $\background$-trajectories that are evolved upward (towards increasing scales $k$) are shown at the corresponding moments.
  The one passing the point $P_1$ ($P_2$) is split-symmetry violating (restoring).
  The symmetry restoring trajectory starts its upward evolution close to $P_2$, the position of the running UV attractor \cite{daniel2}.
As long as $k>k_{\text{sing}}$, which is assumed here to avoid a topology change, this trajectory never leaves the shaded area, and thus its $\cfunc_k$-function is strictly monotone.
 This is different for the   trajectory through $P_1$:
 Attracted by the running UV-attractor, it is pulled into the shaded regime, thus unavoidably crossing the boundary of $\mathcal{T}_+^{\background}(k)$, which causes a sign flip of $\partial_k \cfunc_k$,  rendering $\cfunc_k$ non-monotone.
 }\label{fig:snapsIa}
\end{figure}

\begin{figure}[pt!]
\centering
\psfrag{B}[tc]{${\scriptscriptstyle  }$}
\psfrag{C}[tc]{${\scriptscriptstyle  }$}
\psfrag{g}[tc]{${\scriptscriptstyle \tg^{\background} }$}
\psfrag{l}[c]{${\scriptscriptstyle \KkB }$}   
 \subfloat{%
  \psfrag{B}[tc]{${\scriptscriptstyle P_1 }$}
  \psfrag{C}[tc]{${\scriptscriptstyle P_2 }$}
\label{fig:snapsIIaA}\includegraphics[width=0.35\textwidth]{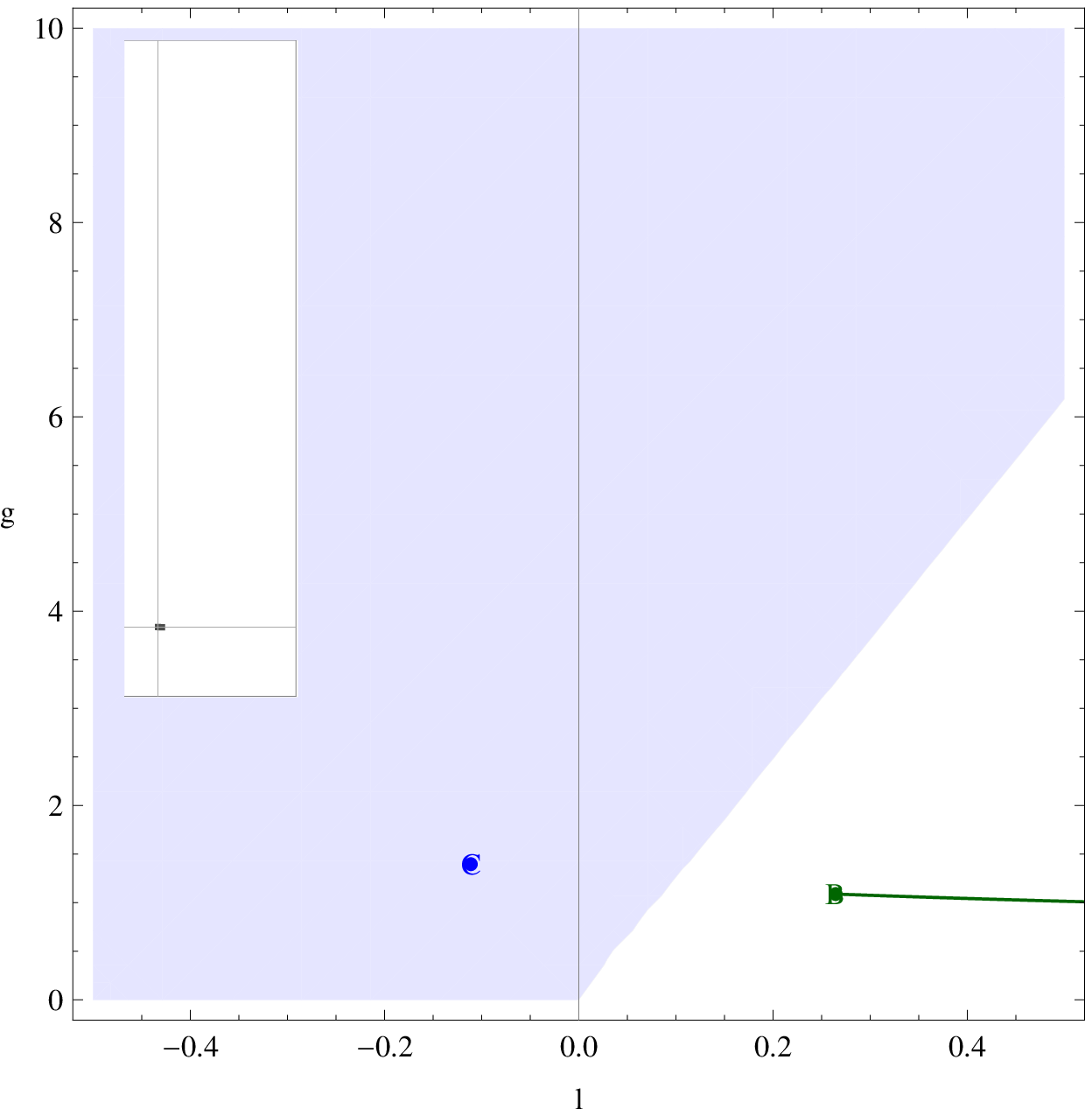}}
\hspace{0.05\textwidth}
 \subfloat{%
\label{fig:snapsIIaB}\includegraphics[width=0.35\textwidth]{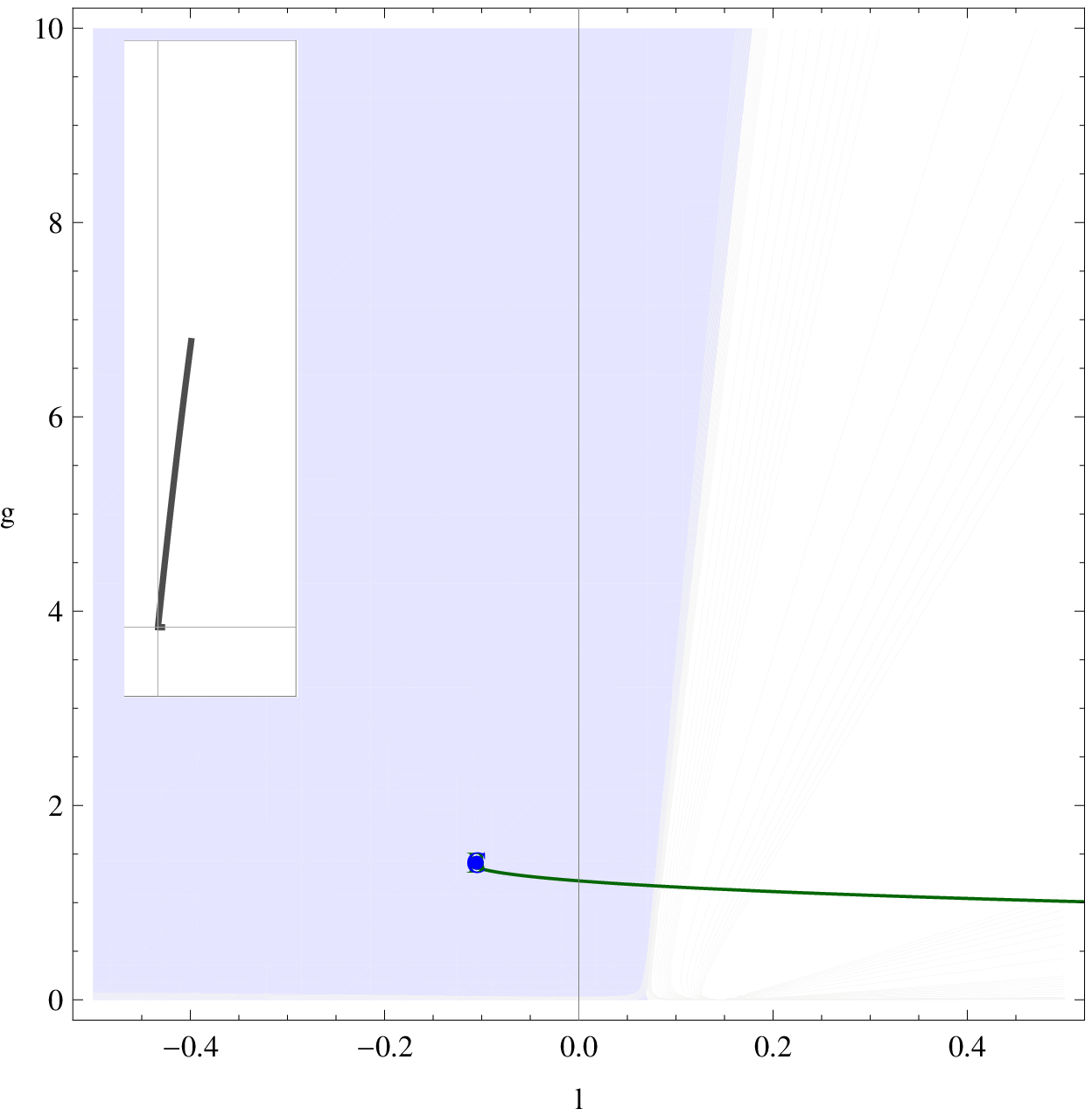}}
\hspace{0.05\textwidth}
 \subfloat{%
\label{fig:snapsIIaC}\includegraphics[width=0.35\textwidth]{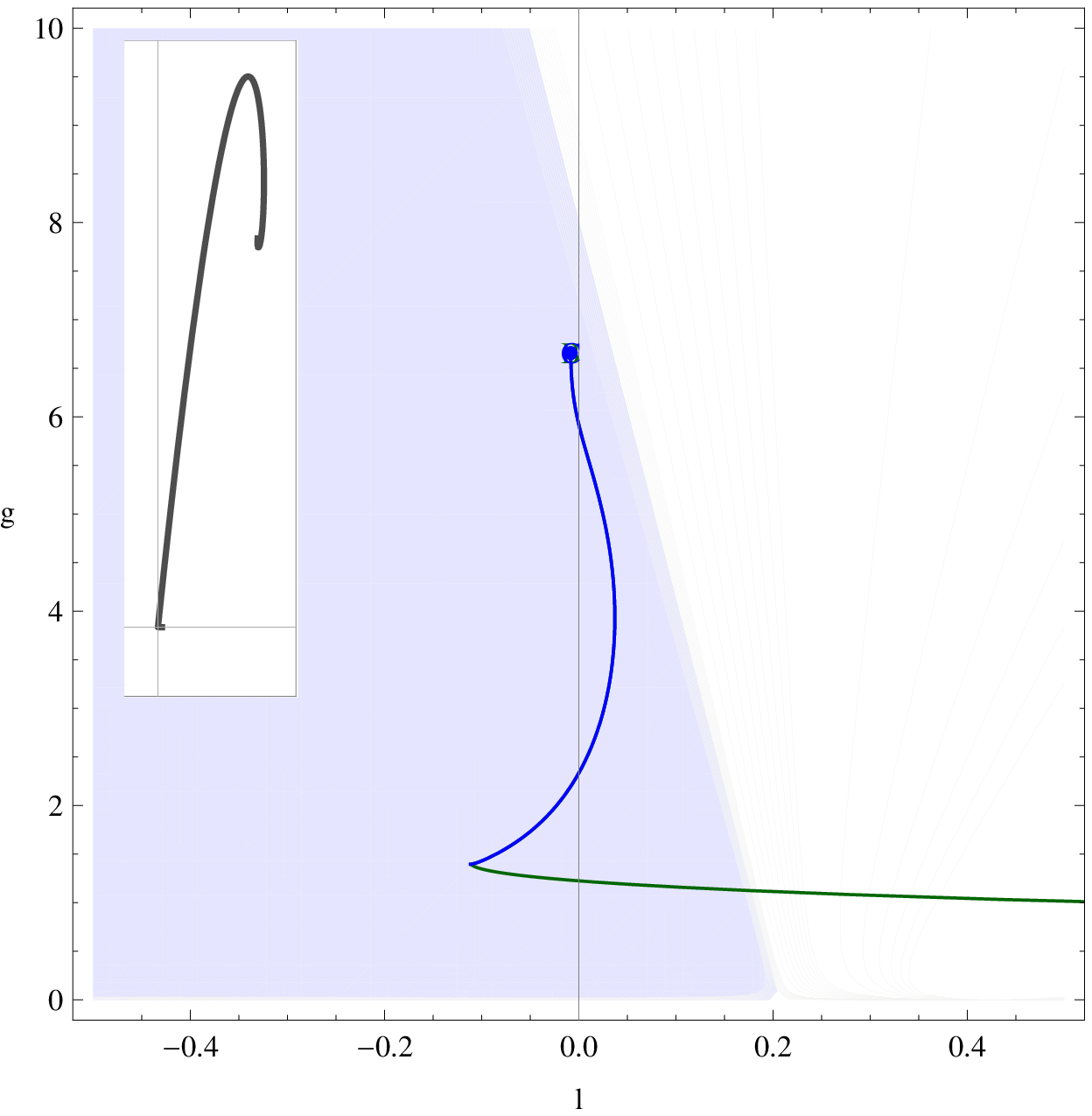}}
\hspace{0.05\textwidth}
 \subfloat{%
\label{fig:snapsIIaD}\includegraphics[width=0.35\textwidth]{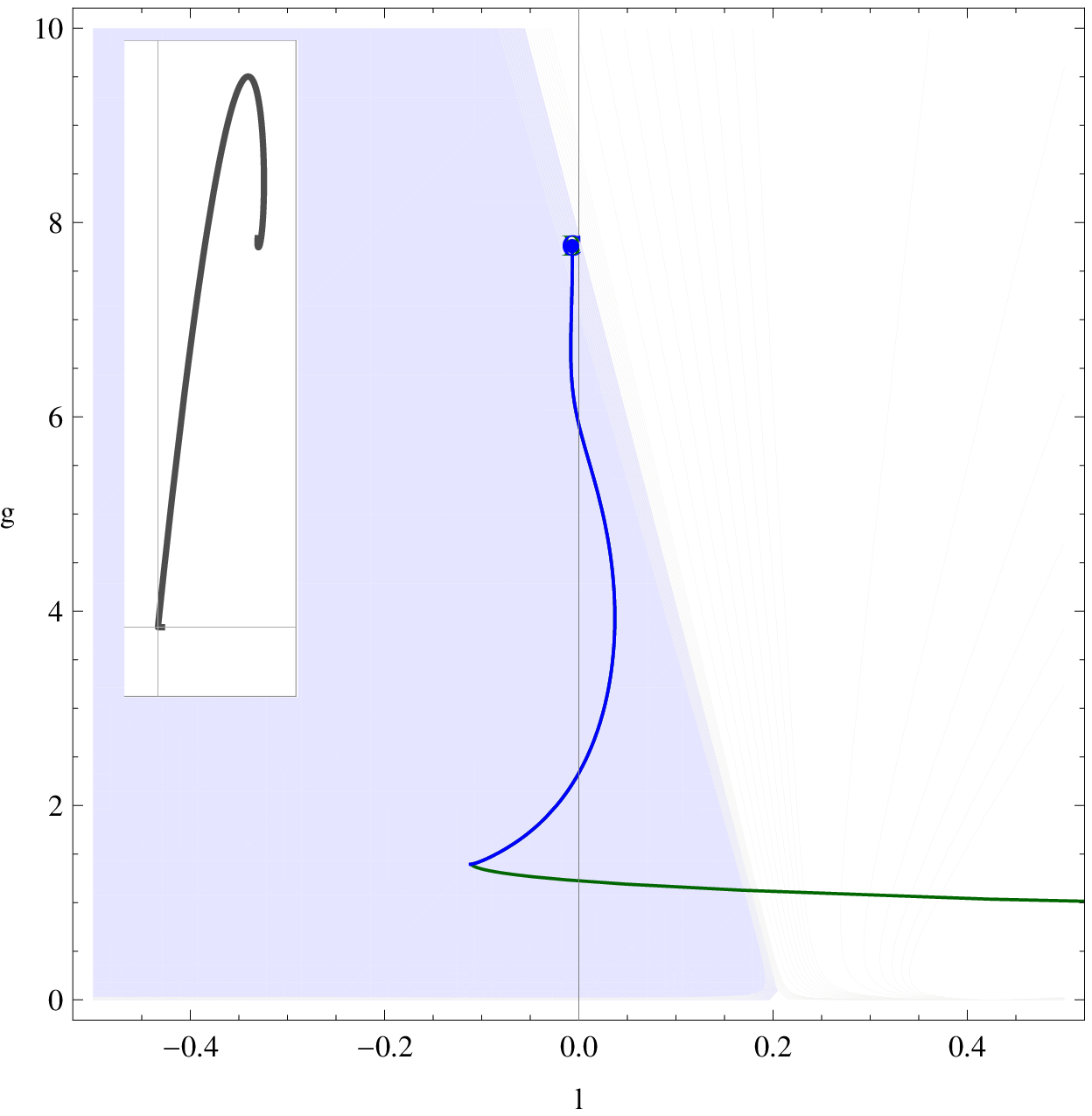}}
 \caption{A series of snapshots as in Fig. \ref{fig:snapsIa}, but for the {\it type \Rmnum{2}a} trajectory. The results are similar to those in subsection \ref{sec:3-03} for the type \Rmnum{3}a trajectories.}\label{fig:snapsIIa}
\end{figure}

In a series of snapshots, Figs. \ref{fig:snapsIa} and \ref{fig:snapsIIa} show the evolution of the RG trajectories in the background sector, from the IR to the UV, on the basis of typical \Rmnum{1}a and \Rmnum{2}a dynamical trajectories, respectively.
The shaded (white) regions partition the $\tg^{\background}$-$\KkB$-plane into subsets of positive (negative) slope $k\partial_k \cF_k$.
A crossing of the  corresponding boundary indicates a violation of the monotonicity of $\cfunc_k$.
While in the case of the separatrix the requirement of split-symmetry restoration in the IR  is sufficient to assure this condition, a more careful study is needed for type \Rmnum{1}a trajectories.
There $\Kk^{(1)}$ turns negative at finite scales $k$, and thus makes eq. \eqref{eq:cm_nr01} inapplicable.
In any case,  trajectories that break split-symmetry at $k=0$ are more vulnerable to monotonicity violation than those restoring it.
 \end{appendix}
 
\end{document}